\documentclass[12pt,fleqn]{article}
\usepackage{cite}
\usepackage{subeqn,amssymb,epsfig,graphicx}
\usepackage{times}
\newcommand{\newsection}[1]{\section{#1}\setcounter{equation}{0}}
\renewcommand{\thefootnote}{\fnsymbol{footnote}}

\def\unit{\leavevmode\hbox{\small1\kern-3.6pt\normalsize1}}
\def \text{\mathrm}

\def \klpn{\klong\to\p^0\n\bar{\n}} 
\def \klmumu{\klong\to \m^+\m^-} 
\def \kpnn{K^+\to \p^+ \n\bar{\n}}

\newcommand{\Li}[1]{{\rm Li}_2{\left(#1\right)}}
\def \be{\begin{equation}}
\def \ee{\end{equation}}
\def \bs{\begin{subequations}}
\def \es{\end{subequations}}
\def \bea{\begin{eqnarray}}
\def \eea{\end{eqnarray}}
\def \ben{\begin{enumerate}}
\def \een{\end{enumerate}}
\def \bit{\begin{itemize}}
\def \eit{\end{itemize}}
\def \baR{\begin{array}}
\def \eaR{\end{array}}
\def \cp{\text{CP}}
\def \nl{\text{NL}}

\def \pole{\text{pole}}
\def \sd{\text{SD}}

\def \vckm{V_{\text{CKM}}}
\def \B{\bar{B}}
\def \klong{K_L}
\def \chargino{\tilde{\chi}}
\def \gluino{\tilde{g}}
\def \sneutrino{\tilde{\n}}
\def \tsquark{\tilde{t}}

\def \GeV{{\text{GeV}}}

\def \MeV{{\text{MeV}}}
\def \TeV{{\text{TeV}}}
\def \ps{{\text{ps}}}
\def \alqcd{\a_s}

\def \bm{\boldmath}
\def \bracket#1#2#3{\langle #1|#2| #3\rangle}
\def \branch{{\mathcal B}\,}
\def \cl#1{{#1\%\ \mathrm{C.L.}}}

\def \diag{{\mathrm{diag}}}
\def \dis{\displaystyle}
\def \ea{{\it et al.}}

\def \eq#1{Eq.~(\ref{#1})}
\def \eqs#1#2{Eqs.~(\ref{#1})--(\ref{#2})}
\def \fig#1{Fig.~\ref{#1}}
\def \figs#1#2{Figs.~\ref{#1}--\ref{#2}}
\def \hc{\mathrm{H.c.}}
\def \heff{{\mathcal H}_{\text{eff}}}
\def \Im{{\text{Im}}\,}
\def \imlt{\Im\la_t}
\def \Lms{\La_{\ol{\text{MS}}}}
\def \msbar{\ol{\text{MS}}}
\def \nnu{\nonumber}

\def \Oi{{\mathcal O}}
\def \ol#1{\overline{#1}}

\def \sm{\mathrm{SM}}
\def \Re{{\text{Re}}\,}
\def \relc{\Re\la_c}
\def \relt{\Re\la_t}
\def \rf{Ref.~\cite}
\def \rfs{Refs.~\cite}
\def \Sec#1{Sec.~\ref{#1}}

\def \mqhat{\hat{m}_q}
\def \miss{\hat{E}}
\def \a{\alpha}
\def \b{\beta}
\def \D{\Delta}
\def \g{\gamma}
\def \G{\Gamma}

\def \k{\kappa}
\def \la{\lambda}
\def \La{\Lambda}
\def \m{\mu}
\def \n{\nu}

\def \p{\pi}
\def \r{\rho}
\def \s{\sigma}

\def \t{\tau}
\def \pslash{p\hspace{-0.45em} /}

\def\euro#1#2#3{\emph{Eur.~Phys.~J.}~{\bf C#1} (#2) #3}

\def\ibid#1#2#3{{\it ibid.\/}~{\bf#1} (#2) #3}

\def\jhep#1#2#3{\emph{J.~High~Energy~Phys.}~{\bf #1} (#2) #3}

\def\nc#1#2#3{\emph{Nuovo Cimento}~{\bf #1A} (#2) #3}
\def\np#1#2#3{\emph{Nucl.~Phys.}~{\bf B#1} (#2) #3}

\def\pl#1#2#3{\emph{Phys. Lett.} {\bf B#1} (#2) #3}
\def\plold#1#2#3{\emph{Phys.~Lett.}~{\bf #1B} (#2) #3}
\def\pr#1#2#3{\emph{Phys.~Rev.}~{\bf #1} (#2) #3}
\def\prd#1#2#3{\emph{Phys.~Rev.} {\bf D#1} (#2) #3}
\def\prl#1#2#3{\emph{Phys.~Rev.~Lett.}~{\bf #1} (#2) #3}
\def\prp#1#2#3{\emph{Phys.~Rep.}~{\bf #1} (#2) #3}

\def\ppnp#1#2#3{\emph{Prog.~Part.~Nucl.~Phys.}~{\bf #1} (#2) #3}
\def\rmp#1#2#3{\emph{Rev.~Mod. Phys.}~{\bf #1} (#2) #3}
\def\zpc#1#2#3{\emph{Z.~Phys.}~{\bf C#1} (#2) #3}
\begin{document}
\begin{titlepage}
\begin{flushright}
TUM-HEP-448/01 \\
hep-ph/0112305 \\
December 2001
\end{flushright}
\vskip 0.2in
\begin{center}
\setlength{\baselineskip}{0.35in} 
{\bm\bf\Large QCD Corrections to  $\B \to X_{d,s}\nu\bar{\nu}$, 
$\B_{d,s}\to l^+l^-$, $K\to \pi\nu\bar{\nu}$\\
 and $K_L\to \m^+ \m^-$ in the MSSM}\\[1cm]
\setlength {\baselineskip}{0.2in}
{\bf Christoph Bobeth\footnote{E-mail address: bobeth@ph.tum.de},
         Andrzej J. Buras\footnote{E-mail address: aburas@ph.tum.de},
         Frank  Kr\"uger\footnote{E-mail address: fkrueger@ph.tum.de},
         J\"org Urban\footnote{E-mail address: urban@ph.tum.de}}\\[1mm]
\emph{Physik Department, Technische Universit\"at M\"unchen,\\
     D-85748 Garching, Germany}\\[0.5cm]
\end{center}
\begin{abstract}
We compute for the first time QCD corrections to the rare decays 
$\B\to X_{d,s}\nu\bar{\nu}$, $\B_{d,s}\to l^+l^-$, 
$K\to \pi\nu\bar{\nu}$ and $K_L\to \m^+ \m^-$, where $l=e$ or $\m$,  in 
the context of a supersymmetric extension of the Standard Model (SM)
with minimal flavour violation and new operators, in addition to those 
present in the SM.
Assuming that the gluino is heavy, we consider an effective theory which 
consists of charged and neutral Higgs particles, charginos and squarks.~We  
evaluate the QCD corrections to box and $Z^0$-penguin diagrams with 
top-quark, charged  Higgs boson, chargino and squark exchanges, 
as well as to  
neutral Higgs boson penguin diagrams. We provide a compendium of analytic formulae for the Wilson coefficients, which are valid for arbitrary values of 
$\tan\b$ (the ratio of the vacuum expectation values of the two Higgs 
fields) except for the case of the neutral Higgs-boson contributions.~These 
contributions have been obtained 
at large $\tan\b$, which may compensate for the 
inevitable suppression by the masses of the light leptons in decays based 
on the $b\to s(d) l^+l^-$ transition.
We investigate the dependence
of the various branching ratios on the renormalization scale $\m$,
which is the main theoretical uncertainty in the short-distance 
calculation. We find that the 
$\m$ dependence of the branching ratios is considerably reduced once
the QCD corrections are taken into 
account. The contributions of new operators are found to be dominant at large $\tan\b$ in $\B_{d,s}\to \m^+\m^-$ while they are subleading in 
$\B\to X_{d,s} \n\bar\n$ and completely negligible in kaon decays. 
\end{abstract}
\end{titlepage}
%
%
\renewcommand{\thefootnote}{\arabic{footnote}}
\setcounter{footnote}{0}

\newsection{Introduction}\label{into}
The rare decays $\B\to X_{d,s}\nu\bar{\nu}$, $\B_{d,s}\to l^+l^-$,
$K^+\to \pi^+\nu\bar{\nu}$ and $\klpn$, 
where $l$ denotes a lepton, are very promising probes 
of flavour physics within the Standard Model (SM) and 
possible extensions, since they are governed essentially by short-distance
interactions. ~(For recent reviews of rare $K$ and $B$ decays, see 
\rfs{Bur01,Buchalla:rev,littenberg,gino:rev:01}.) 
These decays are sensitive to the quantum structure of flavour dynamics and 
can at the same time be computed to an exceptionally high degree of 
precision. This is due to the fact that (a) the required low energy 
hadronic matrix elements are 
just the matrix elements of quark currents between hadron states, which can be 
extracted from semileptonic decays, including isospin-breaking effects \cite{marciano:zarsa}; (b) other long-distance 
contributions have been found to be negligible \cite{LD}; and (c)  
the contributions of higher dimensional operators turn out to be tiny in the 
case of $\klpn$ \cite{GBGI}, and are below $5\%$ of the charm 
contribution in the case of $K^+\to\pi^+\nu\bar\nu$ \cite{FalkLP}. 

As a consequence, the scale ambiguities inherent in  
perturbative QCD constitute the dominant theoretical uncertainties present
in the analysis of the above decay modes. 
Within the SM, these theoretical uncertainties have been 
 considerably  reduced through the inclusion of the next-to-leading order 
QCD corrections \cite{BB1,BB2,MU98,BB98,BB3}. 

In the decay $\klmumu$, on the other hand, there are sizable long-distance
contributions  associated with a two-photon 
intermediate state, in addition to the short-distance interaction 
$d \bar s \to \m^+\m^-$.~While the absorptive contribution with real photons can 
be determined by means of the measured $K_L\to \g\g$ branching ratio 
\cite{lms}, the numerical predictions for the dispersive part of the 
amplitude due to off-shell photons
suffer from theoretical uncertainties inherent in the models for the
form factors \cite{LD:KLmumu}. We therefore 
content ourselves with the short-distance part of $\klmumu$.

In the context of the minimal supersymmetric standard model (MSSM)
\cite{mssm:original,haber:kane:feyn:rules}, 
the branching ratios for rare $B$-meson and kaon decays such 
as $\B\to X_s \n\bar\n$, $\B_s\to \m^+\m^-$, $\klpn$, $\kpnn$ and 
$\klmumu$ have been considered by many authors 
(see, e.g., 
\rfs{grossman:etal,YY97,rare:K:B:decays,BGGJS,bertolini:etal,CMW96,first:papers:Bmumu,me:bqll:SUSY,bobeth:etal:0401,Bmumu:susy:epsilon,gino:retico:01});
however, these decay modes have hitherto been studied without 
the inclusion of QCD corrections. Working in the framework of the SM 
operator basis, 
a compendium of the relevant formulae in the limit of 
degenerate squarks of the first two generations, and valid only for 
small $\tan\b$ (the ratio of the vacuum expectation 
values of the two Higgs fields), has been presented  in \rf{BGGJS}.~Taking into account all presently available constraints on the supersymmetric parameters, 
the phenomenological analysis of  \cite{BGGJS} shows that the 
supersymmetric contributions to the above-mentioned processes 
can be quite substantial. Indeed, denoting the MSSM prediction 
for a given decay normalized to the SM result by $R$, and setting all the 
SM parameters that are unaffected by supersymmetric 
contributions at their central values, one finds the following
ranges\footnote{Note that these ranges become larger once the 
SM parameters are varied.}
\be
0.73 \leqslant R(\B\to X_s\nu\bar{\nu}) \leqslant 1.34, \quad
0.68 \leqslant R(\B_s\to \mu^+\mu^-) \leqslant 1.53,
\ee
\be
0.65 \leqslant R(\kpnn) \leqslant 1.02, \quad 
0.41 \leqslant R(\klpn) \leqslant 1.03.
\ee
Clearly, in view of 
large parametric uncertainties related to the SM parameters 
and the masses of the supersymmetric particles, precise predictions for the 
above branching ratios are not possible at present.~Nevertheless, 
the situation may change considerably in this decade due to the improved 
determination of the 
SM parameters in forthcoming $B$- and $K$-physics dedicated 
experiments, as well as through an anticipated discovery of supersymmetric 
particles. Besides, all the branching ratios considered in this paper 
could conceivably be measured in this decade.
We think that these prospects, together with the theoretical cleanliness of 
these decays, justify a more accurate computation of the relevant branching 
ratios in the MSSM. 

In this paper, we extend existing calculations in two ways:
\ben
\item We compute for the first time QCD 
corrections to the branching ratios of  $\B\to X_{d,s}\nu\bar{\nu}$,
$\B_{d,s}\to l^+l^-$,  $K^+\to \pi^+\nu\bar{\nu}$, $\klpn$ and  $\klmumu$,
within supersymmetry (SUSY).

\item We also include the contributions of new operators beyond  
those present in the SM.
\een
While such an enterprise would have been a formidable task 
ten years ago, it is a relatively straightforward computation by means of 
analytic computer programs developed during the last five years.

It is the main objective of the present work to investigate
the dependence 
of the various branching ratios on  the renormalization scale $\mu$
within SUSY, 
taking into account QCD corrections. As we shall see, the inclusion of 
these corrections will significantly reduce
the unphysical renormalization scale dependence of the branching ratios
originating in the scale dependence of the running quark and squark masses.

In the quantitative analysis, we
focus on the two distinctly different regions
\be\label{tanbeta:regime}
2\leqslant \tan\b \leqslant 5, \quad 40\leqslant \tan\b \leqslant  60,
\ee
which we refer to as the low and high $\tan\b$ regime, respectively.
Our results for the Wilson coefficients 
that will be presented below
are valid for arbitrary values of $\tan\b$, except for the neutral 
Higgs-boson contributions, which are sizable only in the high $\tan\b$ 
regime for the decays under study.  

The outline of this paper is as follows. In \Sec{mssm}, we touch on 
the elements of the MSSM and define our notation. 
Section \ref{effective:Hamiltonian} is a compendium of the relevant effective 
Hamiltonians and the corresponding Wilson coefficients
including $O(\alqcd)$ corrections, in the context of the MSSM. 
In \Sec{calculation:details}, we give technical   
details of our calculation of the QCD corrections.
In \Sec{branching:ratios}, we present a collection of the 
branching ratios in question. 
Section \ref{numerical:analysis} is devoted to the numerical analysis, 
which is based on the assumption of minimal flavour violation and performed in 
the low as well as the high $\tan\b$ regime. 
Some comments on the 
SUSY  QCD corrections to the $b$ quark Yukawa coupling, relevant at large 
$\tan\b$, will be made  
in \Sec{SUSY:QCD:Yukawa}. Finally, in \Sec{conclusions}, we summarize and 
conclude.
A compilation of the  various loop functions is given in the Appendices. 

\newsection{Couplings and mixing matrices in the MSSM}\label{mssm}
We start by introducing the relevant mass and mixing 
matrices in the context of SUSY 
with minimal particle content and $R$-parity conservation, 
which we will call the minimal supersymmetric standard model (MSSM). 
In the numerical analysis, we confine ourselves  to a version of the model 
with minimal flavour violation, i.e., we assume that 
flavour mixing is due exclusively to the  
Cabib\-bo-Ko\-ba\-ya\-shi-Maskawa (CKM) matrix. In addition, we take the 
down squark mass-squared matrix to be flavour diagonal, so that there are no
neutralino contributions. 

The MSSM is usually embedded in a grand unified theory (GUT) in order to 
solve the problem with too large contributions to flavour-changing neutral 
currents and to further reduce the vast number of unknown parameters.
This leads to the minimal supergravity inspired model in which one assumes 
universality of the soft terms at
the scale of gauge coupling unification, say, $M_{\mathrm{GUT}}$.  
Renormalization group effects then induce flavour off-diagonal entries, e.g., 
in the squark mass-squared matrices at the electroweak scale. In our  
numerical analysis, however,  we do not relate the soft SUSY-breaking 
parameters to 
some common high scale, but rather take them at the electroweak 
scale while discarding flavour off-diagonal 
terms in the squark mass-squared matrix.  

\subsection{Chargino mass matrix}
The chargino mass matrix is given by
\be\label{chargino:mass-matrix}
M_{\chargino}=
\left(
\begin{array}{cc}
M_2 & \sqrt{2} M_W \sin\b \\ 
\sqrt{2}M_W \cos\b & \m
\end{array}
\right),
\ee
where $M_2$ and $\m$ are the $W$-ino and Higgsino mass parameters, 
respectively. 
This matrix can be cast in diagonal form by means of a biunitary
transformation
\be\label{chargino:uv:def}
U^{\ast} M_{\chargino}V^{\dagger}=\diag(M_{\tilde{\chi}_1^{}}, M_{\tilde{\chi}_2^{}}),
\ee
$M_{\tilde{\chi}_{1,2}^{}}$ being the chargino masses with 
$M^2_{\tilde{\chi}_1^{}}< M^2_{\tilde{\chi}_2^{}}$.~The analytic 
expressions for $M_{\tilde{\chi}_{1,2}^{}}$ and the matrices $U$, $V$ 
can be found in \rf{KR00}.

\subsection{Squark mass matrix}  
It proves convenient to work in the super-CKM basis \cite{super-ckm}, 
in which the quark mass matrices are diagonal, and 
both quarks and squarks are rotated simultaneously.
Denoting up-type squarks by $\tilde U$,  the $6\times 6$ 
mass-squared matrix in the $(\tilde{U}_L,\tilde{U}_R)$ basis 
is given by
\bea\label{squark:massmatrix}
M^2_{\tilde U}=
\left(\begin{array}{cc}
M_{\tilde U_L}^2 + M_U^2  + M_Z^2\cos2\b(\frac{1}{2} - \frac{2}{3} \sin^2\theta_W)\unit  
&M_U(A_{U}^{\ast}-\m\cot\b\unit)\\
\left[M_U(A_{U}^{\ast}-\m\cot\b\unit)\right]^\dagger
& M_{\tilde U_R}^2 + M_U^2 + \frac{2}{3}M_Z^2\cos 2\b\sin^2\theta_W\unit
\end{array}\right),\nnu\\
\eea
where $\theta_W$ is the Weinberg angle, $M_{\tilde U_{L,R}}$ 
are the soft SUSY breaking up-type squark mass matrices, 
$M_U\equiv \diag(m_u, m_c,m_t)$, $A_U$ contains the trilinear soft 
SUSY-breaking parameters  and $\unit$ represents 
the unit matrix. The matrix $M^2_{\tilde U}$
can be diagonalized by a unitary matrix $\Gamma^U$ such that 
\be
\Gamma^U M_{\tilde U}^2 {\Gamma^U}^{\dagger}= 
\diag(m_{\tilde{u}_1}^2, m_{\tilde{u}_2}^2, \dots,m_{\tilde{u}_6}^2). 
\ee
For later use it is convenient to define the $6\times 3$ matrices 
\be\label{def:stt}
(\G^{U_L})_{ai}=(\Gamma^U)_{ai},\quad  
(\G^{U_R})_{ai}=(\Gamma^U)_{a, i+3}.
\ee


\subsection{Slepton  mass matrices}
Similarly to the mixing matrix for the squarks, we define the matrices 
$\Gamma^E$ in the charged slepton sector through
\be
\Gamma^E M^2_{\tilde l} {\Gamma^E}^{\dagger}
=\diag(m_{\tilde{l}_1}^2, m_{\tilde{l}_2}^2, \dots,m_{\tilde{l}_6}^2), 
\ee
with the mass-squared matrix, in the super-CKM basis,
\bea\label{mass-squared:matrix:slepton}
M^2_{\tilde l}=
\left(\begin{array}{cc}
M_{\tilde l_L}^2 + M_E^2  - \frac{1}{2}M_Z^2\cos2\b(1-2 \sin^2\theta_W)\unit  
&M_E(A_l^{\ast} - \m\tan\b\unit)\\
\left[M_E(A_l^{\ast} - \m\tan\b\unit)\right]^\dagger& M_{\tilde l_R}^2 +M_E^2 - 
M_Z^2\cos 2\b\sin^2\theta_W\unit\end{array}\right),
\eea
where $M_{\tilde l_{L,R}}$ are the soft SUSY breaking charged 
slepton mass matrices,
$M_E\equiv \diag(m_e, m_\mu,m_\t)$ and $A_l$ contains the 
soft SUSY breaking trilinear couplings.
%
%
As before, we introduce the $6\times 3$ mixing matrices 
\be\label{def:sl}
(\G^{E_L})_{ai}=(\Gamma^E)_{ai},\quad  
(\G^{E_R})_{ai}=(\Gamma^E)_{a, i+3}.
\ee

Finally, if we assume the neutrinos to be massless, we are left with the 
$3\times 3$ mixing matrix $\Gamma^N$ in the sneutrino sector, which is 
defined by
\be\label{sneutrino:rnu}
\Gamma^N M_{\sneutrino}^2 {\Gamma^N}^{\dagger}= 
\diag(m_{\sneutrino_1}^2, m_{\sneutrino_2}^2,m_{\sneutrino_3}^2),
\ee
where the sneutrino mass-squared matrix reads 
\be
M_{\sneutrino}^2=M_{\tilde l_L}^2+\frac{1}{2}M_Z^2\cos 2\b \unit.
\ee

\subsection{Interactions within the MSSM}
Recalling  
\bea
&& P_{L,R}= (1\mp \g_5)/2, \quad g=e/\sin\theta_W, \quad g_s^2=4\p \a_s,\nnu\\
&&M_U\equiv \diag(m_u, m_c,m_t), \quad M_E\equiv \diag(m_e, m_\mu,m_\t),
\quad M_D\equiv \diag(m_d, m_s,m_b),
\eea
the relevant interaction vertices of the 
MSSM with decoupled gluinos and massless neutrinos in the super-CKM basis  
can be written as\footnote{For simplicity of presentation,  
we suppress the generation 
indices. Furthermore, we define the charginos $\chargino_i^-$ ($i=1,2$) as particles, 
contrary to \rfs{haber:kane:feyn:rules,janusz:feyn:rules,super-ckm}, with 
$\chargino_i^+ \equiv (\chargino_i^-)^c$.} 
\be {\cal L}_H = \frac{g}{\sqrt 2 M_W}[
 \cot\beta (\bar{u} M_U \vckm P_L d)+ \tan\beta (\bar{u}\vckm M_D P_R d)] 
H^+ + \hc,
\ee
\bea
{\cal L}_{\tilde\chi} &=&\sum_{i=1}^2\Bigg\{\ol{\tilde{\chi}_i^-}
[\tilde\nu^\dagger(X_i^{N_L}P_L + X_i^{N_R}P_R)l
+\tilde{u}^\dagger(X_i^{U_L} P_L + X_i^{U_R} P_R) d]\nnu\\
&+& \ol{\chargino_i^+}[\tilde{l}^\dagger (X_i^{E_L}P_L + X_i^{E_R}P_R)
\nu]\Bigg\}+\hc,
\eea
\be
{\cal L}_{\tilde\chi\tilde\chi Z}=\frac{g \sec\theta_W}{2} \sum_{i,j=1}^2 
\ol{\tilde{\chi}_i^-}\gamma^{\mu}(U_{i1} U_{j1}^{\ast}P_L +
V_{i1}^{\ast} V_{j1}  
P_R + \cos 2\theta_W \delta_{ij}) 
\tilde{\chi}_j^- Z_{\mu}^0,
\ee
\be\label{ch:squ:qu:ver}
{\cal L}_{\tilde u\tilde u Z} = - \frac{i g\sec\theta_W}{2} 
\Bigg[(\Gamma^{U_L}{\Gamma^{U_L}}^{\dagger})_{ab}-\frac{4}{3} 
\sin^2\theta_W \delta_{ab} \Bigg](\tilde{u}_a^{\ast} 
\stackrel{\leftrightarrow}{\partial^{\mu}}\tilde{u}_b)Z_{\mu}^0,
\ee
\be\label{quartic:squarks}
{\cal L}_4 \equiv {\cal L}_{\tilde u\tilde u\tilde u\tilde u}^{g_s}
=  -\frac{1}{2}g_s^2 (\tilde u^* P_U T^c \tilde u )^2, 
\ee
where
\be\label{mu:dep:XUL}
X_i^{U_L} = -g \Bigg[a_g V_{i1}^{\ast}\Gamma^{U_L} 
- a_Y V_{i2}^{\ast} \Gamma^{U_R} \frac{M_U}{\sqrt2 M_W \sin\beta}\Bigg]\vckm,
\ee
\be\label{mu:dep:XUR}
X_i^{U_R} = g a_Y  U_{i2}  \Gamma^{U_L}\vckm\frac{M_D}{\sqrt2 M_W 
\cos\beta},
\ee
\be
X_i^{N_L} = -g V_{i1}^{\ast} \Gamma^N,\quad
X_i^{N_R} =  g U_{i2} \Gamma^N \frac{M_E}{\sqrt2 M_W \cos\beta},
\ee
\be
X_i^{E_L} =-g\Bigg[U_{i1}^\ast \Gamma^{E_L}
-U_{i2}^\ast \Gamma^{E_R}\frac{M_E}{\sqrt2 M_W \cos\beta}\Bigg],\quad X_i^{E_R}=0,
\ee
\be\label{def:pu}
P_U\equiv \Gamma^{U_L}\Gamma^{U_L
    \dagger}-\Gamma^{U_R}\Gamma^{U_R \dagger}.
\ee
In these equations, $\vckm$ is the CKM matrix
and $T^c$ ($c=1,\dots, 8$) are the generators of colour gauge symmetry.  
The mixing matrices $U, V$, $\G^{U_{L,R}}$ and $\G^{E_{L,R}}$
are defined  through  Eqs.~(\ref{chargino:uv:def}), (\ref{def:stt}) and 
(\ref{def:sl}), respectively, and 
the $3\times 3$ sneutrino mixing 
matrix $\Gamma^N$ is defined via \eq{sneutrino:rnu}. 
 
The effect of the decoupled gluino, with mass $M_{\gluino}$, is contained in the functions 
$a_g$ and $a_Y$ \cite{BoMU00}, which are given by
\be\label{decoupled:gluino} 
a_g = 1-\frac{\alqcd(\mu)}{4\pi}\Bigg[ \frac{7}{3} 
+ 2\ln\Bigg(\frac{\mu^2}{M_{\tilde{g}}^2}\Bigg)\Bigg], \quad
a_Y = 1+\frac{\alqcd(\mu)}{4\pi}\Bigg[1+ 
2\ln\Bigg(\frac{\mu^2}{M_{\tilde{g}}^2}\Bigg)\Bigg],
\ee
$\m$ being the matching scale which will be discussed in more detail 
at the end of \Sec{effective:Hamiltonian}.\footnote{Equation (\ref{decoupled:gluino}) can be obtained from the result of 
\rf{CDGG98} by performing the limit $M_{\gluino}\to\infty$.}  

As for the couplings of the neutral Higgs bosons, we refer to 
\cite{janusz:feyn:rules}.

\subsection{Tree-level relations within the MSSM}
In calculating the neutral Higgs-boson contributions, we shall exploit the tree-level relations
\be
M_{A^0}^2=M_H^2-M_W^2, 
\ee
\be\label{tree-level:h0}
M^2_{h^0,H^0}=\frac{1}{2}\{M_{A^0}^2+M_Z^2\mp 
[(M_{A^0}^2+M_Z^2)^2-4M_Z^2M_{A^0}^2
\cos^2 2\b]^{1/2}\},
\ee
\be\label{relation:tree-level}
\sin 2\a = -\sin 2\b \left(\frac{M^2_{H^0}+M^2_{h^0}}{M^2_{H^0}-M^2_{h^0}}
\right),
\ee
where $M_H$ and $M_{A^0}$ are the masses of the charged and
$\cp$-odd Higgs boson, respectively, and $M_{h^0,H^0}$  and $\a$
are the masses and mixing angle in the $\cp$-even Higgs sector. 
Thus, at the tree-level, 
the Higgs sector of the MSSM is described by merely two 
free parameters that we choose to be $\tan\b$ and $M_H$.
Since the results that will be presented in the next section have been 
obtained at leading order 
in the electroweak couplings, 
the tree-level relations are quite adequate for our purposes.
%
\newsection{Effective Hamiltonians and Wilson coefficients}\label{effective:Hamiltonian}
\subsection{Notation}
Within the context of the SM and the two-Higgs-doublet model (2HDM),
as well as the MSSM, at low $\tan\b$,  only a small number of operators 
contribute to the
decays under study, and the Wilson coefficients can be
expressed in terms of only two functions usually denoted by $X$ and $Y$ 
\cite{Bur01,BB1,BB2,MU98,BB98,BB3} 
(see Appendix \ref{diff:notation} for the SM result).

For large values of $\tan\b$, on the other hand, significant contributions 
of new operators are conceivable and such a convenient  notation is no longer applicable. In fact, since a given $Z^0$-penguin or box diagram with a 
certain particle exchange $(W^\pm, H^\pm, \chargino^\pm)$ 
can contribute to the 
coefficients of several  operators,  it is useful to 
proceed as follows:
\bit
\item All $Z^{0}$-penguin contributions
are included in the functions $C^{\nu\bar\nu}$ and $C^{l\bar l}$ which 
describe the processes $b\rightarrow q  \nu \bar\nu$ and $b\rightarrow q l^+ l^-$ ($q=d,s$), respectively.
Likewise, the box-diagram contributions are expressed in terms of 
the functions $B^{\nu\bar\nu}$ and $B^{l\bar l}$. 
\item In the case of the $b\to q l^+l^-$ transition, 
there may be sizable contributions due to neutral Higgs boson penguin 
diagrams at large $\tan\b$, leading to the function $N^{l\bar l}$. 
\item Owing to quartic squark vertices, 
there are additional contributions 
to $Z^0$-penguin, box- and penguin diagrams with neutral Higgs bosons.
Indeed, when the running mass of the squarks in the modified minimal 
subtraction ($\msbar$) scheme is used, diagrams involving
quartic squark couplings have to be included.
As we shall discuss in \Sec{calculation:details}, 
these contributions vanish 
after renormalization when the pole mass of  the squarks is used instead. 
Below, these contributions will be labelled by the index $J=4$. 
\item The results obtained in the $B$ 
sector can be easily altered to apply to the kaon system.
\item Including QCD corrections, and denoting  the 
$\sm$, charged Higgs-boson, chargino and `quartic' contributions by 
$J=\sm,H, \chargino,4$, the functions 
mentioned above have the structure \footnote{Recall that in our
  scenario (i) the gluino decouples, and (ii) there are no neutralino
  contributions since we assume the down squark mass-squared matrix to
  be flavour diagonal.}
\begin{subequations}\label{WC:var:contr}
\be
[C^{\nu\bar\nu}_I]_J=
[C^{\nu\bar\nu}_I]^{(0)}_J + 
\frac{\alqcd}{4\pi} [C^{\nu\bar\nu}_I]^{(1)}_J, \quad
[C^{l\bar l}_I]_J=[C^{l\bar l}_I]^{(0)}_J + 
\frac{\alqcd}{4\pi} [C^{l\bar l}_I]^{(1)}_J, 
\ee
\be
[B^{\nu\bar\nu}_I]_J=[B^{\nu\bar\nu}_I]^{(0)}_J + 
\frac{\alqcd}{4\pi} [B^{\nu\bar\nu}_I]^{(1)}_J,\quad
[B^{l\bar l}_I]_J=[B^{l\bar l}_I]^{(0)}_J + 
\frac{\alqcd}{4\pi} [B^{l\bar l}_I]^{(1)}_J,
\ee
\be
[N^{l\bar l}_I]_J =[N^{l\bar l}_I]^{(0)}_J + 
\frac{\alqcd}{4\pi} [N^{l\bar l}_I]^{(1)}_J,
\ee
\end{subequations}
where the subscript $I$ indicates the various operators, which will be 
discussed shortly.
Note that there are no sizable neutral Higgs-boson contributions within
the $\sm$ as they are invariably suppressed by the light 
fermion masses, and hence can be safely neglected.
\item Finally, we define the mass ratios
\begin{subequations}\label{shorthand:notation}
\be
  x=\frac{m_t^2}{M_W^2}, \quad 
  y=\frac{m_t^2}{M_H^2}, \quad
  z=\frac{M_H^2}{M_W^2},
\ee
\be
x_{ij} = \frac{M^2_{\tilde{\chi}_i^{}}}{M^2_{\tilde{\chi}_j^{}}}, \quad
  y_{ai} = \frac{m^2_{\tilde{u}_a}}{M^2_{\tilde{\chi}_i^{}}},    \quad
  z_{bi} = \frac{m^2_{\tilde{l}_b}}{M^2_{\tilde{\chi}_i^{}}},    \quad
  v_{fi} = \frac{m^2_{\tilde{\nu}_f}}{M^2_{\tilde{\chi}_i^{}}},
\ee
and introduce the abbreviations
\be
s_W\equiv \sin\theta_W,\quad
L_t \equiv  \ln \Bigg(\frac{\m^2}{m_t^2}\Bigg),\quad
L_{\tilde u_a} \equiv  \ln \Bigg(\frac{\m^2}{m_{\tilde u_a}^2}\Bigg),\quad
\kappa_q \equiv \frac{1}{8\sqrt2 G_F e^2  V_{tb}^{} V_{tq}^*}.
\ee
\end{subequations}
\eit

\subsection{The \bm$b\to s (d) \nu\bar\nu$ transition}\label{btoq:transition}
Let us start with the effective Hamiltonian 
describing the
$b\to q\nu_f\bar\nu_f$ transition, where $q=d, s$ and $f=e,\m,\t$. 
Within the SM, it is given by\footnote{Throughout 
this paper, we suppress the neutrino flavour 
index $f$ carried by the Wilson coefficients.} 
\be\label{genereal:amplitude:bsnunu}
{\cal H}_{\rm eff}^f = \frac{4 G_F}{\sqrt{2}}\frac{\a}{2\p
  \sin^2\theta_W} \sum_{i=u,c,t} V_{ib}^{}V_{iq}^* 
\,{\tilde c}_L^{i} (m_i)
(\bar q \g_\mu P_L b)( \bar \nu_f \g^\mu P_L\nu_f),
\ee
where ${\tilde c}_L^{i}(m_i)$ represent the contributions from 
internal quarks $i=u,c,t$, and $m_i$ are the corresponding masses.
Using the  unitarity of the CKM matrix, 
the effective Hamiltonian can be written as
\bea\label{heff:gim:canc}
{\cal H}_{\rm eff}^f &=& \frac{4 G_F}{\sqrt{2}}\frac{\a}{2\p
  \sin^2\theta_W} \sum_{i=c,t} V_{ib}^{}V_{iq}^*
[{\tilde c}_L^{i} (m_i)-{\tilde c}_L^{u} (m_u)]
(\bar q \g_\mu P_L b)( \bar \nu_f \g^\mu P_L\nu_f)\nnu\\
&=& \frac{4 G_F}{\sqrt{2}}\frac{\a}{2\p
  \sin^2\theta_W} \sum_{i=c,t} V_{ib}^{}V_{iq}^*\, c_L^i (m_i)
(\bar q \g_\mu P_L b)( \bar \nu_f \g^\mu P_L\nu_f),
\eea
with $c_L^i (0)=0$. [Note that in the second line of \eq{heff:gim:canc}
we have taken the limit $m_u\to 0$.] 

Exploiting the fact that
\be\label{SD:O}
\frac{c_L^c(m_c)}{c_L^t(m_t)}\sim O(10^{-3}), \quad 
\left|\frac{V_{cb}^{}V_{cq}^*}{V_{tb}^{}V_{tq}^*}\right|
\sim  O(1) \quad (q=d,s),
\ee
the charm-quark contribution to the quark-level process 
$b\to q\n_f\bar\n_f$ can be safely neglected, and we arrive at 

\be\label{example:heff:sm}
{\cal H}_{\rm eff}^f = \frac{4 G_F}{\sqrt{2}}\frac{\a}{2\p
  \sin^2\theta_W} V_{tb}^{}V_{tq}^* \,c^t_L (m_t)
(\bar q \g_\mu P_L b)( \bar \nu_f \g^\mu P_L\nu_f).
\ee
The extension of this result to a non-standard operator 
basis is straightforward. Henceforth, we will omit the superscript $t$ of the 
Wilson coefficient in \eq{example:heff:sm}.

Assuming that the neutrinos are essentially massless, and 
hence purely left-handed, the effective Hamiltonian in the presence of 
new operators has 
the particularly simple form
\be\label{eff:ham:bqnunu}
{\cal H}_{\rm eff} 
= \frac{4 G_F}{\sqrt{2}}\frac{\a}{2\p \sin^2\theta_W} 
 V_{tb}^{}V_{tq}^*\sum_{f=e,\m,\t}[c_{L} \Oi_{L} + c_{R} \Oi_{R}],
\ee
where 
\be\label{ops:bqnunu}
\Oi_{L}= (\bar q \g_\mu P_L b)( \bar \nu_f \g^\mu P_L\nu_f),\quad 
\Oi_{R}=(\bar q \g_\mu P_R b)( \bar \nu_f \g^\mu P_L\nu_f),
\ee
and the short-distance coefficients
\be\label{wilson:coeffs::bqnunu}
c_{L}= \sum_{J=\sm,H,\chargino,4}\{[C^{\nu\bar\nu}_{L}]_J +
[B^{\nu\bar\nu}_{L}]_J\},\quad
c_{R}= \sum_{J=\sm,H,\chargino, 4}\{[C^{\nu\bar\nu}_{R}]_J +
         [B^{\nu\bar\nu}_{R}]_J\}.
\ee
Note that the Wilson coefficients of tensor and scalar
operators such as $(\bar{q} \sigma^{\mu\nu} P_{L,R} b)(\bar{\n} 
\sigma_{\mu\nu} P_{L,R} \n)$ and $(\bar{q} P_{L,R} b)(\bar{\n} 
P_{L,R} \n)$ vanish for massless neutrinos in the final state.

\subsubsection{\bm$Z^0$-penguin contributions}\label{subsubsec:zpengnunu} 
\begin{figure}
\begin{center}
\epsfig{file=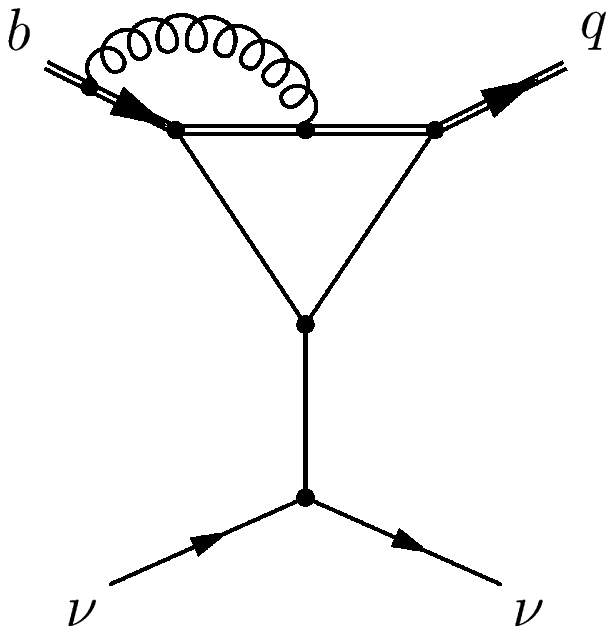,height=1.5in}\hspace{3em}
\epsfig{file=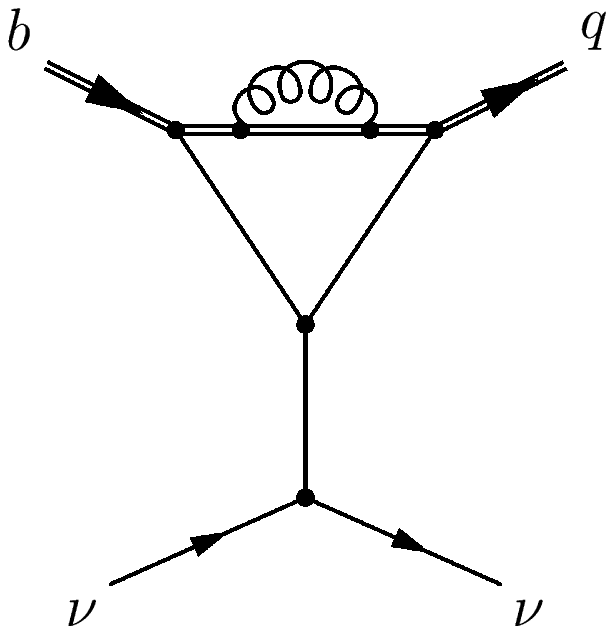,height=1.5in}\hspace{3em}
\epsfig{file=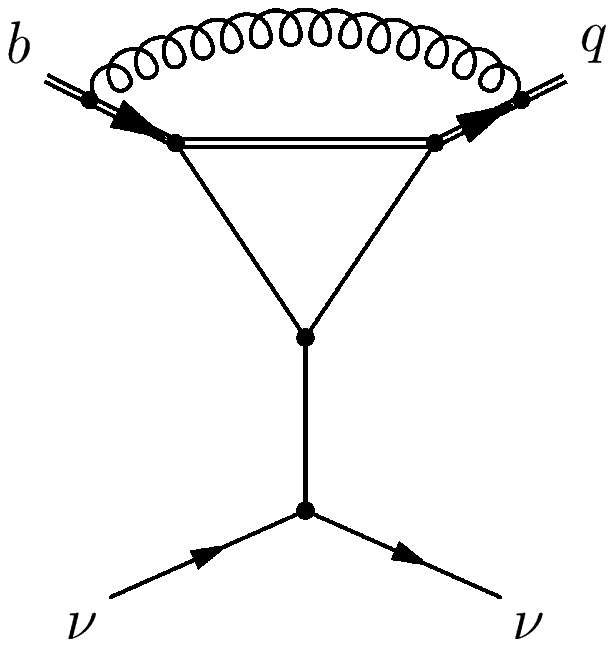,height=1.5in}\vspace{2em}
\epsfig{file=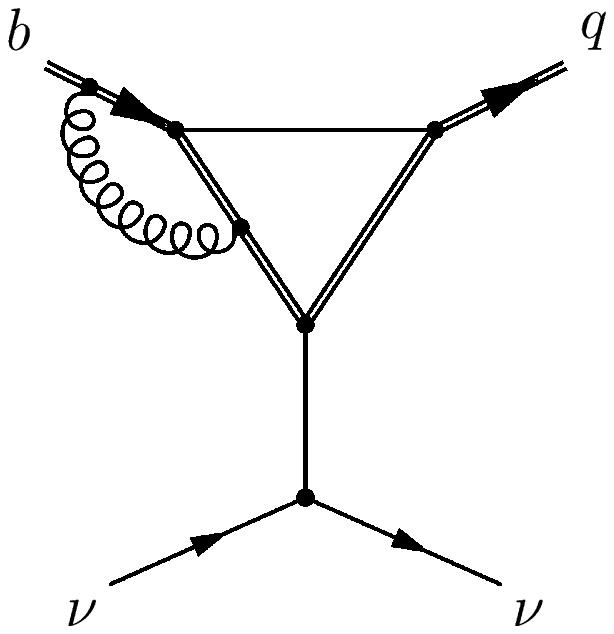,height=1.5in}\hspace{3em}
\epsfig{file=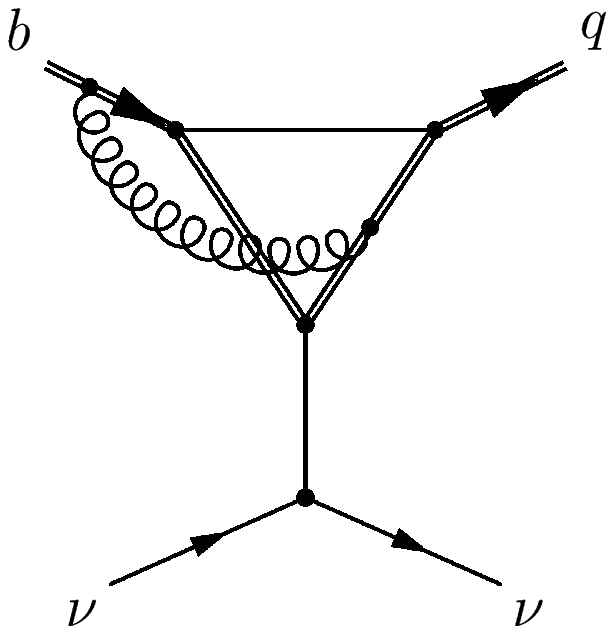,height=1.5in}\hspace{3em}
\epsfig{file=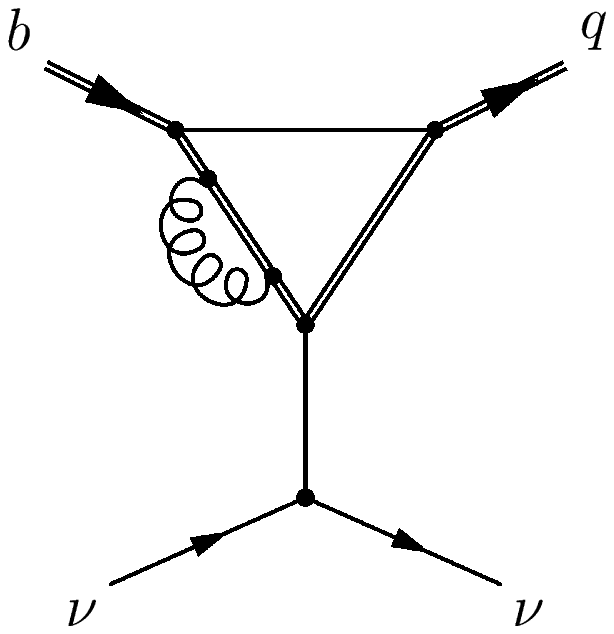,height=1.5in}\vspace{2em}
\epsfig{file=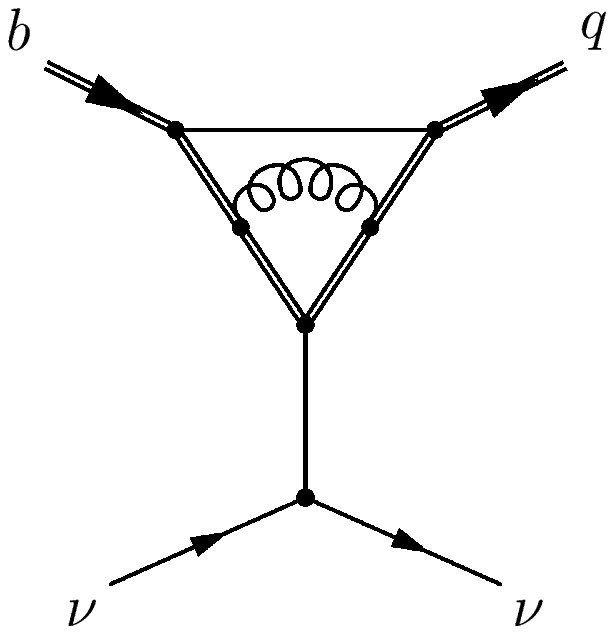,height=1.5in}\hspace{3em}
\epsfig{file=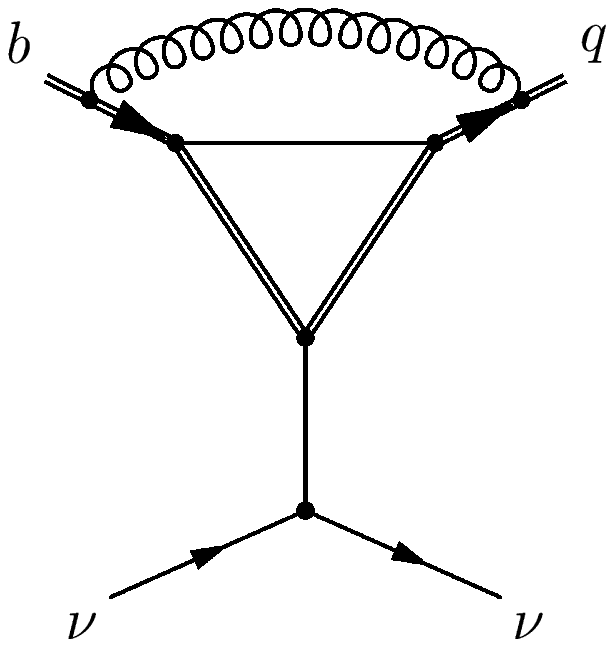,height=1.5in}\vspace{2em}\\
\epsfig{file=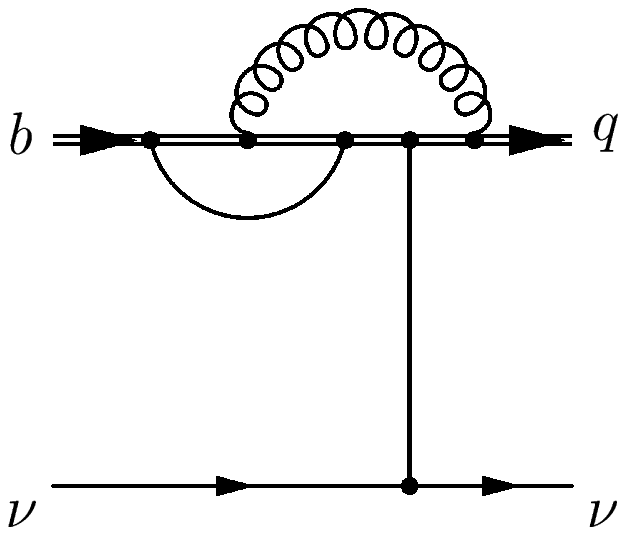,height=1.5in}\hspace{3em}
\epsfig{file=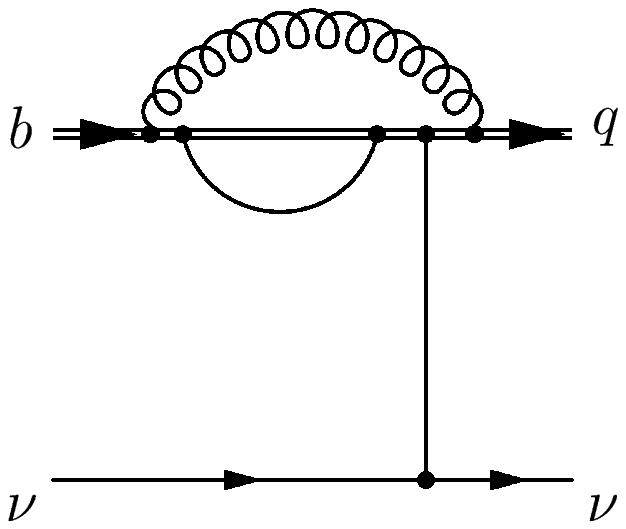,height=1.5in}
\caption{Penguin diagrams contributing to the transition 
$b\rightarrow q \nu \bar\nu$ ($q=d,s$) at order $\alqcd$.
The vertical and curly lines denote $Z^0$ bosons and gluons, respectively, 
while the coloured particles 
(i.e.~quarks and their superpartners) are represented by double lines. 
The diagrams for $b\to q l^+l^-$ may be obtained by replacing 
$\n\to l$ and by taking into account neutral Higgs and 
would-be-Goldstone bosons, in addition to the $Z^0$ boson. 
The corresponding symmetric diagrams are not shown here.}
\label{fig:peng}
\end{center}
\end{figure}

\begin{figure}
\begin{center}
\epsfig{file=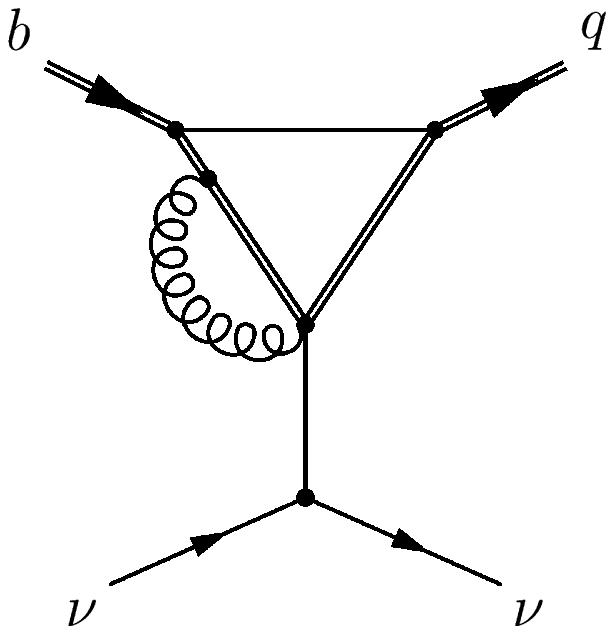,height=1.6in}\hspace{3em}
\epsfig{file=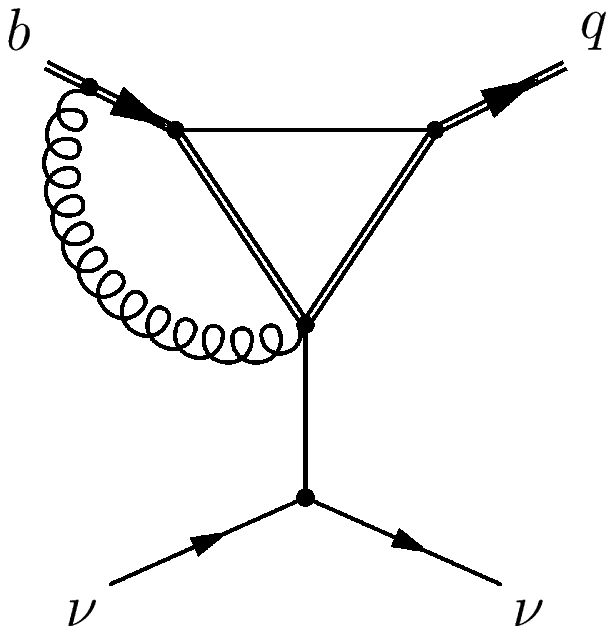,height=1.6in}
\caption{Feynman 
diagrams with quartic couplings contributing to the decay 
$b\to q \n\bar{\n}$, $q=d,s$, where the vertical line corresponds  to the 
$Z^0$ boson.
In the case of the $b\to q l^+l^-$ transition, 
one simply replaces $\n\to l$. We refrain from showing the symmetric 
vertex corrections here.} 
\label{fig:peng:quartic}
\end{center}
\end{figure}

\begin{figure}
\begin{center}
\epsfig{file=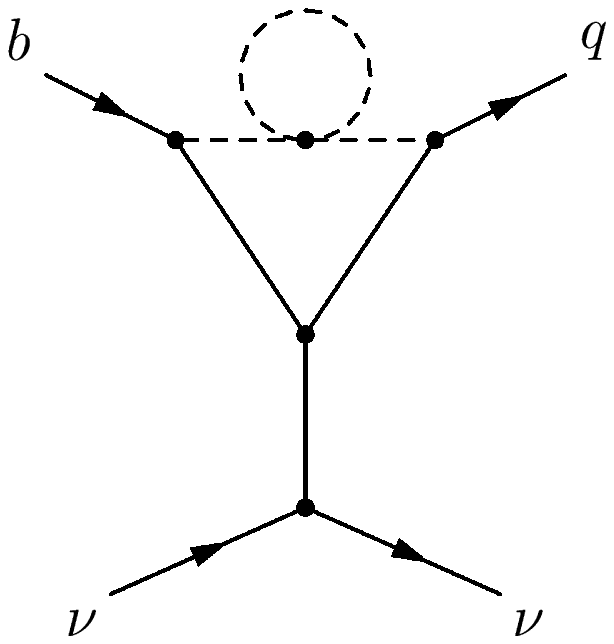,height=1.6in}\hspace{3em}
\epsfig{file=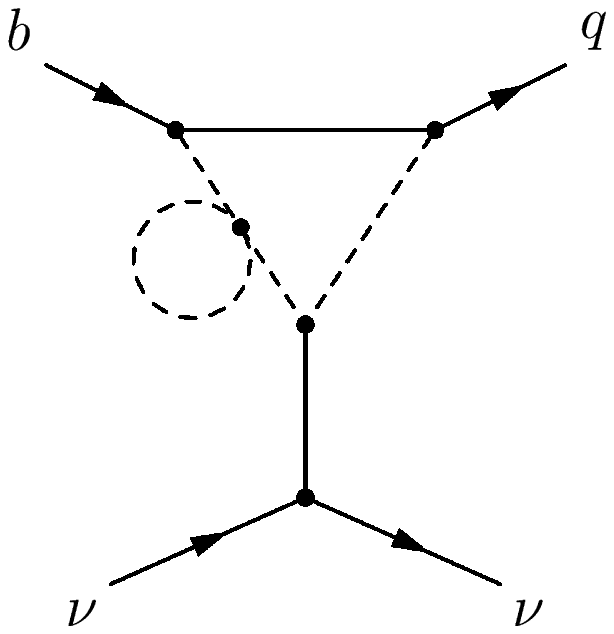,height=1.6in}\hspace{3em}
\epsfig{file=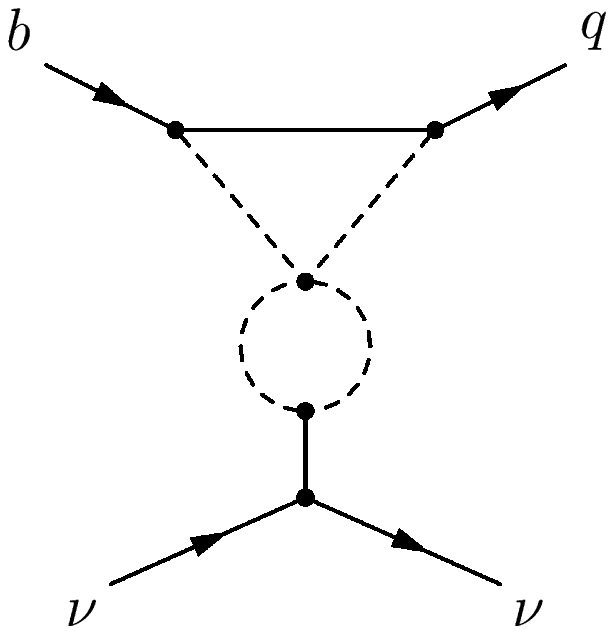,height=1.6in}\vspace{2em}
\epsfig{file=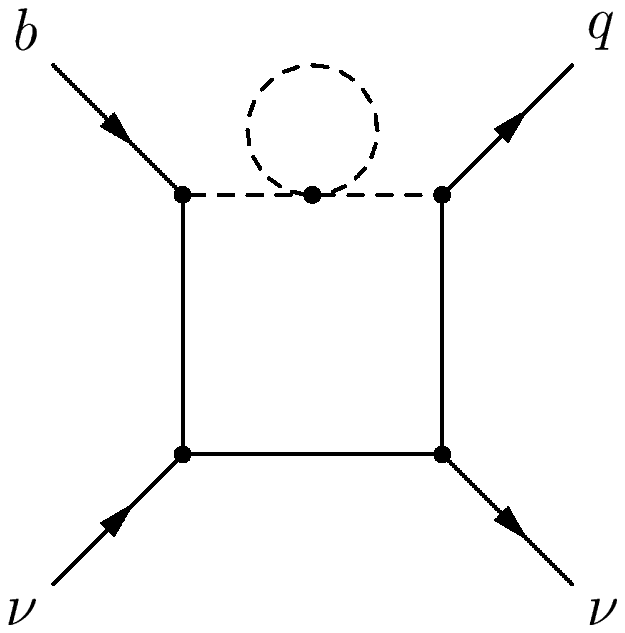,height=1.6in}
\caption{The contributions to the $b\to q \n\bar\n$ ($q=d,s$) transition
due to quartic squark couplings.
The dashed lines denote scalar quarks while the solid lines
represent charginos, scalar leptons, and the $Z^0$ boson. 
As for the diagrams  describing the $b\to q l^+l^-$ transition, 
we refer to the captions of Figs.~\ref{fig:peng} and \ref{fig:box}.}  
\label{fig:peng:quartsquark}
\end{center}
\end{figure}

The QCD corrections to the Wilson coefficients are calculated from the 
Feynman diagrams in Figs.~\ref{fig:peng} and \ref{fig:peng:quartic}, 
%
%
while those involving the quartic squark couplings are depicted 
in \fig{fig:peng:quartsquark}.~Recalling the definitions in Eqs.~(\ref{shorthand:notation}),
we find
%
%
%
%
\be\label{bsnunu:SM}
[C^{\nu\bar\nu}_{L}]_{\mathrm{SM}} =
    \frac{1}{4} 
     \Bigg\{f^{(0)}_1(x)+\frac{\alpha_s}{4\pi}\Bigg[f^{(1)}_1(x)
+ 8x \frac{\partial}{\partial x}f^{(0)}_1(x) L_t \Bigg]\Bigg\},
\ee
\be
 [C^{\nu\bar\nu}_{R}]_{\mathrm{SM}} = 0,
\ee
\be\label{bsnunu:2HDM}
[C^{\nu\bar\nu}_{L}]_H  = -\frac{M_H^2 \cot^2 \beta}{8M_W^2}
   \Bigg\{y f^{(0)}_2(y)+\frac{\alpha_s}{4\pi}\Bigg[y f^{(1)}_2(y)+ 8
    y \frac{\partial}{\partial y} [y f^{(0)}_2(y)] L_t\Bigg]
\Bigg\},
\ee
\be\label{bsnunu:2HDM:CR}
[C^{\nu\bar\nu}_{R}]_H = \frac{m_b m_q\tan^2\beta}{8M_W^2}
\Bigg\{f^{(0)}_2(y)+\frac{\alpha_s}{4\pi}\Bigg[ f^{(1)}_2(y)+
8\Bigg(1+y \frac{\partial}{\partial y}\Bigg)f^{(0)}_2(y) L_t\Bigg]\Bigg\},
\ee
\bea\label{CLnunu:chargino}
[C^{\nu\bar\nu}_{L}]_{\tilde\chi}&=& -\frac{\k_q e^2}{4M_W^2}
\sum_{i,j=1}^2 \sum_{a,b=1}^6 (X^{U_{L}\dagger}_j)_{qb}(X^{U_{L}}_i)_{a3} 
\nnu\\
&&\hspace{-2cm}\mbox{}\times\Bigg\{2\sqrt{x_{ji}} \Bigg\{f_3^{(0)} (x_{ji}, y_{ai})
+\frac{\alpha_s}{4\pi}\Bigg[ f^{(1)}_3(x_{ji}, y_{ai})+4\left(1+
    y_{ai}\frac{\partial}{\partial
    y_{ai}}\right) f_3^{(0)}(x_{ji}, y_{ai})  L_{\tilde u_a}
  \Bigg]\Bigg\} U_{j1}U_{i1}^\ast \delta_{ab}
\nnu\\
&&\hspace{-2cm}\mbox{}-\Bigg\{f_4^{(0)} (x_{ji}, y_{ai})+\frac{\alpha_s}{4\pi}
\Bigg[
  f^{(1)}_4(x_{ji}, y_{ai})+4\left(1+
    y_{ai}\frac{\partial}{\partial
    y_{ai}}\right) f_4^{(0)}(x_{ji}, y_{ai}) L_{\tilde u_a}
  \Bigg]\Bigg\}V_{j1}^\ast V_{i1} \delta_{ab}\nnu\\
&&\hspace{-2cm}\mbox{}+\Bigg\{f_4^{(0)} (y_{ai}, y_{bi})+\frac{\alpha_s}{4\pi} 
\left[f^{(1)}_5(y_{ai}, y_{bi})+4\left(1+y_{ai}
    \frac{\partial}{\partial y_{ai}} + y_{bi}\frac{\partial}{\partial
      y_{bi}}\right) f_4^{(0)} (y_{ai}, y_{bi})L_{\tilde u_a}\right]
\Bigg\}\nnu\\
&&\hspace{-2cm}\mbox{}\times (\Gamma^{U_{L}} {\Gamma^{U_{L}\dagger}})_{b a} \delta_{ij}\Bigg\},
\eea
\bea\label{CRnunu:chargino}
[C^{\nu\bar\nu}_{R}]_{\tilde\chi}&=& 
[C^{\nu\bar\nu}_{L}]_{\tilde\chi}\left(X^{U_L}\rightarrow X^{U_R};
  U\rightarrow V^*; V\rightarrow U^*; 
  \Gamma^{U_L} \Gamma^{U_L\dagger}\rightarrow-\Gamma^{U_R} \Gamma^{U_R\dagger}\right),
\eea
\bea\label{tadpole:cl}
\lefteqn{[C^{\nu\bar\nu}_{L}]_4=
\frac{\alpha_s}{4 \pi}\frac{\k_q e^2}{3M_W^2}
\sum_{i,j=1}^2\sum_{a, \dots, e,g,k=1}^6 (X^{U_{L}\dagger}_j)_{qd}(X^{U_{L}}_i)_{a3} 
[(P_U)_{ek} y_{ki} (P_U)_{kg}
(1+L_{\tilde u_k})]}
\nnu\\
&\times&
\Bigg\{2\sqrt{x_{ji}} f_6^{(0)} (x_{ji}, y_{ai}, y_{di})
U_{j1}U_{i1}^\ast \delta_{ae} \delta_{gd} -
f_5^{(0)} (x_{ji}, y_{ai}, y_{di}) V_{j1}^\ast V_{i1} \delta_{ae}\delta_{gd}\nnu\\
&+&f_5^{(0)} (y_{ai}, y_{bi}, y_{ci})
(\Gamma^{U_{L}} {\Gamma^{U_{L}\dagger}})_{b c}
\delta_{ij}\delta_{ae}\delta_{bg} \delta_{cd} +
f_5^{(0)} (y_{ai}, y_{ci}, y_{di})
(\Gamma^{U_{L}} {\Gamma^{U_{L}\dagger}})_{b c}
\delta_{ij}\delta_{ab}\delta_{ce}\delta_{dg}\Bigg\},
\eea
\bea\label{tadpole:cr}
[C^{\nu\bar\nu}_{R}]_4&=& 
[C^{\nu\bar\nu}_{L}]_4\left(X^{U_L}\rightarrow X^{U_R};
  U\rightarrow V^*; V\rightarrow U^*; 
  \Gamma^{U_L} \Gamma^{U_L\dagger}\rightarrow-\Gamma^{U_R} \Gamma^{U_R\dagger}\right),
\eea
where the index $q$ carried by the $X^{U_{L,R}}_i$ matrices is  
\be
q=\left\{\begin{array}{l}1\quad \text{for}\quad b\to d,\\ 
2\quad \text{for}\quad b\to s,\end{array}\right.
\ee
while $P_U$ has already been defined in \eq{def:pu}. 
We should stress that all quark and scalar quark masses 
entering the above formulae, as well as the expressions that 
follow, are running masses in the $\msbar$ scheme, 
i.e., $m_b\equiv {m}_b(\m)$, 
$m_q\equiv {m}_q(\m)$, and so on (see \Sec{RGE} below).
The loop functions, $f_p^{(0)}$, $f_{p}^{(1)}$, are listed  
in  Appendix \ref{loop-functions:app}.

\subsubsection{Box-diagram contributions}
\begin{figure}
\begin{center}
\epsfig{file=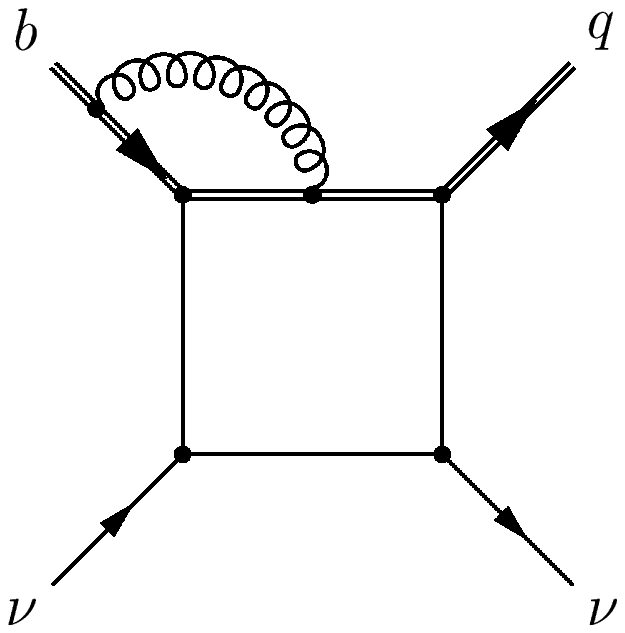,height=1.6in}\hspace{3em}
\epsfig{file=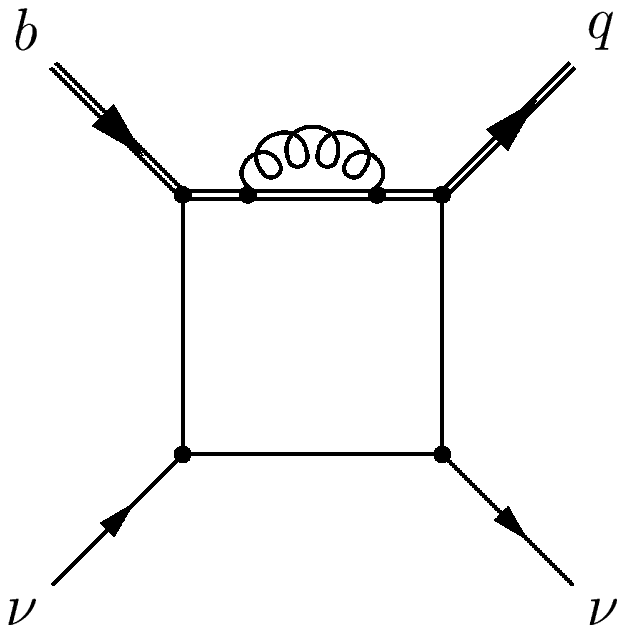,height=1.6in}\hspace{3em}
\epsfig{file=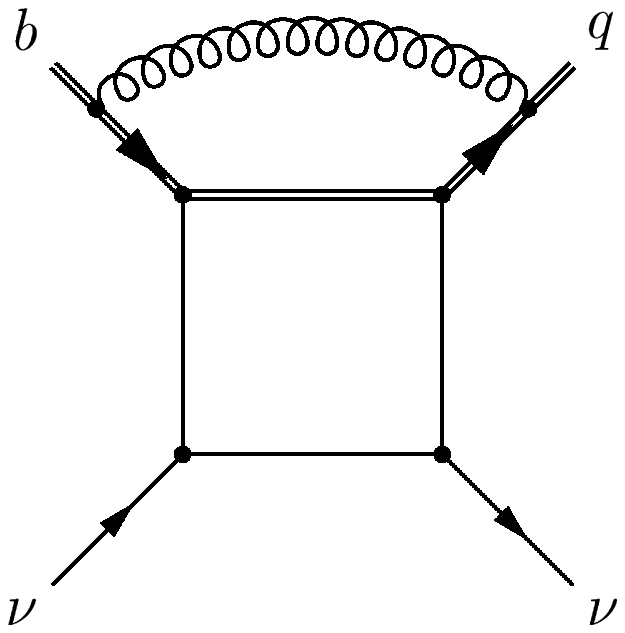,height=1.6in}
\caption{Box diagrams contributing to the decay 
$b\rightarrow q \nu \bar\nu$, $q=d,s$,  at $O(\alqcd)$. 
The curly  and double lines represent gluons and quarks 
(or squarks), respectively. The remaining internal 
lines denote gauge bosons, leptons, charged Higgs bosons, scalar leptons and 
charginos. The diagrams for $b\to q l^+l^-$ can be obtained by 
the replacements $l\leftrightarrow \n$ and 
$\tilde l\to \tilde\n$. The symmetric diagrams are not 
shown here explicitly.}\label{fig:box}
\end{center}
\end{figure}

The relevant Feynman diagrams, which contribute at $O(\alqcd)$, 
are depicted in Figs.~\ref{fig:peng:quartsquark} and \ref{fig:box}. 
We find
%
%
\bea
 {[B_{L}^{\nu\bar\nu}]}_{\sm} = 
      -f^{(0)}_2 (x) - \frac{\alpha_s}{4\pi}\Bigg[f^{(1)}_6 (x)
+ 8x \frac{\partial}{\partial x}f^{(0)}_2(x) L_t\Bigg],
\eea
\be
{[B_{R}^{\nu\bar\nu}]}_{\sm} =  0,
\ee
\be\label{BsubL}
{[B_{L}^{\nu\bar\nu}]}_{H}  = 0,
\ee
\be\label{box:yukawa:lepton}
{[B_{R}^{\nu\bar\nu}]}_{H} = 
      -\frac{m_bm_q m_f^2 \tan^4\beta}{16 M_H^2 M_W^2} 
      \Bigg\{f^{(0)}_2 (y)+ \frac{\alpha_s}{4\pi}\Bigg[f^{(1)}_7 (y)
+ 8\Bigg(1+ y \frac{\partial}{\partial y}\Bigg)f^{(0)}_2(y)L_t\Bigg]\Bigg\},
\ee
\bea\label{BLnunu:chargino}
 {[B_{L}^{\nu\bar\nu}]}_{\tilde\chi} &=&\frac{\kappa_q s_W^2}{2}
      \sum_{i,j=1}^2\sum_{a,b=1}^6
      \frac{1}{M^2_{\tilde{\chi}_i^{}}}(X_j^{E_L})_{bf} (X_i^{E_L\dagger})_{fb}
           (X_j^{U_L\dagger})_{qa} (X_i^{U_L})_{a3}  
    \Bigg\{f_5^{(0)}(x_{ji},y_{ai},z_{bi})\nnu\\
&+& \frac{\alpha_s}{4\pi}\Bigg[ f^{(1)}_8 (x_{ji},y_{ai},z_{bi})
+4 \Bigg(1+ y_{ai}\frac{\partial}{\partial y_{ai}}\Bigg)f^{(0)}_5(x_{ji},y_{ai},z_{bi}) 
L_{\tilde u_a}\Bigg]\Bigg\},
\eea
\bea\label{BRnunu:chargino}
{[B_{R}^{\nu\bar\nu}]}_{\tilde\chi} &=&-\kappa_q s_W^2 
      \sum_{i,j=1}^2  \sum_{a,b=1}^6
      \frac{\sqrt{x_{ji}}}{M^2_{\tilde{\chi}_i^{}}}
      (X_j^{E_L})_{bf} (X_i^{E_L\dagger})_{fb} 
(X_j^{U_R\dagger})_{qa}(X_i^{U_R})_{a3} 
     \Bigg\{f_6^{(0)}(x_{ji},y_{ai},z_{bi})\nnu\\
&+&\frac{\alpha_s}{4\pi}\Bigg[ f^{(1)}_{9} (x_{ji},y_{ai},z_{bi})
+4 \Bigg(1+ y_{ai}\frac{\partial}{\partial y_{ai}}
\Bigg)f^{(0)}_6(x_{ji},y_{ai},z_{bi}) 
L_{\tilde u_a}\Bigg]\Bigg\},
\eea
\bea\label{tadpole:bl}
 {[B_{L}^{\nu\bar\nu}]}_4 &=&
-\frac{\alpha_s}{4 \pi}\frac{2 \kappa_q s_W^2}{3}
\sum_{i,j=1}^2 \sum_{a, \dots, d =1}^6
 \frac{1}{M^2_{\tilde{\chi}_i^{}}}(X_j^{E_L})_{cf} (X_i^{E_L\dagger})_{fc}
(X_j^{U_L\dagger})_{qb} (X_i^{U_L})_{a3} \nnu\\ 
&\times& [(P_U)_{ad} y_{di}(P_U)_{db}
(1+L_{\tilde u_d})]f_{9}^{(0)}(x_{ji},y_{ai},y_{bi},z_{ci}),
\eea
\bea\label{tadpole:br}
{[B_{R}^{\nu\bar\nu}]}_4 &=&
 \frac{\alpha_s}{4 \pi}\frac{4\kappa_q s_W^2}{3}
\sum_{i,j=1}^2\sum_{a, \dots, d=1}^6 
\frac{\sqrt{x_{ji}}}{M^2_{\tilde{\chi}_i^{}}}
(X_j^{E_L})_{cf} (X_i^{E_L\dagger})_{fc} 
(X_j^{U_R\dagger})_{qb}(X_i^{U_R})_{a3}  \nnu\\ 
&\times& [(P_U)_{ad} y_{di}(P_U)_{db}
(1+L_{\tilde u_d})]f_{10}^{(0)}(x_{ji},y_{ai},y_{bi},z_{ci}),
\eea
where, as before, $f=e,\m,\t$ and $q=1$ (2) for the 
$b\to d$ $(b\to s)$ transition.

\subsection{The \bm$b\to s (d)l^+l^-$ transition}\label{btoqll:transition}
The part of the effective Hamiltonian describing the transition 
$b\to q l^+l^-$ ($q=d,s$) that is relevant for the decay 
$\B_q\to l^+l^-$ has the form
\bea\label{eff:ham:bqll}
\heff &=&-\frac{2G_F}{\sqrt{2}}\frac{\a}{2\p\sin^2\theta_W}V_{tb}^{}V_{tq}^{*}
[c_{A} \Oi_{A} + c_{A}' \Oi_{A}'+ c_S \Oi_S + c_S' \Oi_S' + c_P \Oi_P
+ c_P' \Oi_P'], 
\eea
where 
\be\label{ops:bsll:av}
{\Oi}_A =(\bar{q} \gamma^\mu P_L b) (\bar{l}\g_\m\gamma_5 l),
\quad
{\Oi}'_A = (\bar{q} \gamma^\mu P_R b) (\bar{l} 
\g_\m\gamma_5 l),
\ee
\be\label{ops:bsll:s}
{\Oi}_S = m_b(\bar{q} P_R b) (\bar{l}l),\quad
{\Oi}'_S = m_q(\bar{q} P_L b) (\bar{l}l),
\ee
\be\label{ops:bsll:p}
{\Oi}_P = m_b(\bar{q} P_R b) (\bar{l}\gamma_5 l),\quad
{\Oi}'_P =m_q(\bar{q} P_L b) (\bar{l} \gamma_5 l).
\ee 
Again, the effective Hamiltonian [\eq{eff:ham:bqll}] has been obtained 
by exploiting the  Glashow-Iliopoulos-Maiani (GIM) mechanism \cite{GIM}, 
as discussed in \Sec{btoq:transition}.

As far as  additional operators like 
$(\bar{q} \gamma^{\mu} P_{L,R} b)(\bar{l} \gamma_{\mu} l)$ and
$(\bar{q} \sigma^{\mu\nu} P_{L,R} b)(\bar{l} \sigma_{\mu\nu} P_{L,R} l)$
are concerned,
they also receive contributions within SUSY. However, as we shall see in 
\Sec{branching:ratios:Bmumu}, the  hadronic matrix elements of these vector 
and tensor operators vanish.~Moreover, the corresponding 
Wilson coefficients do not mix with the remaining short-distance coefficients.

The Wilson coefficients can then be written as 
\begin{subequations}\label{wilson:coeffs:general}
\be 
c_{A} = \sum_{J=\sm,H,\chargino,4} \{{[C^{l\bar{l}}_{A}]}_J 
+ {[B^{l\bar{l}}_{A}]}_J\}, \quad
c_{A}'= \sum_{J=\sm,H,\chargino,4} \{{[C^{l\bar{l}}_{A'}]}_J 
+ {[B^{l\bar{l}}_{A'}]}_J\},
\ee
\be
  c_S = \sum_{J= H,\chargino,4} \{ 
    {[N_{S}^{l\bar l}]}_J + {[B^{l\bar{l}}_{S}]}_J\}, \quad
   c_S' = \sum_{J= H,\chargino,4} \{ 
    {[N_{S'}^{l\bar l}]}_J + {[B^{l\bar{l}}_{S'}]}_J\},
\ee
\be
  c_P = \sum_{J= H,\chargino,4} \{ 
    {[N_{P}^{l\bar l}]}_J + {[B^{l\bar{l}}_{P}]}_J\}, \quad
   c_P' = \sum_{J= H,\chargino,4} \{ 
    {[N_{P'}^{l\bar l}]}_J + {[B^{l\bar{l}}_{P'}]}_J\}.
\ee
\end{subequations}%
There are several points to be made here. (a) The $\sm$ contribution to 
scalar and 
pseudoscalar operators is suppressed by $m_l m_{b,q}/M_W^2$, and so can be 
neglected. (b) 
The $Z^0$ boson as a vector particle does not contribute to 
scalar and pseudoscalar operators. (c) Because of their 
scalar-type couplings, the neutral Higgs bosons do not contribute to the 
Wilson coefficients $c_{A}$ and $c'_{A}$.

\subsubsection{\bm$Z^0$-penguin contributions}
Evaluation of the diagrams with $Z^0$ exchange results in 
\be\label{cRll}
 {[C^{l\bar l}_{A}]}_J  =  -[C_{L}^{\n\bar\n}]_J,\quad
 {[C^{l\bar l}_{A'}]}_J  = -[C_{R}^{\n\bar\n}]_J\qquad
(J=\sm, H, \chargino,4),
\ee
with the functions $[C_{L}^{\n\bar\n}]_J$ and $[C_{R}^{\n\bar\n}]_J$ 
as given in \eqs{bsnunu:SM}{tadpole:cr}.
 
\subsubsection{Box-diagram contributions}
\be
 {[B^{l\bar{l}}_{A}]}_{\sm} =
    \frac{1}{4} \Bigg\{ f_2^{(0)}(x) + 
    \frac{\alpha_s}{4\pi} \left[f_{10}^{(1)} (x)+8 x
      \frac{\partial}{\partial x} f_2^{(0)}(x) L_t\right]\Bigg\},
\ee
\be
{[B^{l\bar{l}}_{A'}]}_{\sm}= 0,
\ee
\be
{[B^{l\bar{l}}_{S,P}]}_\sm= {[B^{l\bar{l}}_{S',P'}]}_\sm=0,
\ee
\be
 {[B^{l\bar{l}}_{A}]}_H = 0,
\ee
\be
 {[B^{l\bar{l}}_{A'}]}_H =
    -\frac{m_b m_q m_l^2 \tan^4\beta}{16 M_W^2 M_H^2} 
    \Bigg\{f_{2}^{(0)}(y) + \frac{\alpha_s}{4 \pi}
    \left[f_{7}^{(1)}(y)+8\left(1+y\frac{\partial}{\partial y}\right)
      f_{2}^{(0)}(y) L_t\right]\Bigg\},
\ee

\be
{[B^{l\bar{l}}_{S,P}]}_H = 
    \pm\frac{m_l\tan^2\beta}{4 M_W^2} 
    \Bigg\{ f_7^{(0)}(x,z) + \frac{\alpha_s}{4 \pi} \left[f_{11}^{(1)}(x,z)+8 x
      \frac{\partial}{\partial x} f_7^{(0)}(x,z) L_t\right]\Bigg\},
\ee
\be
{[B^{l\bar{l}}_{S',P'}]}_H =
    \frac{m_l\tan^2\beta}{4 M_W^2} 
    \Bigg\{ f_7^{(0)}(x,z) + 
\frac{\alpha_s}{4 \pi} \left[f_{11}^{(1)}(x,z)+8 x
      \frac{\partial}{\partial x} f_7^{(0)}(x,z) L_t\right]\Bigg\},
\ee
\bea
 {[B^{l\bar{l}}_{A}]}_{\tilde{\chi}} &=& 
    \kappa_q s_W^2
    \sum_{i,j=1}^2 \sum_{f=1}^3 \sum_{a=1}^6 
    \frac{(X_j^{U_L\dagger})_{qa}(X_i^{U_L})_{a 3}}{M_{\tilde\chi_i^{}}^2}
\Bigg\{ \frac{1}{2}
  \Bigg\{f^{(0)}_5(x_{ji}, y_{ai}, v_{fi})\nnu\\ 
&&\hspace{-1.2cm}\mbox{}+ \frac{\alpha_s}{4\pi}\left[f^{(1)}_8(x_{ji}, y_{ai},
  v_{fi})+4\left(1+y_{ai}\frac{\partial}{\partial y_{ai}}
\right)f^{(0)}_5(x_{ji}, y_{ai}, v_{fi})L_{\tilde u_a}\right]\Bigg\}
(X_i^{N_L\dagger})_{lf} (X_j^{N_L})_{f l}\nnu\\
&&\hspace{-1.2cm}\mbox{}+ \sqrt{x_{ji}}
      \Bigg\{f^{(0)}_6(x_{ji}, y_{ai}, v_{fi}) + 
      \frac{\alpha_s}{4\pi} \Bigg[f^{(1)}_{9}(x_{ji}, y_{ai}, v_{fi})
+4\left(1+y_{ai}\frac{\partial}{\partial y_{ai}}\right)f^{(0)}_6(x_{ji},
       y_{ai}, v_{fi})L_{\tilde u_a}\Bigg]\Bigg\}\nnu\\
&&\hspace{-1.2cm}\mbox{}\times(X_i^{N_R\dagger})_{lf} (X_j^{N_R})_{f l}\Bigg\},
\eea
\be
  {[B^{l\bar{l}}_{A'}]}_{\tilde{\chi}} = - 
  {[B^{l\bar{l}}_{A}]}_{\tilde{\chi}}(X^{U_L}\rightarrow X^{U_R}; 
X^{N_L}\leftrightarrow X^{N_R}),
\ee
\bea
{[B^{l\bar{l}}_{S,P}]}_{\tilde{\chi}} &=& \pm  
    \frac{\kappa_q s_W^2}{m_b}
    \sum_{i,j=1}^2 \sum_{f=1}^3  \sum_{a=1}^6
    \frac{(X_j^{U_L\dagger})_{qa}(X_i^{U_R})_{a 3}}{M_{\tilde\chi_i^{}}^2}
\Bigg\{ 
      \Bigg\{f^{(0)}_5(x_{ji}, y_{ai}, v_{fi})\nnu\\
&+&\frac{\alpha_s}{4\pi}
\left[f^{(1)}_{12}(x_{ji}, y_{ai},
  v_{fi})+4y_{ai}\frac{\partial}{\partial y_{ai}}f^{(0)}_5(x_{ji},
  y_{ai}, v_{fi})L_{\tilde u_a}\right] \Bigg\}
(X_i^{N_R\dagger})_{lf}(X_j^{N_L})_{f l}\nnu\\
&\pm& \sqrt{x_{ji}}
    \Bigg\{f^{(0)}_6(x_{ji}, y_{ai}, v_{fi}) + 
    \frac{\alpha_s}{4\pi} \left[f^{(1)}_{13}(x_{ji}, y_{ai}, v_{fi})
    +4 y_{ai}\frac{\partial}{\partial y_{ai}}f^{(0)}_6(x_{ji},
  y_{ai}, v_{fi})L_{\tilde u_a}\right]\Bigg\}\nnu\\
&\times &(X_i^{N_L\dagger})_{lf}(X_j^{N_R})_{f l}\Bigg\},
\eea
\be
{[B^{l\bar{l}}_{S',P'}]}_{\tilde{\chi}} = \pm
{[B^{l\bar{l}}_{S,P}]}_{\tilde{\chi}} (
{X^{U_L}\leftrightarrow X^{U_R}; X^{N_L}\leftrightarrow X^{N_R};
m_b\rightarrow m_q}).
\ee
Here $l$ denotes the external leptons (not summed over) and $f$ labels the 
flavour of the intermediate neutrino and its scalar partner. 
The loop functions 
$f_p^{(0)}$, $f_{p'}^{(1)}$ are given in Appendix \ref{loop-functions:app}.

Turning to the contributions from the quartic couplings, 
we obtain
\bea
 {[B^{l\bar{l}}_{A}]}_4 &=&- 
 \frac{\alpha_s}{4 \pi} \frac{4\kappa_q s_W^2}{3}
 \sum_{i,j=1}^2 \sum_{f=1}^3  \sum_{a, b, c =1}^6
 \frac{(X_j^{U_L\dagger})_{qb}(X_i^{U_L})_{a 3}}{M_{\tilde\chi_i^{}}^2}
 [(P_U)_{ac} y_{ci}(P_U)_{cb}
 (1+L_{\tilde u_c})]\nnu\\
&&\hspace{-1.5cm}\mbox{}
\times\Bigg\{ \frac{1}{2}
    f^{(0)}_{9}(x_{ji}, y_{ai}, y_{bi}, v_{fi})
    (X_i^{N_L\dagger})_{lf} (X_j^{N_L})_{f l}
+ \sqrt{x_{ji}} f^{(0)}_{10}(x_{ji}, y_{ai}, y_{bi}, v_{fi})
    (X_i^{N_R\dagger})_{lf} (X_j^{N_R})_{f l}\Bigg\},\nnu\\
\eea
\be
  {[B^{l\bar{l}}_{A'}]}_4 = - 
  {[B^{l\bar{l}}_{A}]}_4(X^{U_L}\rightarrow X^{U_R}; 
X^{N_L}\leftrightarrow X^{N_R}),
\ee
\bea
{[B^{l\bar{l}}_{S,P}]}_4 &=& \mp \frac{\alpha_s}{4 \pi}
  \frac{4\kappa_q s_W^2}{3m_b}
 \sum_{i,j=1}^2 \sum_{f=1}^3  \sum_{a, b, c =1}^6 
 \frac{(X_j^{U_L\dagger})_{qb}(X_i^{U_R})_{a3}}{M_{\tilde\chi_i^{}}^2}
 [(P_U)_{ac} y_{ci}(P_U)_{cb}
 (1+L_{\tilde u_c})]\nnu\\
&&\hspace{-1cm}\mbox{}
\times\Bigg\{ f^{(0)}_{9}(x_{ji}, y_{ai}, y_{bi}, v_{fi})
(X_i^{N_R\dagger})_{lf}(X_j^{N_L})_{f l}
\pm \sqrt{x_{ji}} f^{(0)}_{10}(x_{ji}, y_{ai}, y_{bi}, v_{fi}) 
      (X_i^{N_L\dagger})_{lf}(X_j^{N_R})_{f l}\Bigg\},\nnu\\
\eea
\be
{[B^{l\bar{l}}_{S',P'}]}_4 = \pm
{[B^{l\bar{l}}_{S,P}]}_4 (
{X^{U_L}\leftrightarrow X^{U_R}; X^{N_L}\leftrightarrow X^{N_R};
m_b\rightarrow m_q}).
\ee

\subsubsection{Neutral Higgs boson penguin diagrams}
\label{sec:neuthiggs}
As we have already mentioned, the contributions of neutral Higgs bosons are 
relevant only for large values of 
$\tan\beta$. In this case, we obtain 
\be
 {[N^{l\bar{l}}_{S,P}]}_H = 
    \mp\frac{m_l \tan^2\beta}{4 M_W^2} 
    \Bigg\{x f_3^{(0)}(x,z) + \frac{\alpha_s}{4 \pi}
    \left[f_{14}^{(1)}(x,z) + 8 x \frac{\partial}{\partial x} [x
        f_3^{(0)}(x,z)] L_t\right]\Bigg\}, 
\ee
\bea
{[N^{l\bar{l}}_{S',P'}]}_{H}
&=&\pm [N^{l\bar{l}}_{S,P}]^*_{H},
\eea
\bea\label{neuthiggscharg}
[N^{l\bar{l}}_{S,P}]_{\tilde{\chi}}= &\pm &
\frac{m_l  \tan^2\beta}{M_W (M_H^2 - M_W^2)} 
    \sum_{i,j=1}^2 \sum_{a,b=1}^6 \sum_{m,n=1}^3 
    \Gamma_{imn}^a (\Gamma^{U_L})_{bm} U_{j2}a_Y\nnu\\
&\times &\Bigg[ a_0^{S,P} + a_1 \tan \b 
+ \frac{\alqcd}{4\p} a_2 m_q^2 \tan^2\b\Bigg],
\eea
\bea\label{neuthiggscharg:prime}
{[N^{l\bar{l}}_{S',P'}]}_{{\tilde{\chi}}}
&=&\pm [N^{l\bar{l}}_{S,P}]^*_{{\tilde{\chi}}}
(m_q\rightarrow m_b,\la_{mn}^*\to \la_{nm}),
\eea
where 
\be\label{susy:result:gamma}
\G^a_{imn}\equiv \frac{1}{2\sqrt{2}}
[\sqrt{2}M_W V_{i1}(\Gamma^{U_L\dagger})_{na}a_g -(M_U)_{nn}V_{i2}
     (\Gamma^{U_R\dagger})_{na}a_Y]\la_{mn},
\ee 
\be
\la_{mn}\equiv \frac{V_{mb}^{}V_{nq}^{\ast}}{V_{tb}^{}V_{tq}^{\ast}}, 
\ee
with $a_g, a_Y$ defined in \eq{decoupled:gluino}. 
The coefficients $a_0^{S,P}, a_1, a_2$ in \eq{neuthiggscharg} are given by  
\bea
\lefteqn{a_0^{S,P}= \mp \Bigg[ \sqrt{x_{ij}}
      \Bigg\{ f_3^{(0)}(x_{ij}, y_{aj}) + \frac{\alpha_s}{4\pi}\Bigg[
      f_{18}^{(1)}(x_{ij}, y_{aj}) +4 y_{ai} \frac{\partial}{\partial
        y_{ai}}f_3^{(0)}(x_{ij}, y_{aj}) L_{\tilde u_a}\Bigg]\Bigg\} 
      U_{i2} V_{j1}}\nnu\\
&\pm &\Bigg\{ f_4^{(0)}(x_{ij}, y_{aj}) + \frac{\alpha_s}{4\pi}
    \Bigg[f_{19}^{(1)}(x_{ij}, y_{aj}) +4 y_{ai}
      \frac{\partial}{\partial y_{ai}}f_4^{(0)}(x_{ij}, y_{aj})L_{\tilde u_a}\Bigg]\Bigg\} 
      U_{j2}^* V_{i1}^* \Bigg]
    \delta_{ab}\nnu \\
&+&\frac{(\D^\pm_i)_{ab}}{M_W}
\Bigg\{ f_3^{(0)}(y_{ai}, y_{bi}) + \frac{\alpha_s}{4\pi}\Bigg[
    f_{17}^{(1)}(y_{ai}, y_{bi}) +4 \Bigg(1+y_{ai}
      \frac{\partial}{\partial y_{ai}} + y_{bi} \frac{\partial}{\partial y_{bi}}\Bigg)
    f_3^{(0)}(y_{ai},y_{bi})L_{\tilde u_a}\Bigg]\Bigg\}\delta_{ij}\nnu\\
&+&\frac{\alpha_s}{4\pi} 
\Bigg[\frac{4\Gamma_{imn}^{a*}}{M_W\left(\Gamma^{U_L}\right)_{bm} 
    \dis \la_{mn}^*U_{j2}}\Bigg] 
f_{15}^{(1)}(y_{ai})\delta_{ij}\delta_{ab}\delta_{mn},
\eea
\bea
a_1=\frac{M_{\tilde{\chi}_i^{}}}{\sqrt2 M_W} 
    \Bigg\{ f_8^{(0)}(y_{ai}) +
    \frac{\alpha_s}{4\pi}\Bigg[f_{16}^{(1)}(y_{ai})+
    4 y_{ai} \frac{\partial}{\partial y_{ai}} f_8^{(0)}(y_{ai})L_{\tilde u_a}\Bigg]\Bigg\}\delta_{ij}\delta_{ab},
\eea
\bea
a_2= \frac{(\Gamma^{U_L\dagger})_{mb} 
    \lambda_{mn}U_{j2}^*}{2M_W\Gamma_{imn}^a}f_{15}^{(1)}(y_{ai})\delta_{ij}\delta_{ab}\delta_{mn},
\eea
with 
\be\label{dela:pm:i}
(\D^\pm_i)_{ab}\equiv  \sum_{f=1}^3 
\frac{(M_U)_{ff}}{\sqrt2 M_{\tilde{\chi}_i^{}}} 
[\mu^* (\Gamma^{U_R})_{af}(\Gamma^{U_L\dagger})_{fb} 
\pm \mu   (\Gamma^{U_L})_{af} (\Gamma^{U_R\dagger})_{fb}].
\ee
It is interesting to note that the contribution in 
Eqs.~(\ref{neuthiggscharg}) and (\ref{neuthiggscharg:prime}) 
that is proportional to 
$a_1\tan^3\b $ comes from the counterterm of the electroweak wave function 
renormalization. Since it is not suppressed by the mass of the light 
quarks, it gives by far the dominant contribution in the high $\tan\b$ 
regime. 
(Further details may be found in \rf{bobeth:etal:0401}.) 

Finally, the calculation of the diagrams involving quartic squark couplings 
yields 
\bea
 {[N^{l\bar{l}}_{S,P}]}_4 &=& 
    \mp \frac{\alpha_s}{4 \pi} \frac{4  m_l  \tan^2\beta}{3M_W^2 (M_H^2 
- M_W^2)}
    \sum_{i,j=1}^2 \sum_{m,n=1}^3 \sum_{a, \dots,e,g,k =1}^6     
    \Gamma_{imn}^a (\Gamma^{U_L})_{dm} U_{j2}a_Y
\nnu\\
&\times&\Bigg\{(P_U)_{ek} y_{kj}(P_U)_{kg}
     (1+L_{\tilde u_k})
    \Bigg\{\tan\beta \frac{M_{\tilde\chi_i^{}}}{\sqrt 2}
    f_3^{(0)}(y_{ai},y_{di}) \delta_{ij} \delta_{ae} \delta_{gd}\nnu\\
&+&(\D^\pm_i)_{bc}\Bigg[\delta_{ae}\delta_{gb}\delta_{cd}
f_6^{(0)}(y_{ai}, y_{bi}, y_{ci})
+\delta_{ab}\delta_{ce}\delta_{gd}
f_6^{(0)}(y_{ai}, y_{ci}, y_{di})\Bigg]\delta_{ij} \nnu\\
&\mp& M_W \Bigg[ \sqrt{x_{ij}}
      f_6^{(0)}(x_{ij}, y_{aj}, y_{dj})  
      U_{i2} V_{j1}\pm f_5^{(0)}(x_{ij}, y_{aj}, y_{dj})  
      U_{j2}^* V_{i1}^* \Bigg]
    \delta_{ae}\delta_{dg} \Bigg\}\nnu\\
&-&(P_U)_{ae} 
     [1+L_{\tilde u_g}-f^{(0)}_{11}(y_{ej},y_{gj})](P_U)_{gd}
     (\D^\pm_i)_{eg}\delta_{ij}f_3^{(0)}(y_{ai}, y_{di})\Bigg\},
\label{quartneuthiggs}
\eea
\bea\label{quartneuthiggs:prime}
{[N^{l\bar{l}}_{S',P'}]}_4
&=&\pm [N^{l\bar{l}}_{S,P}]^*_4(\la_{mn}^*\to \la_{nm}).
\eea
Notice that there is again a $\tan^3\beta$ enhancement in
Eqs.~(\ref{quartneuthiggs}) and (\ref{quartneuthiggs:prime}), 
due to the counterterm contributions.

\subsection{The \bm$\bar s\to \bar  d\nu\bar\nu$ transition}\label{SD:k:sector}
The results for the $\bar s\to \bar d\nu\bar \n$ transition related to
the coupling $V_{ts}^*V_{td}^{}$ can easily be obtained from the results of 
\Sec{btoq:transition} through the appropriate replacements of flavours. 
However, unlike the $b\to s(d) \n\bar \n$ transition,
the internal charm-quark contribution to 
$\bar s\to \bar d\nu\bar \n$ cannot be neglected, since 
$|V_{cs}^*V_{cd}^{}|/|V_{ts}^*V_{td}^{}|\sim 70$, which 
 partially compensates for the suppression of the charm-quark  
relative to the top-quark contributions
due to $m_c^2\ll m_t^2$ [cf.~\eq{SD:O}]. 
Yet, it turns out that the new-physics
contributions proportional to $V_{cs}^*V_{cd}^{}$ are small.
Accordingly, the charm contribution is completely described by the 
SM for which next-to-leading-order corrections are known from 
\cite{BB98,BB3}.

The effective Hamiltonian may then be written as 
\bea\label{hkpn} 
\heff= \frac{4G_F}{\sqrt{2}}\frac{\a}{2\p\sin^2\theta_W}
\sum_{f}[V^{\ast}_{cs}V_{cd}^{} X^f_\nl\Oi_L +
V^{\ast}_{ts}V_{td}^{}(c_L\Oi_L + c_R\Oi_R)],\nnu\\
\eea
where $f=e,\m,\t$ and  
\be\label{ops:kpn}
\Oi_L= (\bar{s}\gamma_\m P_L d)(\bar{\nu}_f\gamma^{\m} P_L \nu_f),\quad
\Oi_R= (\bar{s}\gamma_\m P_R d)(\bar{\nu}_f\gamma^{\m} P_L \nu_f).
\ee
The coefficients $c_L$ and $c_R$ are given by   
\eqs{wilson:coeffs::bqnunu}{tadpole:br} if we make the replacements
\be\label{replacements}
m_bm_q\to m_dm_s, \quad \k_q\to \bar \k=\frac{1}{8\sqrt{2}G_F e^2 
V_{td}^{}V_{ts}^*}.
\ee
Additionally, the indices of the 
matrices $X^{U_{L,R}}_i$ in \eqs{CLnunu:chargino}{tadpole:cr} 
and (\ref{BLnunu:chargino})--(\ref{tadpole:br}) 
should be changed as follows: $q\to 2$, $3\to 1$.
Note that the results of the 
previous subsections suggest that the contributions 
of the Wilson coefficient $c_R$ can be neglected in  the case of 
the $s\to d$ transition.
Indeed, since $c_R$ involves always the factors 
$m_s m_d/M_W^2$, it is far too small to give an appreciable 
contribution.

As for the function $X^f_\nl$ in \eq{hkpn}, it 
results from the next-to-leading order calculation \cite{BB3}, and is
given explicitly in \rf{BB98}, where numerical values of $X^f_\nl$ for
$\m = m_c$ and different choices of $\Lms^{(4)}$ and $m_c$ may  be found. 
Note that there is typically a $30\%$ suppression of the charm contribution 
due to the QCD corrections.

\subsection{The \bm$\bar s\to \bar  d l^+ l^-$ transition}
The results for this transition can be readily obtained from those given 
in \Sec{btoqll:transition}. As in the $\bar s\to \bar d\nu\bar
\nu$ transition, the charm contribution proportional to $V_{cs}^*V_{cd}^{}$ cannot be 
neglected. However, it turns out that 
the new-physics effects in the charm sector 
are numerically small, and thus the charm contribution is dominated by 
that of the SM.

The effective Hamiltonian is then given by
\bea\label{klmumu}
\heff = &-&\frac{2G_F}{\sqrt{2}}\frac{\a}{2\p \sin^2
\theta_W}\Bigg[V_{cs}^*V_{cd}^{} Y_{\nl}\Oi_A \nnu\\
&+&
V_{ts}^*V_{td}^{}(c_{A} \Oi_{A} + c_{A}' \Oi_{A}'+ 
c_S \Oi_S + c_S' \Oi_S' + c_P \Oi_P + c_P' \Oi_P')\Bigg],
\eea
where the function $Y_{\nl}$ (the analogue of $X_\nl^f$) has been 
calculated in \rfs{BB98,BB3}. The remaining short-distance coefficients
are given in   
Eqs.~(\ref{wilson:coeffs:general})  with the replacements
(\ref{replacements}), together with 
\be
\la_{mn}\to \bar \la_{mn}\equiv 
\frac{V_{md}^{} V_{ns}^{\ast}}{V_{td}^{}V_{ts}^{\ast}},
\ee
while the operators can be obtained from 
\eqs{ops:bsll:av}{ops:bsll:p} through the appropriate changes of 
flavours.

\subsection{Renormalization group evolution and scale dependence}\label{RGE}
The renormalization scale dependence of the Wilson coefficients 
$\vec{C}(\m)$ is governed by
\be
\frac{d}{d\ln\mu} \vec{C}(\mu) = \hat\gamma^{\text{T}} \vec{C}(\mu),
\ee
where $\hat\gamma$ is the anomalous dimension matrix.\footnote{For a 
detailed discussion see, e.g., \rfs{BBL:review,AJB:leshouches}.}~Since 
the anomalous dimensions for all operators in Eqs.~(\ref{ops:bqnunu}),
(\ref{ops:bsll:av})--(\ref{ops:bsll:p}) and (\ref{ops:kpn}) vanish, the Wilson coefficients
must be independent of $\m$. Otherwise the physical amplitudes would depend 
on the renormalization scale.

We now demonstrate that all contributions from the 
$Z^0$-penguin diagrams, as well as from box and penguin diagrams with  
neutral Higgs bosons, are separately independent of $\m$ when 
$O(\alqcd)$ corrections are taken into account. 

In the leading-order expressions of the Wilson coefficients, there are two 
sources of scale dependence. Firstly, the running quark mass
\be\label{gamma0:beta0}
m_q(\m) = m_q(\m_0)\left[\frac{\alqcd(\m)}{\alqcd(\m_0)}
                 \right]^{\gamma_m^{(0)}/(2\beta_0)},\quad
\g^{(0)}_m=8, \quad \b_0= 11-\frac{2}{3}n_f,
\ee
where $m_q(\m_0)$ is the value of the quark mass at the scale $\m_0$ 
and $n_f$ is the number of active flavours.
To first order in $\alqcd$, the above expression can be written as
\be
m_q(\mu) = m_q(\m_0) \Bigg[1-\frac{\gamma_m^{(0)}}{2} 
\frac{\alqcd(\m_0)}{4\pi}\ln\Bigg(\frac{\mu^2}{\m_0^2}\Bigg) \Bigg].
\ee
Secondly, the running squark mass
\be\label{running:squark:masses} 
m_{\tilde q}(\mu) = m_{\tilde q}(\m_0) 
\left[\frac{\alqcd(\mu)}{\alqcd(\m_0)}
\right]^{\gamma_{\widetilde m}^{(0)}/(2\beta_0)}, \quad
\gamma_{\widetilde m}^{(0)} = 4,
\ee
which to first order in $\alqcd$ can be written as
\be\label{running:squark:masses:approx}
m_{\tilde q}(\mu) = m_{\tilde q}(\m_0) \Bigg[1-
\frac{\gamma_{\widetilde m}^{(0)}}{2} \frac{\alqcd(\m_0)}{4\pi} 
\ln\Bigg(\frac{\mu^2}{\m_0^2}\Bigg) \Bigg].
\ee
The $\m$ dependence of the top-quark mass in the leading-order  terms in both 
the SM and the 2HDM coefficients is cancelled by the $\alqcd$ corrections of the Wilson coefficients that involve the derivatives of the leading functions
[see, e.g., Eqs.~(\ref{bsnunu:SM}) and (\ref{bsnunu:2HDM})].
The same applies to the $\m$ dependence of the squark masses in the supersymmetric contributions [see, e.g., \eq{CLnunu:chargino}].

As can be seen, for example, from \eq{bsnunu:2HDM:CR}, 
there is an additional $\m$ dependence in the leading contributions of the 
new operators generated by charged 
Higgs-boson exchanges that is related to the light quark masses 
present in the Yukawa 
couplings. This dependence on $\m$ is cancelled by the $O(\alqcd)$
corrections proportional to the leading functions.

Finally, we observe that in the chargino sector there are additional $\m$ 
dependencies in the chargino-squark-quark vertices, which are given in 
Eqs.~(\ref{mu:dep:XUL}) and (\ref{mu:dep:XUR}). These $\m$
dependencies arise not only from the $\m$ dependence of $M_U$ and
$M_D$  but also from the coefficients $a_g$ and $a_Y$ given 
in \eq{decoupled:gluino}. Taking into account both $\m$ 
dependencies, the effective 
$\m$ dependence of the vertex $X_i^{U_{L,R}}$ is given by
\be
  X_i^{U_{L,R}}(\mu) = 
  X_i^{U_{L,R}}(M_{\tilde g}) \left[1- 2\frac{\alpha_s(\mu)}{4\pi}
\ln\left(\frac{\mu^2}{M_{\tilde g}^2}\right)\right].
\ee
This $\m$ dependence is cancelled by the corresponding dependence 
in the $O(\alqcd)$ corrections proportional to 
the leading functions [see, e.g., \eq{CLnunu:chargino}]. 
We observe that the $\m$ dependence of 
$a_g$ and $a_Y$ is essential to obtain $\m$-independent chargino 
contributions.
The $\mu$ dependence in $a_g$ and $a_Y$ is related to the fact 
that the effective theory does not contain gluinos, which have been 
integrated out at the scale $O(M_{\tilde g})$. 
Thus, in the 
effective theory the gaugino and Higgsino couplings renormalize 
differently from the ordinary gauge and Yukawa
couplings. If one expresses the supersymmetric couplings at 
$\mu\ll M_{\tilde g}$ in terms of the $\sm$ couplings such as $g$ and the 
top Yukawa coupling, this difference has to be taken into account in order
to obtain correct results.

An additional $\mu$ dependence arises in the MSSM from the 
$O(\alpha_s)$ contributions due to the quartic squark vertex.
Note that only the $\mu$ dependence from gluonic corrections
has been taken into account in Eqs.~(\ref{running:squark:masses})
and (\ref{running:squark:masses:approx}). The inclusion of the contribution 
from the quartic squark coupling
results in a modification of the anomalous dimension
$\gamma_{\widetilde m}^{(0)}$. This additional contribution
would then cancel the
$\mu$ dependence present in the two-loop diagrams with quartic squark
couplings denoted by the subscript $J=4$ in Eqs.~(\ref{wilson:coeffs::bqnunu}) and (\ref{wilson:coeffs:general}).
In the next section, we will give a recipe how to avoid the appearance
of the quartic squark coupling using an on-shell prescription for
the squark mass. In this sense Eqs.~(\ref{running:squark:masses})
and (\ref{running:squark:masses:approx}) can be considered
as complete.

\newsection{Details of the calculation}\label{calculation:details}
In this section we present the details of the
calculation.~Readers who are not interested in these technical issues 
should proceed to \Sec{branching:ratios}. 

For processes that take
place at scales much lower than $M_W$, models such as the SM, 2HDM and MSSM 
can be replaced by an effective theory by means of integrating out all
particles heavier than $O(M_W)$. Our aim is to find
the QCD corrections to the Wilson coefficients of the relevant
operators in the effective theory. The operators of interest 
are given in Eqs.~(\ref{ops:bqnunu}) and 
(\ref{ops:bsll:av})--(\ref{ops:bsll:p}) for the quark-level transition
$b\rightarrow s (d) \nu\bar\nu$ and $b\rightarrow s (d) l^+ l^-$,
respectively. (The presentation given below applies to $b\to s l^+l^-$
transitions and with obvious changes to  $b\to s \nu\bar\nu$.)  

The simplest way of finding the Wilson coefficients is to require equality
of one-particle-irre\-du\-cible (1PI) amputated Green's functions 
calculated in the full and effective theory. The former 
requires the calculation of box-
and penguin-type diagrams. The different possible topologies
can be found in \figs{fig:peng}{fig:box}. 
Note that photonic penguin diagrams do not contribute
to the processes under consideration. Within the SM, the 
neutral Higgs boson penguin contributions are negligibly small, since 
they are suppressed by the light lepton and quark masses
and not enhanced by $\tan\beta$. All together we have
16 box diagrams in the SM, additional 20 box diagrams in the 
2HDM and 4 box diagrams in
the chargino sector of the MSSM. Further,  the number of $Z^0$-penguin
diagrams amounts to 40 in the SM, additional 16 in the 2HDM and 
20 in the
chargino sector. Note that the topology shown in \fig{fig:peng:quartic}
is present only in the chargino sector. We have 240 diagrams
containing neutral Higgs bosons in the 2HDM and 64 neutral Higgs 
penguin diagrams containing charginos and squarks.  
While counting the diagrams, different quark and squark flavours, as
well as different chargino generations, are not taken into
account. Note that the up- and charm-quark
contributions to the SM and 2HDM Wilson coefficients are included 
via the GIM mechanism, as explained  in 
\Sec{btoq:transition}. The sum
over all squarks and charginos is shown explicitly in our results.

The calculation has been performed
within a covariant $R_{\xi_g}$ and $R_{\xi_W}$ gauge for the gluon and
the $W$ boson, respectively, which
provides a useful check of our computation. The contributions of the
penguin and box diagrams to the Wilson coefficients are
separately  \emph{gauge independent} with respect to the 
gluon gauge, but in general \emph{gauge dependent} with respect to the 
$W$-boson
gauge. Thus, the results presented in the previous sections are
given in the 't Hooft-Feynman gauge for the $W$ boson. We have checked
that the sum of penguin and box diagrams contributing to 
the Wilson coefficients is independent of $\xi_W$. 

We are interested
in dimension six four-fermion operators, so the external momenta can be set to zero
from the very beginning. The masses of the light quarks can be safely
neglected as long as they are not enhanced by $\tan\beta$. That is, we
neglect $m_b,m_{d,s}$ and $m_l$ in the propagator, but keep them 
in the Yukawa couplings when multiplied by a factor of $\tan\beta$. Since
we are interested in on-shell results, it is pertinent to ask whether
we are allowed to keep light masses in the Yukawa coupling while neglecting 
external momenta.
In fact, higher-order terms in external momenta would enlarge our 
operator basis, but also give contributions to Wilson coefficients of 
dimension-six
operators after applying the equation of motion.~However,
these terms are not enhanced by $\tan\beta$ as they come solely from
an expansion of the propagators, and therefore are negligible. 

Setting the light quark masses to zero in the propagator 
produces infrared (IR) divergencies. We regularize
 the IR and ultraviolet (UV) divergencies simultaneously in 
$D=4 - 2 \varepsilon$
dimensions. At one-loop level, the UV divergencies in the penguin diagrams
can be removed by the electroweak renormalization of the wave
function. We have chosen an on-shell prescription, following the approach 
of \rf{bobeth:etal:0401}.
The QCD renormalization, which becomes necessary at the two-loop level, 
is performed
in the $\overline{\rm MS}$ scheme. 

As mentioned earlier, the results for the Wilson coefficients presented in 
\Sec{effective:Hamiltonian} are given in terms of the running $\msbar$ 
quark and squark masses $m_q\equiv {m}_q(\m)$ and  
$m_{\tilde q}\equiv {m}_{\tilde q}(\m)$,
respectively. Alternatively, one can work with the pole masses,
in which case the following steps should be performed:

{\bf Step 1} Remove the contributions due to quartic squark couplings
(i.e.~the contributions with the index $J=4$).

{\bf Step 2} Make a shift from the $\msbar$ scheme to the corresponding 
pole masses, namely:\footnote{Note that only in the case of light quarks
($d,s,b$) it is necessary to resum the large logarithms.}
\bea\label{top-quark:onshell}
{m}_t (\m) = m^{\rm pole}_t \Bigg\{1-\frac{\alpha_s(m^{\rm pole}_t)}{4\pi}\Bigg[\frac{16}{3}-4\ln \Bigg(\frac{m^{\rm pole}_t}{\m}\Bigg)^2
\Bigg]\Bigg\},
\eea
\bea\label{light:quark:onshell}
 m_q (\m) &=& m^{\rm pole}_q \Bigg[1-\frac{\alpha_s(m^{\rm pole}_q)}{4
    \pi}\frac{16}3\Bigg] \Bigg[\frac{\alpha_s(\mu)}{\alpha_s(m^{\rm
      pole}_q)}\Bigg]^{\gamma_m^{(0)}/(2 \beta_0)}\nnu\\
&\times &
    \Bigg\{1+\Bigg[\frac{\gamma_m^{(1)}}{2 \beta_0}-\frac{\beta_1
        \gamma_m^{(0)}}{2\beta_0^2}\Bigg]\frac{\alpha_s(\mu)-\alpha_s(m^{\rm pole}_q)}{4 \pi}\Bigg\} \quad (q=d,s,b),
\eea
\bea\label{squark:onshell}
{m}_{\tilde q}(\m) = 
m^{\rm pole}_{\tilde q} \Bigg\{1-\frac{\alpha_s(m^{\rm pole}_{\tilde q})}{4\pi}\Bigg[\frac{14}{3}-2\ln \Bigg(\frac{m^{\rm pole}_{\tilde q}}{\m}\Bigg)^2\Bigg]
\Bigg\},
\eea
where
\be
\g_m^{(1)}=\frac{404}{3}-\frac{40}{9}n_f,\quad \b_1=102-\frac{38}{3}n_f,
\ee
and $\g_m^{(0)}, \b_0$ are given in \eq{gamma0:beta0}.
Observe that the shift in \eq{squark:onshell} involves only the 
gluonic corrections, since  the
contributions due to quartic squark couplings have  
already been considered in step 1. 
In this context, we would like to remark that the absence of the
`quartic' contributions in an on-shell scheme is
related to the  renormalization of the squark mass and mixing angle. 
(For details, we refer the reader to \rf{djouadi:etal}.)

After the proper renormalization, the left-over divergencies  in the
full and effective theory must be of infrared character, and they must
cancel each other in the matching procedure. Note that in
order to obtain this cancellation, the treatment of light masses must be
identical in the full and the effective theory. 
As the light masses are zero, all
unrenormalized loop diagrams in the effective theory vanish because of 
a cancellation of IR and UV divergencies, which considerably simplifies
our calculation. The UV counterterms in  the
effective theory  must therefore exactly reproduce the  IR divergencies of
the full theory, which provides a further check of the
calculation. We wish to emphasize that owing to the dimensional
regularization of the IR divergencies, the matching procedure must be
performed in $D$ dimensions \cite{BKP}. 

In intermediate steps of the calculation structures like
\be
(\gamma_{\alpha_1} \gamma_{\alpha_2} \gamma_{\alpha_3} P_A) \otimes
(\gamma^{\alpha_1} \gamma^{\alpha_2}  \gamma^{\alpha_3}P_B)
\ee
occur with $P_{A,B}$ being either $P_L$ or $P_R$. They cannot be reduced
using  $D$ dimensional Dirac algebra, due to
the appearance of the matrix $\gamma_5$. Only after the matching all
divergencies cancel and the limit $D\rightarrow 4$ can be taken.
Consequently, evanescent operators must be introduced in the
effective theory. These operators appear in box diagrams in the SM
and in the MSSM. For example, in the SM we define the evanescent 
operator for $b\to s l^+l^-$ as follows :
\be\label{evanops1}
{\cal O}_1^E = (\bar{s}\gamma_{\alpha_1} \gamma_{\alpha_2} 
\gamma_{\alpha_3} P_L b)(\bar{l}\gamma^{\alpha_3} \gamma^{\alpha_2}  
\gamma^{\alpha_1} P_L l) - 
               4(\bar s \g_\m P_L b)(\bar l \g^\m P_L l). 
\ee
(Further details can be found in \cite{MU98}.)
In the chargino sector of the MSSM, operators with a different spinor
ordering show up in box diagrams:
\begin{subequations}\label{mssmbasis}
\be
\tilde{{\cal O}}_{SAB} = (\bar{s} P_A l)(\bar{l} P_B b),
\ee
\be
\tilde{{\cal O}}_{VAB} = (\bar{s}\gamma_\alpha P_A l)(\bar{l}\gamma^{\alpha} P_B b),
\ee
\be
\tilde{{\cal O}}_{TAA} = (\bar{s}\sigma_{\alpha\beta} P_A l)
(\bar{l}\sigma^{\alpha\beta} P_A b).
\ee
\end{subequations}
We are not allowed to project these
operators onto the operators given in
Eqs.~(\ref{ops:bsll:av})--(\ref{ops:bsll:p}). 
For such a projection we would have to apply Fierz
identities which cannot be continued to $D$ dimensions.
For this reason, we have to define the following so-called `Fierz-vanishing'
evanescent operators \cite{BMU00,jamin:pich:94}:
\begin{subequations}
\be
\tilde{\cal O}_{SLL}^E = \tilde{\cal O}_{  SLL}+\frac{1}{2}{\cal O}_{ 
  SLL}-\frac{1}{8} {\cal O}_{  TLL},
\ee
\be
\tilde{\cal O}_{SLR}^E = \tilde{\cal O}_{  SLR}+\frac{1}{2} {\cal O}_{  VRL},
\ee
\be
\tilde{\cal O}_{  VLL}^E = \tilde{\cal O}_{  VLL} - {\cal O}_{  VLL},
\ee
\be
\tilde{\cal O}_{  VLR}^E = \tilde{\cal O}_{  VLR}+2 {\cal O}_{  SRL},
\ee
\be
\label{evanescent:ops}
\tilde{\cal O}_{  TLL}^E = \tilde{\cal O}_{  TLL}-6 {\cal
  O}_{   SLL} -\frac{1}{2} {\cal O}_{TLL},
\ee
\end{subequations}
as well as operators which can be obtained by an interchange of 
$P_L\leftrightarrow P_R$. The operators without tilde are identical to the
ones given in Eqs.~(\ref{mssmbasis}), but with exchanged $l$ and $b$ spinors.
Moreover $\sigma_{\mu\nu}=[\gamma_\mu,\gamma_\nu]/2$. Due to a finite
mixing into the physical operators  the `Fierz-vanishing' evanescent
operators contribute at next-to-leading order.
A similar situation was  described in \cite{MU98}.
From  \eq{evanescent:ops} it is evident that we cannot neglect the 
contributions to the tensor operator $\tilde{\Oi}_{TAA}$ as it affects 
the Wilson coefficients of the scalar operators. 
Furthermore the following evanescent operators are necessary at
intermediate steps:
\bea
\label{evanops2}
\tilde{\cal O}_1^E &=&
  (\bar{s}\gamma_{\alpha_1} \gamma_{\alpha_2} \gamma_{\alpha_3} P_L l)
  (\bar{l}\gamma^{\alpha_1} \gamma^{\alpha_2}  \gamma^{\alpha_3} P_L b) - 
  16 \tilde{\cal O}_{VLL},\nnu\\
\tilde{\cal O}_2^E &=&
  (\bar{s}\gamma_{\alpha_1} \gamma_{\alpha_2} \gamma_{\alpha_3} P_L l)
  (\bar{l}\gamma^{\alpha_1} \gamma^{\alpha_2}  \gamma^{\alpha_3} P_R b) - 
  4 \tilde{\cal O}_{VLR}. 
\eea

\newsection{Branching ratios}\label{branching:ratios}
\subsection{Rare \bm$B$  decays}
\subsubsection{\bm$\B\to X_{d,s}\nu\bar\nu$}
The decays $\B\to X_q\nu\bar\nu$ ($q=d,s$) are the cleanest theoretically
in the field of rare $B$ decays. 
Since the neutrinos escape detection, 
these decays are probed by requiring very large missing energy, 
$\miss_q \equiv (E_B-E_{X_q})/m_b$
(for a full discussion see, e.g.,  \rf{grossman:etal}).

Using the effective Hamiltonian given in \eq{eff:ham:bqnunu},
we obtain the differential decay rate
\bea\label{diff:branch:bsnunu}
&&\frac{d \branch(\B \to  X_q\nu\bar\nu)}{d\miss_q} 
= 
\frac{4 \a^2}{\p^2\sin^4\theta_W}\,
\frac{|V_{tb}^{}V_{tq}^*|^2}{|V_{cb}|^2}\frac{\branch(\B\to X_c e \bar \n_e)\k(0)}{ f(\hat m_c )\k(\hat m_c)}[(1-\miss_q)^2-\mqhat^2]^{1/2}\sum_{f}
\nnu\\
&&\mbox{}\times\Bigg\{
(|c_{L}|^2 + |c_{R}|^2)[(1-\miss_q)(4\miss_q-1)+\mqhat^2(1-3\miss_q)] + 6
 \mqhat (1-2 \miss_q-\mqhat^2)\Re (c_L^* c_R^{})\Bigg\},\nnu\\
\eea
where the sum runs over the flavour of the three neutrinos,
$\hat m_i \equiv m_i/m_b$, and the various 
input parameters are listed  in Table \ref{table:input}.
%
%
\begin{table}
\begin{center}
\caption{Input parameters used in our numerical analysis. 
The isospin-breaking 
corrections $r_K$ and the charm contributions 
$X_\nl \equiv (2 X^e_\nl + X^\tau_\nl)/3,Y_\nl$ have been taken 
from Refs.~\cite{marciano:zarsa} and \cite{BB98}, respectively. Note that 
$m_c$ is the running charm quark mass in the $\msbar$ scheme 
normalized at $m_c$.
For the CKM matrix, we use the standard parametrization \cite{CKM:standard},
with four independent parameters  $s_{12},s_{13}, s_{23},\delta$. 
As for the remaining parameters, we utilize the values 
compiled by  the Particle Data Group \cite{PDG}. \label{table:input}}
\vskip0.2cm

\begin{tabular}{lcc}\hline\hline
Quantity && Value \\ \hline
$\sin^2\theta_W$ && $0.23$\\
$\a_s (M_Z)$ && $0.118$\\
$s_{12}$ & & $0.222$\\
$s_{13}$ & & $3.49\times 10^{-3}$\\ 
$s_{23}$ & & $0.041$\\ 
$\delta$ & & $57^\circ$\\ 
$|V_{td}|$ & & $7.81\times 10^{-3}$\\ 
$f_{B_s}$ & & $230\ \MeV$\\ 
$m_b^\pole$ & & $4.8\ \GeV$\\
$m_t^\pole$ & & $174.3\ \GeV$\\
$m_c$ & & $1.3\ \GeV$\\    
$\a$ && $1/129$\\
$r_1$ & & $1.17\times 10^{-4}$\\
$r_2$ & & $0.24$\\
$r_3$ & & $13.17$\\
$r_{K^+}$ && $0.901$\\
$r_{K_L}$ && $0.944$\\
$X_\nl$ && $9.78\times 10^{-4}$\\
$Y_\nl$ && $3.03\times 10^{-4}$\\
$\branch(\B\to X_c e\bar \n_e)$ & & $10.58\%$\\
\hline\hline
\end{tabular}
\end{center}
\end{table}
%
%
\begin{table}
\begin{center}
\caption{Numerical values for the running quark masses 
$m_i\equiv  m_i(\m)$ employed in our 
analysis.\label{table:quark-masses}}
\vskip0.2cm

\begin{tabular}{lcccc}\hline\hline
Scale&$m_t\ [\GeV] $& $m_s \ [\MeV]$ & $m_b\ [\GeV]$\\ \hline
$\m=m_t^\pole$ &$166$ &$61$& $2.9$\\ 
$\m=m_b^\pole$ &$-$ &$90$ &$4.4$\\ 
$\m=2\ \GeV$ &$-$ &$110$&$-$\\ 
\hline\hline
\end{tabular}
\end{center}
\end{table}
The expression for the missing energy spectrum is equivalent to   
the result presented in \rf{grossman:etal}.
In writing  \eq{diff:branch:bsnunu}, we have 
neglected non-perturbative corrections  of $O(1/m_b^2)$ 
\cite{mb:corrections} and $O(1/m_c^2)$ 
\cite{mc:corr:buchalla:etal,mc:corrections} which have been found to be small over most of the Dalitz plot.\footnote{Unlike the 
$b\to q l^+l^-$ transition, there is no virtual photon contribution
in $b\to q \n\bar\n$, so that non-perturbative 
corrections due to charm quarks are further suppressed \cite{mc:corr:buchalla:etal}.} 
Furthermore, neglecting non-perturbative corrections, 
we have used the inclusive semileptonic decay rate
\be
\G(\B\to X_c e \bar \n_e)=\frac{G_F^2 m_b^5}{192\p^3}|V_{cb}|^2
f(\hat m_c )\k(\hat m_c)
\ee
in order to remove the uncertainties due to an overall factor of $m_b^5$.~The 
functions $f(\hat m_c )$ and $\k(\hat m_c)$ represent 
the phase-space and the one-loop QCD corrections, respectively 
\cite{cabibbo:maiani}:
\be\label{phase:space}
f(\hat m_c)=1-8\hat m_c^2 +8\hat m_c^6-\hat m_c^8-24\hat m_c^4 \ln \hat m_c,
\ee
\be\label{kappa:exact}
\k(\hat m_c) = 1+ \frac{\alqcd(m_b)}{\p}\frac{A_0(\hat m_c)}{f(\hat m_c)},
\ee
where $A_0$ can be found in 
\rf{kappa:exact}. Expanding $\k(\hat m_c)$ in 
\eq{kappa:exact} around $\hat m_c=0.3$ results in  
\be\label{kappa:approx}
\k(\hat m_c)\simeq 1-\frac{\alqcd(m_b)}{\p}\Bigg[1.670 +2.027(0.3-\hat m_c)
+2.152 (0.3-\hat m_c)^2\Bigg],
\ee
which is  accurate to better than $1\%$, and hence perfectly  
adequate for our purposes. Further, 
\be\label{kappa:zero}
\k(0)=  1+ \frac{\alqcd(m_b)}{\p}\Bigg[\frac{25}{6}-\frac{2}{3}\p^2\Bigg]
=0.830
\ee
represents the QCD correction to the matrix element of $b\to q \n\bar\n$ 
due to virtual and bremsstrahlung contributions \cite{BBL:review}.

Integration of \eq{diff:branch:bsnunu} 
over $(1-\mqhat^2)/2\leqslant \miss_q \leqslant (1-\mqhat)$
then yields the branching ratio   
\bea\label{branch:bsnunu}
\branch(\B\to X_q\nu\bar\nu) 
&=& 
\frac{\a^2}{4 \p^2\sin^4\theta_W}\,
\frac{|V_{tb}^{}V_{tq}^*|^2}{|V_{cb}|^2}\frac{\branch(\B\to X_c e \bar \n_e)
\k(0)}{ f(\hat m_c )\k(\hat m_c)}\nnu\\
&\times&\sum_{f}\Bigg\{(|c_{L}|^2 + |c_{R}|^2)f(\mqhat)-4\Re(c_Lc_R^*)
\mqhat \tilde f (\mqhat)\Bigg\},
\eea
where 
\be
\tilde f (\mqhat)=1+9\mqhat^2-9\mqhat^4-\mqhat^6+12\mqhat^2(1+\mqhat^2)
\ln \mqhat,
\ee
and $f(\mqhat)$  is given in \eq{phase:space}. 

Thus far, no attempt has been made to search for the inclusive  
$b\to d\n\bar \n$ decay, and so we concentrate on the $b\to s$ transition.~The best upper limit has been set by the 
ALEPH Collaboration 
\cite{bsnunu:aleph}:\footnote{A similar upper limit has been derived 
by the CLEO Collaboration for the 
branching fraction $\branch(B^\pm \to K^\pm \nu\bar\nu)< 2.4 \times 10^{-4}$ 
at $\cl{90}$ \cite{bsnunu:cleo}. We do not address the issue of 
exclusive decays here (see, e.g., \rfs{bsnunu:excl,GGG01}).}
\be
\branch(\B \to X_s \nu\bar\nu)< 6.4 \times 10^{-4}\quad (\cl{90}).
\ee
Using the numerical values listed in Tables \ref{table:input} and 
\ref{table:quark-masses}, together with 
\eq{kappa:zero}, we obtain
\bea\label{btosnunubar:br}
\branch(\B\to X_s\nu\bar\nu)&=& 5.39\times 10^{-6}\  
\left[\frac{0.53}{f(\hat m_c)}\right]
\left[\frac{0.88}{\kappa(\hat m_c)}\right]
\left[\frac{\branch(\B\to X_c e \bar \n_e)}{10.58\%}\right]
\frac{|V_{ts}|^2}{|V_{cb}|^2}\nnu\\
&\times& \sum_{f}
(|c_{L}|^2 + |c_{R}|^2)
\left[1-0.08\frac{\Re(c_L^{}c_R^*)}{|c_{L}|^2 + |c_{R}|^2}\right],
\eea
which will be used in the subsequent analysis.
We note in passing that, since 
\be
- \frac{1}{2}\leqslant
\frac{\Re(c_L^{}c_R^*)}{|c_{L}|^2 + |c_{R}|^2} \leqslant \frac{1}{2},
\ee
the last term in square brackets in \eq{btosnunubar:br} deviates from 
unity by at most $4 \%$.
The SM result is obtained by summing over the neutrino flavours $f$, 
and 
taking the limit $c_R\to 0$
while retaining only the SM 
contribution in $c_L$. In this case, $c_L=X_{\sm}$, 
with $X_{\sm}$ given in \eq{sm:x} of the Appendix. 

\subsubsection{\bm$\B_{d,s}\to l^+l^-$}\label{branching:ratios:Bmumu}
The decays $\B_q\to l^+l^-$ are after $\B\to X_q\nu\bar\nu$ 
the theoretically cleanest decays in the field of rare $B$ decays. In fact, 
like in the decay $\B\to X_{d,s}\nu\bar\nu$, 
the charm contributions are completely negligible.
These processes, which are dominated by $Z^0$-penguin and box 
diagrams, have been studied by a number of authors 
\cite{first:papers:Bmumu,me:bqll:SUSY,bobeth:etal:0401,Bmumu:susy:epsilon,gino:retico:01,GGG01,me:bqll}
in extensions of the SM, but without QCD corrections.

Let us start by considering the matrix element for the decay 
$\B_q\to l^+l^-$ in the presence of the operators defined in 
\eqs{ops:bsll:av}{ops:bsll:p}, which has the 
general form\footnote{Observe that there is no 
tensor-type interaction as
$\bracket{0}{\bar{q}\s_{\m\n} b}{\B_q(p)}\equiv 0$. 
In fact, it is not possible to construct a 
combination made up of $p^\m$
that is antisymmetric with respect to the index 
interchange $\m\leftrightarrow\n$.} 
\be
{\mathcal M}=-i f_{B_q}\frac{ G_F \a}{\sqrt{2}\p\sin^2\theta_W }V_{tb}^{}V_{tq}^{\ast}
[F_S \bar{l}l + F_P \bar{l}\g_5 l  + 
F_V p^{\m} \bar{l}\g_{\m}l+ F_A p^{\m} \bar{l}\g_{\m}\g_5 l],
\ee 
where the $F_i$'s  are Lorentz-invariant form factors, $p^\m$ is the 
four-momentum of the initial $B$ meson, and $f_{B_q}$ is the 
corresponding decay constant defined via the axial vector 
current matrix element
\be
\bracket{0}{\bar{q}\g_{\m}\g_5 b}{\B_q(p)}=ip_{\mu}f_{B_q},
\ee
while the  matrix element of the vector current vanishes.  
It may also be noted that the form factor $F_V$ 
does not contribute to the decay $\B_q\to l^+ l^-$ since 
$\bar l \pslash l=0$.
Employing the equation of motion, we find for the remaining 
matrix element
\be\label{matrix:element:btoll:II}
\bracket{0}{\bar{q}\g_5 b}{\B_q(p)}=- i f_{B_q} \frac{M_{B_q}^2}{m_b+m_q}.
\ee

Squaring the matrix element and summing over the final lepton spins, 
the branching ratio can be written in a compact form \cite{bobeth:etal:0401}:
\bea\label{BR:bqll}
\branch(\B_q\to l^+ l^-)=\frac{G_F^2 \a^2 M_{B_q}f_{B_q}^2\t_{B_q}}{16 \p^3\sin^4\theta_W}
|V_{tb}^{}V_{tq}^{\ast}|^2 \sqrt{1-\frac{4m_l^2}{M_{B_q}^2}}\Bigg\{
\Bigg(1-\frac{4m_l^2}{M_{B_q}^2}\Bigg)|F_S|^2 +
|F_P + 2 m_l F_A|^2\Bigg\}.\hspace{-2cm}\nnu\\
\eea
Here, $\tau_{B_q}$ is the lifetime of the $B_q$ meson and
\be\label{Fsubis}
F_S=\frac{1}{2}M_{B_q}^2\Bigg[\frac{c_S -c_S' \hat m_q}{1+\hat m_q}\Bigg],\quad
F_P=\frac{1}{2}M_{B_q}^2\Bigg[\frac{c_P -c_P'\hat m_q}{1+\hat m_q}\Bigg],
\quad F_A= \frac{1}{2}(c_A-c_A')
\ee
(remembering that $\hat m_q \equiv m_q/m_b$).
At present, the best upper limit on the above decay modes comes from the 
Collider Detector at Fermilab (CDF)
and has been derived for the $b\to s$ transition
\cite{exp:bstomumu}: 
\be\label{exp:limit:bsmumu}
\branch(\B_s\to \m^+\m^-)<2.6\times 10^{-6} \quad (\cl{95}),
\ee
and so we focus on the $\B_s\to\mu^+\mu^-$ decay. 

Recalling the scalar and pseudoscalar Wilson coefficients in 
\Sec{btoqll:transition}, it turns out that the contributions of
the operators $\Oi_{S,P}$ and $\Oi'_{S,P}$ are of 
comparable size. Thus, the Wilson coefficients $c_{S,P}'$   
in \eq{Fsubis} can be neglected since $\hat m_q \ll 1$ for 
$q=d,s$. 

Introducing the dimensionless Wilson coefficients 
\be
\tilde c_S=M_{B_s}c_S, \quad  \tilde c_P=M_{B_s}c_P,
\ee 
the branching fraction is given by
\bea\label{num:BRbmumu}
  \branch(\B_s\to\mu^+\mu^-)&=&2.32\times 10^{-6}
\Bigg[\frac{\tau_{B_s}}{1.5\ \ps}\Bigg]
  \Bigg[\frac{f_{B_s}}{230\ \MeV}\Bigg]^2 
  \Bigg[\frac{|V_{ts}|}{0.040}\Bigg]^2\nnu\\
&\times&[0.998|\tilde c_S|^2 +
|\tilde c_P + 0.039(c_A-c_A')|^2],
\eea
where we have set $\hat m_s=0$ in \eq{Fsubis} and used the input parameters 
shown in Tables \ref{table:input} and 
\ref{table:quark-masses}.
The SM result for the branching fraction may be obtained 
from \eq{num:BRbmumu} 
by setting $\tilde c_S= \tilde c_P=c_A'=0$ and $c_A=-Y_{\sm}$, with $Y_{\sm}$ 
defined in \eq{sm:y} of the Appendix.

\subsection{Rare \bm $K$ decays}
In this section we adopt a somewhat different notation from that 
of \rfs{BB98,BB3,BBL:review} in order to avoid high powers of 
$|V_{us}|$. That is, we define the ratios
\be\label{kappa:i}
r_1={\alpha^2 \branch(K^+\to\pi^0e^+\nu_e)\over 4\pi^2
\sin^4\theta_W|V_{us}|^2} \frac{\tau_{K_L}}{\tau_{ K^+}},\quad
r_2= \frac{\tau_{K^+}}{\tau_{ K_L}},\quad 
r_3=\frac{\branch(K^+\to\m^+\nu_\m)}{\branch(K^+\to\pi^0e^+\nu_e)},
\ee 
with their numerical values summarized in Table \ref{table:input}.

\subsubsection{\bm$\kpnn$}
Using the effective Hamiltonian in \eq{hkpn}, 
it is straightforward to find the branching ratio
for this decay. Since the matrix element of the operator 
$\Oi_R$ equals the known matrix element of $\Oi_L$, 
we readily obtain the branching ratio from the formula 
given in \rfs{BB98,BB3}: 
\be\label{BR:kpnunu}
\branch(\kpnn)= 2r_1 r_2 r_{K^+}\sum_f \{[{\imlt}X(x)]^2+
[{\relc}X_\nl
+{\relt}X(x)]^2\},
\ee
where the sum is over three neutrino species, $r_{K^+}$ represents an 
isospin correction that one encounters 
when relating 
$K^+\to \p^+ \n\bar \n$ to $K^+\to \p^0 e^+\n_e$ \cite{marciano:zarsa}, and
$\la_i= V_{is}^*V_{id}^{}$. The function
\be
X_\nl \equiv\frac{1}{3}(2 X^e_\nl + X^\tau_\nl) 
\ee
denotes the charm contributions discussed in 
\Sec{SD:k:sector}, and
\be\label{X:NP}
X=c_L+c_R
\ee 
replaces the SM function $X_{\sm}$ given in Appendix 
\ref{diff:notation}. The numerical values of $r_{1,2}$, $r_{K^+}$, and 
$X_\nl$ are given in Table \ref{table:input}.

As far as the current experimental situation is concerned, the first clean 
$\kpnn$ event was found by the E787 Collaboration \cite{E787}.~Recently, 
further evidence for this decay mode has been reported in \rf{e787:recent}.
The updated branching ratio
\be
\branch(\kpnn)=(1.57^{+1.75}_{-0.82})\times 10^{-10},
\ee
being roughly by a factor of two higher than the SM expectation,
provides already a non-trivial lower bound on $V_{td}$ when interpreted
within the SM framework \cite{e787:recent,dambrosio:isidori:kpinunu}.

\subsubsection{\bm$\klpn$}
The decay $\klpn$ is a short-distance dominated process that is largely 
governed by $\cp$-violating contributions \cite{GBGI,YY97,littenberg:89}, 
and thus is  sensitive only to the imaginary parts of the CKM couplings.\footnote{Strictly speaking, this does not pertain to models with 
lepton flavour 
violation, in which the CP-conserving amplitude can  dominate \cite{YY97}.}
As the coupling $V_{cs}^*V_{cd}^{}$ 
is real to an excellent approximation, the internal charm
contributions can be completely neglected.

The $\klpn$ branching ratio may be obtained from 
the usual SM expression \cite{BB98,BB3} by replacing $X_{\sm}$ through
$X$ as defined in \eq{X:NP}. Summing over neutrino flavours, we have
\be\label{klpinunu:branch}
\branch(\klpn)=  2r_1 r_{K_L} \sum_f [{\rm Im}\lambda_t X(x)]^2,
\ee
where the isospin correction factor $ r_{K_L}$ 
is given in Table \ref{table:input}.
The best current upper limit has been set by the KTeV Collaboration
\cite{KTeV:KL}:
\be
\branch(\klpn)< 5.9\times 10^{-7}\ (\cl{90}).
\ee

\subsubsection{\bm$\klmumu$}
As mentioned at the outset of the paper, 
the decay $\klmumu$ suffers from  theoretical 
uncertainties due to the long-distance 
dispersive contribution \cite{LD:KLmumu}, and we therefore concentrate on 
the short-distance effects. In the part 
$\propto V_{ts}^*V_{td}^{}$ only the SM operator $\Oi_A$ has to be kept,
as in the decays $\kpnn$ and $\klpn$, since the Wilson coefficients of 
the remaining operators are completely negligible. Likewise, in the part  
proportional to the CKM elements $V_{cs}^*V_{cd}^{}$, 
non-negligible contributions arise only from $\Oi_A$.

Recalling $\la_i= V_{is}^*V_{id}^{}$ and \eq{kappa:i}, 
the branching ratio may be written as
\be\label{klmumu:branch}
\branch(\klmumu)_{\sd}= 4r_1r_3 [\Re \la_c Y_\nl + 
\Re \la_t Y(x)]^2,
\ee
where $Y_\nl$ represents the
charm contribution obtained in \rf{BB98} and $Y(x)=-c_A$. 
The numerical values of $r_3$, as well as the remaining parameters, are 
listed in Tables \ref{table:input} and \ref{table:quark-masses}.

\newsection{Numerical analysis}\label{numerical:analysis}
In the subsequent analysis we will adopt the following procedure:
\bit
\item We restrict our attention to the low and high 
$\tan\b$ regime, as defined in \eq{tanbeta:regime}. For the decays with 
a dilepton in the final state, we focus on the $\mu^+\m^-$ mode. 
\item In the case of $B$ decays, we will study the $\m$ dependence of 
the various branching fractions only for those decay modes that are 
mediated by the $b\to s$ transition, which essentially involves the 
CKM element $V_{ts}$. It is important to emphasize that the presence of 
new-physics contributions may not only affect the 
decay modes under study but also $B^0_q$--$\bar B^0_q$ mixing and the CP 
violation parameter $\epsilon_K$, and consequently
the extraction of the CKM elements $V_{td}$ and $V_{ts}$. In this 
case, the standard analysis of the unitarity triangle may lead to false 
results. In fact, while to a good approximation
$|V_{ts}|\approx |V_{cb}|$, independent of new-physics effects, the value of 
$V_{td}$ determined using the SM formulae might differ from 
that obtained in the context of SUSY (see, e.g., \rfs{BGGJS,CDGG98,BCRS01}).

Since we are mainly interested in the 
$\m$ dependence of the various branching fractions, rather than on 
their actual values, we fix $|V_{td}|$ to the SM value given 
in Table \ref{table:input}. Note that this treatment 
is different from the analysis 
of \rf{BGGJS}, where the new-physics effects on $V_{td}$
have also been taken into account.  

\item The light quark masses, $m_{s}$ and $m_b$, appearing in the Wilson 
coefficients are 
determined at the high-energy scale, but otherwise are evaluated at the 
low-energy scale (cf.~Table \ref{table:quark-masses}). 
Since the  contributions proportional to the down-quark mass are 
negligibly small,  we may take  $m_d=0$. 

\item Our calculation is based on the 
assumption of minimal flavour violation,  as outlined in \Sec{mssm}.
Furthermore, we assume that there are no new $\cp$-violating phases
in addition to the single phase residing in the CKM matrix.

\item Since we ignore flavour-mixing effects among squarks,
the matrix in \eq{squark:massmatrix} decomposes into three 
$2\times 2 $ matrices. A noticeable feature is that the LR terms are 
proportional to the masses of the up-type quarks. 
Hence, large mixing can occur in the scalar top quark sector, leading  
to a mass eigenstate, say, $\tilde{t}_1$, possibly much lighter than the 
remaining squarks. We therefore keep LR mixing 
only in the stop sector, where the mass matrix is given by
\bea
M^2_{\tsquark}=
\left(\begin{array}{cc}
m_{\tsquark_L}^2 + m_t^2 + \frac{1}{6}M_Z^2\cos2\b(3-4\sin^2\theta_W) 
&m_t(A_t-\m\cot\b)\\
m_t(A_t-\m\cot\b)& m_{\tsquark_R}^2 + m_t^2 + 
\frac{2}{3}M_Z^2\cos 2\b\sin^2\theta_W\end{array}\right),\nnu\\
\eea
where $m_{\tsquark_{L,R}}$ are the soft 
SUSY breaking scalar masses and $A_t$ is the trilinear coupling.
In this framework, the mixing matrices 
$\G^{U_L}$ and  $\G^{U_R}$ [\eq{def:stt}]
take the simple form 
\be\label{stop:gammaULR}
(\G^{U_L})^{\text{T}}=\left(\begin{array}{cccccc}
1&0&0&0&0&0\\
0&1&0&0&0&0\\
0&0& \cos\theta_{\tsquark}&0&0&-\sin\theta_{\tsquark}
\end{array}\right), \quad
(\G^{U_R})^{\text{T}}=\left(\begin{array}{cccccc}
0&0&0&1&0&0\\
0&0&0&0&1&0\\
0&0& \sin\theta_{\tsquark}&0&0&\cos\theta_{\tsquark}
\end{array}\right).  
\ee 
The  physical mass eigenstates are then given by 
\be\label{mass:eigenstates:stop}
\tilde t_1= \cos\theta_{\tsquark}\tilde t_L + 
\sin\theta_{\tsquark}\tilde t_R,\quad
\tilde t_2= -\sin\theta_{\tsquark}\tilde t_L + 
\cos\theta_{\tsquark}\tilde t_R,
\ee
with the mixing angle $(-\p/2\leqslant \theta_{\tsquark} \leqslant \pi/2$)
\be
\sin 2\theta_{\tsquark}=\frac{2m_t (A_t-\m\cot\b)}{m_{\tsquark_1}^2-m_{\tsquark_2}^2}, \
\cos 2\theta_{\tsquark}=\frac{(m_{\tsquark_L}^2 -m_{\tsquark_R}^2)+ \frac{1}{6}M_Z^2\cos 2\b(3-8\sin^2\theta_W)}{m_{\tsquark_1}^2-m_{\tsquark_2}^2}, 
\ee
$m_{\tsquark_{1,2}}$ being the stop masses with $m^2_{\tsquark_1}< m^2_{\tsquark_2}$.  (The remaining up-type squark masses are taken to be equal.)

\item For simplicity, we assume that the scalar partners of the leptons 
are degenerate in mass.  
\item 
For the results presented below we take into account the following 
lower bounds on the SUSY particle masses 
\cite{PDG,manuell:recent,ggh}:
\bea
m_{\tilde t_1, \tilde b_1}\gtrsim & 100\ \GeV,\quad 
m_{\tilde q \neq \tilde t_1, \tilde b_1}\gtrsim260\ \GeV,\quad
m_{\tilde l, \tilde \n} \gtrsim 100\ \GeV, \quad
M_{\chargino_1^{}}\gtrsim 100\ \GeV.
\eea
As far as the lightest neutral Higgs 
boson, $h^0$, is concerned,  
we must ensure that  $M_{h^0} \geqslant 113.5\ \GeV$
\cite{ggh,LEP:Higgs-working-group}, taking into account 
radiative corrections  \cite{one-loop:h0,rad:corr:h0} 
to the tree-level mass defined in \eq{tree-level:h0}.

Further restrictions on the SUSY parameter space are imposed 
by electroweak precision data such as the 
$\r$ parameter \cite{PDG,manuell:recent,rho:parameter}. Also, we take into 
account the constraint on the trilinear coupling, $A_t$, arising from the 
requirement of the absence of colour and 
charge breaking minima \cite{CCB}.
\item  We require 
the various SUSY contributions to be consistent with the 
measured inclusive $b\to s\g$ 
branching fraction \cite{bsgamma:exp:recent}. 
To be specific, we will allow the 
range of $2.0 \times  10^{-4}$ to $4.5 \times  10^{-4}$ for the 
branching ratio $\branch (\B\to X_s\g)$.
%
Besides, we take into account constraints on the 
Wilson coefficients arising from other rare exclusive $B$ decays such as
$\B_s\to \m^+\m^-$ [\eq{exp:limit:bsmumu}].
\eit

To determine the impact of the QCD corrections on the 
various branching ratios, 
we examine their $\mu$ dependence for given points in the SUSY parameter space.
For definiteness, we have chosen the following SUSY parameter sets:
\bea\label{choice:parameters:low:right-handed}
\mbox{$\tan\b=3$:}\quad \left\{\begin{array}{l}
m_{\tilde{t}_1}=200\ \GeV,
\ m_{\tilde{t}_2}=800\ \GeV,
\ \theta_{\tilde t}= -70^\circ,
\ m_{\tilde l,\tilde \n}=100\ \GeV,\\
M_H=300\ \GeV,
\ \mu=-300\ \GeV,
\ M_2=800\ \GeV,
\ M_{\tilde g}=1\ \TeV,\end{array}\right.
\eea
\bea\label{choice:parameters:high:right-handed}
\mbox{$\tan\b=40$:}\quad \left\{\begin{array}{l}
 m_{\tilde{t}_1}=120\ \GeV,
\ m_{\tilde{t}_2}=500\ \GeV,
\ \theta_{\tilde t}= -70^\circ,
\ m_{\tilde l,\tilde \n}=100\ \GeV,\\
M_H=250\ \GeV,
\ \mu=-350\ \GeV,
\ M_2=800\ \GeV,
\ M_{\tilde g}=1\ \TeV,\end{array}\right.
\eea
\bea\label{choice:parameters:high:max-mixing}
\mbox{$\tan\b=40$:}\quad \left\{\begin{array}{l}
 m_{\tilde{t}_1}=500\ \GeV,
\ m_{\tilde{t}_2}=650\ \GeV,
\ \theta_{\tilde t}= -44^\circ,
\ m_{\tilde l,\tilde \n}=100\ \GeV,\\
M_H=200\ \GeV,
\ \mu=-600\ \GeV,
\ M_2=800\ \GeV,
\ M_{\tilde g}=1\ \TeV.\end{array}\right.
\eea
Note that in the case of $\theta_{\tilde t}= -70^\circ$ 
the lighter scalar top quark is predominantly right-handed 
[cf.~\eq{mass:eigenstates:stop}] while the choice 
$\theta_{\tilde t}= -44^\circ$ corresponds to a scenario with 
almost maximal mixing, 
i.e.~$|\sin(2\theta_{\tilde t})|\approx 1$.

\begin{figure}
\begin{center}
\epsfig{file=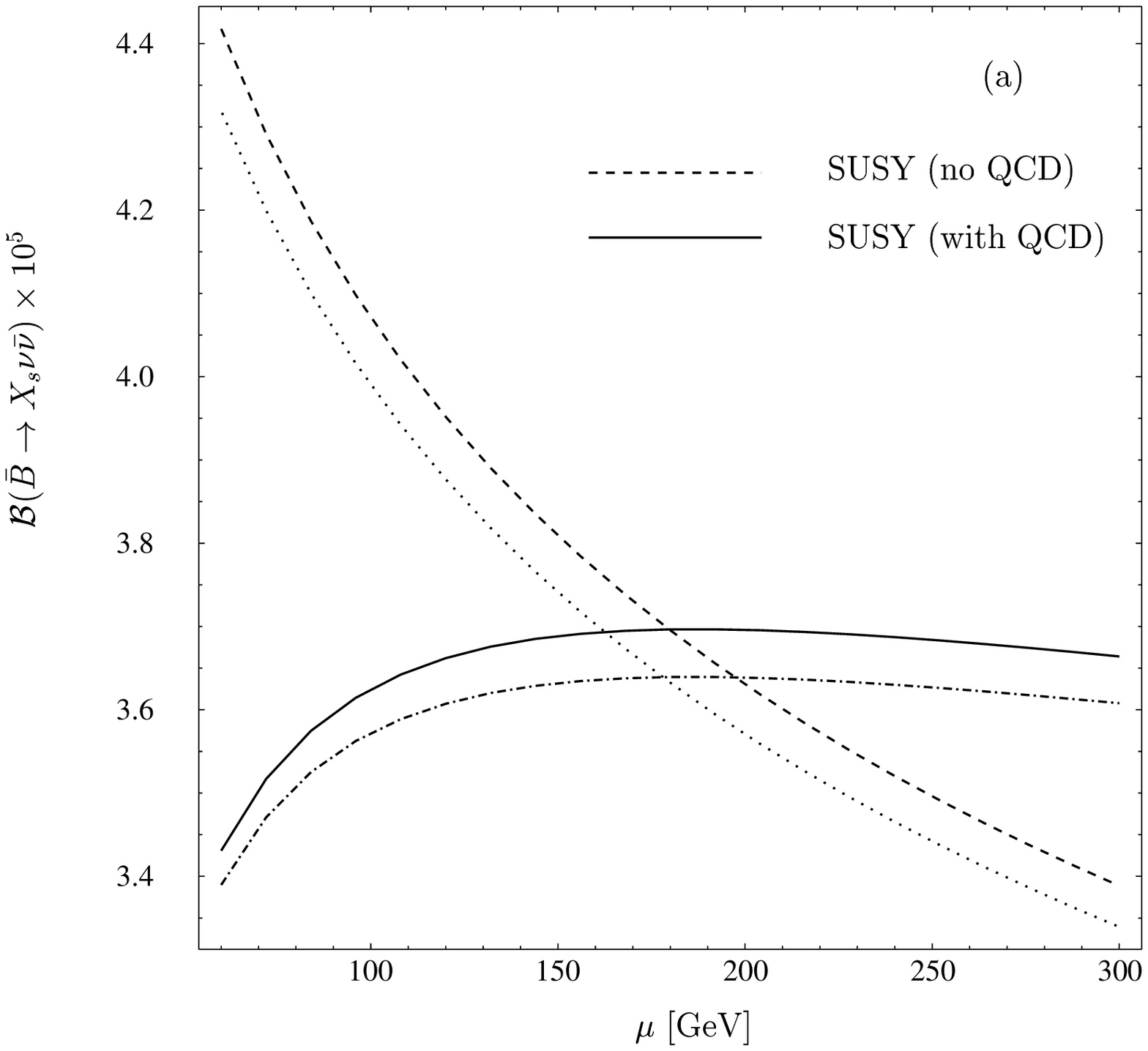,height=2.8in}\hspace{1em}
\epsfig{file=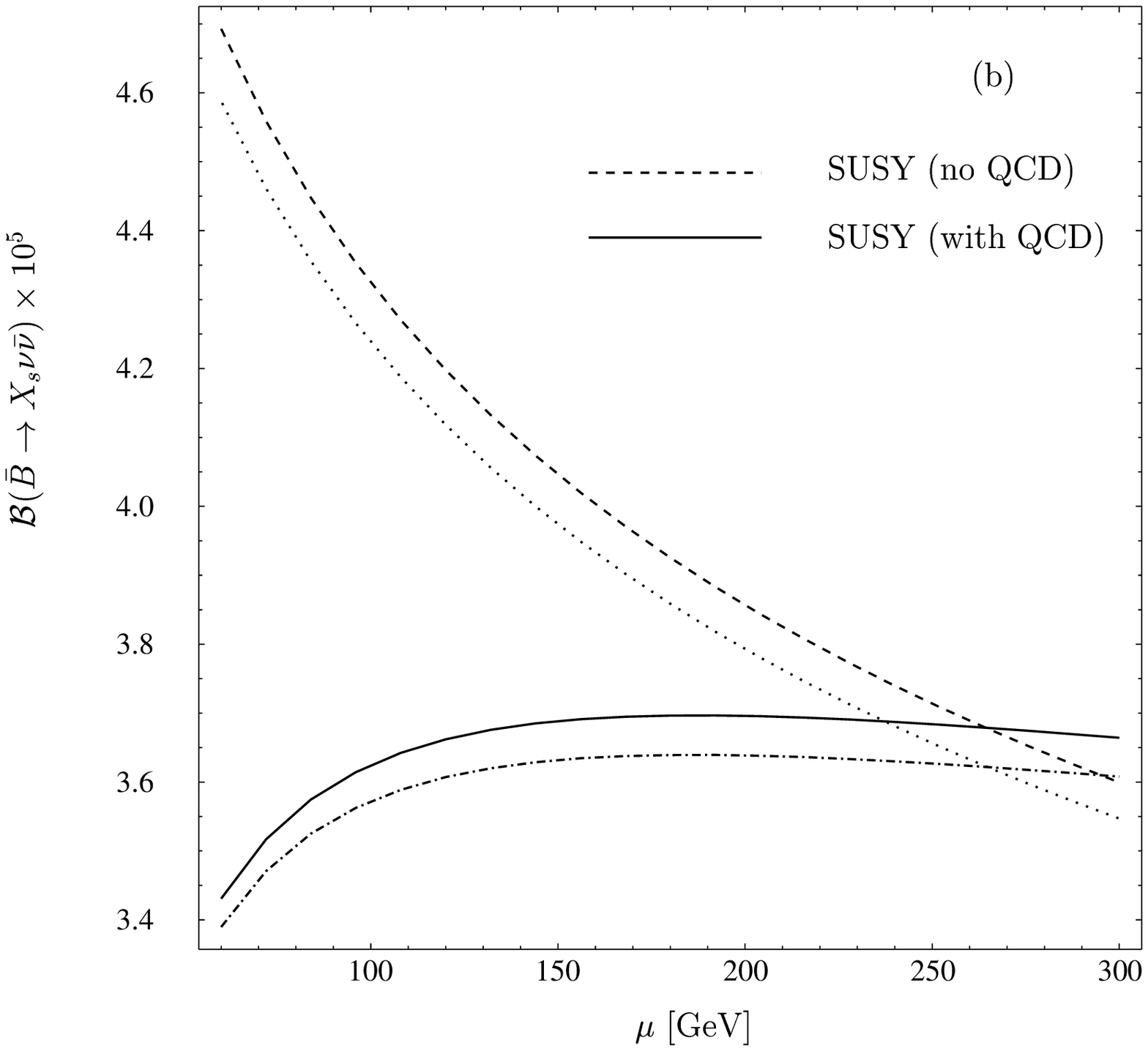,height=2.8in}
\caption{Predictions of the  $\m$ dependence of the $\B\to X_s\n\bar \n$ 
branching ratio obtained  (a) using $\kappa$ as given in 
Eqs.~(\ref{kappa:approx}) and (\ref{kappa:zero}) in the leading order (LO) 
as well as the  next-to-leading order (NLO) branching ratio  
[\eq{branch:bsnunu}], and (b) taking $\kappa=1$ in the LO expression of the 
branching ratio. 
The solid  (dashed)  curves represent the SUSY results 
with (without) QCD corrections. We have chosen  $\tan\b=3$, together 
with the parameter set given in  \eq{choice:parameters:low:right-handed}. 
For comparison, we also show the SM prediction 
with (dash-dotted curve) and without (dotted curve) 
QCD corrections.\label{fig:branching:ratios:BXsnunubar}}
\end{center}
\end{figure}

In \figs{fig:branching:ratios:BXsnunubar}{fig:branching:ratios:K} 
we summarize our results for the $\m$ dependence of the branching ratios 
of $\B\rightarrow X_{s} \nu\bar\nu$, $\bar B_{s}\rightarrow
\mu^+\mu^-$, $K^+\rightarrow\pi^+\nu\bar\nu$, 
$K_L\rightarrow\pi^0\nu\bar\nu$ and
$K_L\rightarrow\mu^+\mu^-$.~Overall we see that the dependence of 
the various branching fractions on the renormalization 
scale is considerably reduced once QCD corrections are taken into 
account. In fact, the $\m$ dependence inherent in 
the leading-order predictions, typically $10$--$20\%$, 
is reduced to a few per cent 
once $O(\alqcd)$ corrections are taken into account.
 
\begin{figure}
\begin{center}
\epsfig{file=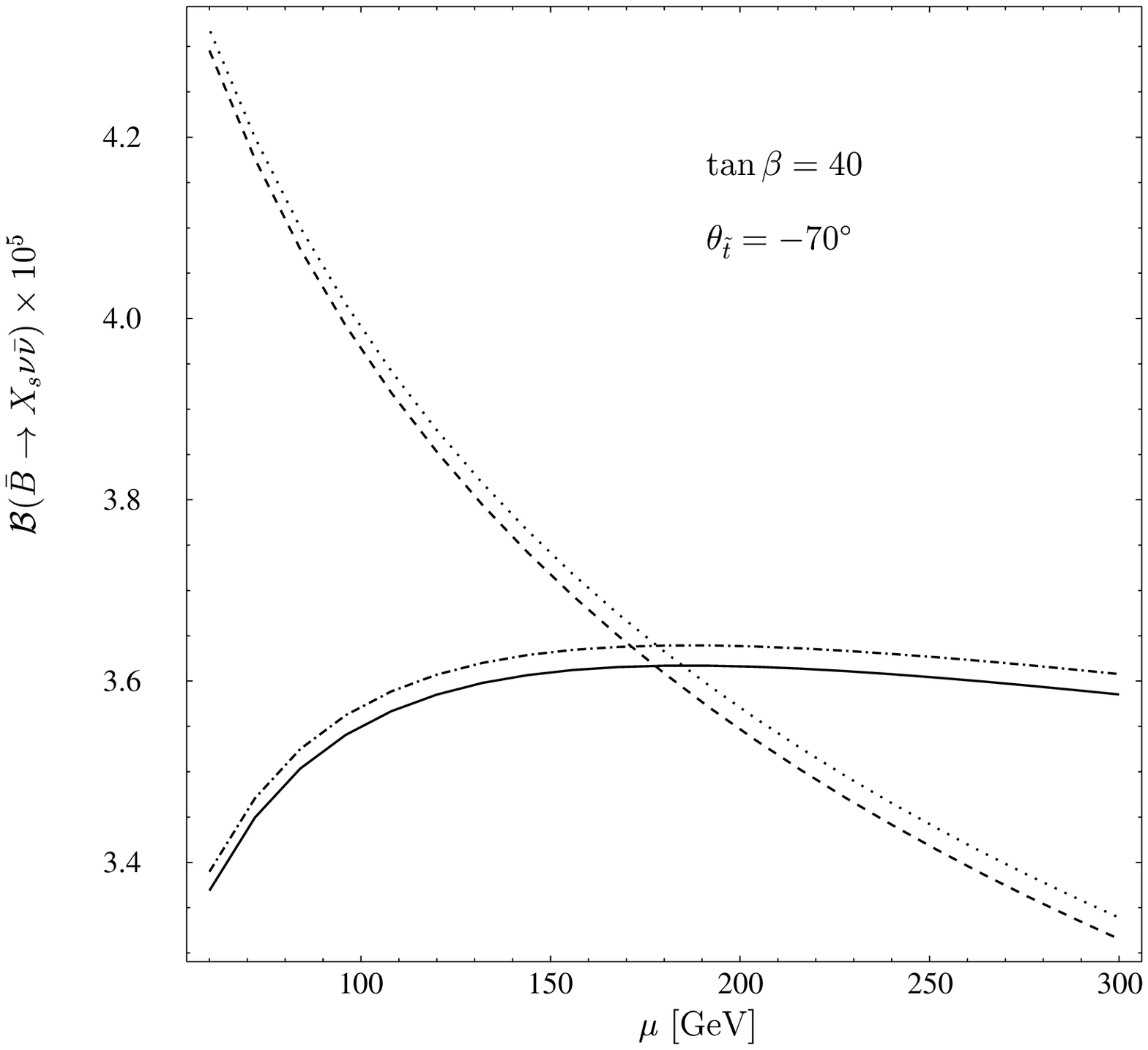,height=2.8in}
\caption{Branching ratio of the $\B\to X_s\n\bar \n$ decay vs 
the renormalization scale $\m$ at  
large $\tan\beta$, as defined in \eq{choice:parameters:high:right-handed}. 
We have used 
$\kappa$ according to Eqs.~(\ref{kappa:approx}) and (\ref{kappa:zero}) in 
both the LO as well as the NLO expression for the branching ratio 
[\eq{branch:bsnunu}].
The legends are the same as in \fig{fig:branching:ratios:BXsnunubar}.
\label{fig:branching:ratios:BXsnunubar:large:tanbeta}}
\end{center}
\end{figure}

Figures \ref{fig:branching:ratios:BXsnunubar} and 
\ref{fig:branching:ratios:BXsnunubar:large:tanbeta} 
show the $\m$ dependence of the 
$\B\rightarrow X_{s} \nu\bar\nu$ branching ratio for the low and high  
$\tan\b$ regime, respectively. Notice that the SUSY contributions interfere 
constructively or destructively with those of the SM, 
depending on our choice of SUSY parameters. 
Another noticeable feature is that the value of $\mu$ at which 
leading order (LO) and next-to-leading order (NLO) corrections coincide 
depends on whether 
one uses $\kappa=1$ or $\kappa$ according to  
Eqs.~(\ref{kappa:approx}) and (\ref{kappa:zero}) in the LO expression for the 
$\B\rightarrow X_{s} \nu\bar\nu$ branching ratio.

\begin{figure}
\begin{center}
\epsfig{file=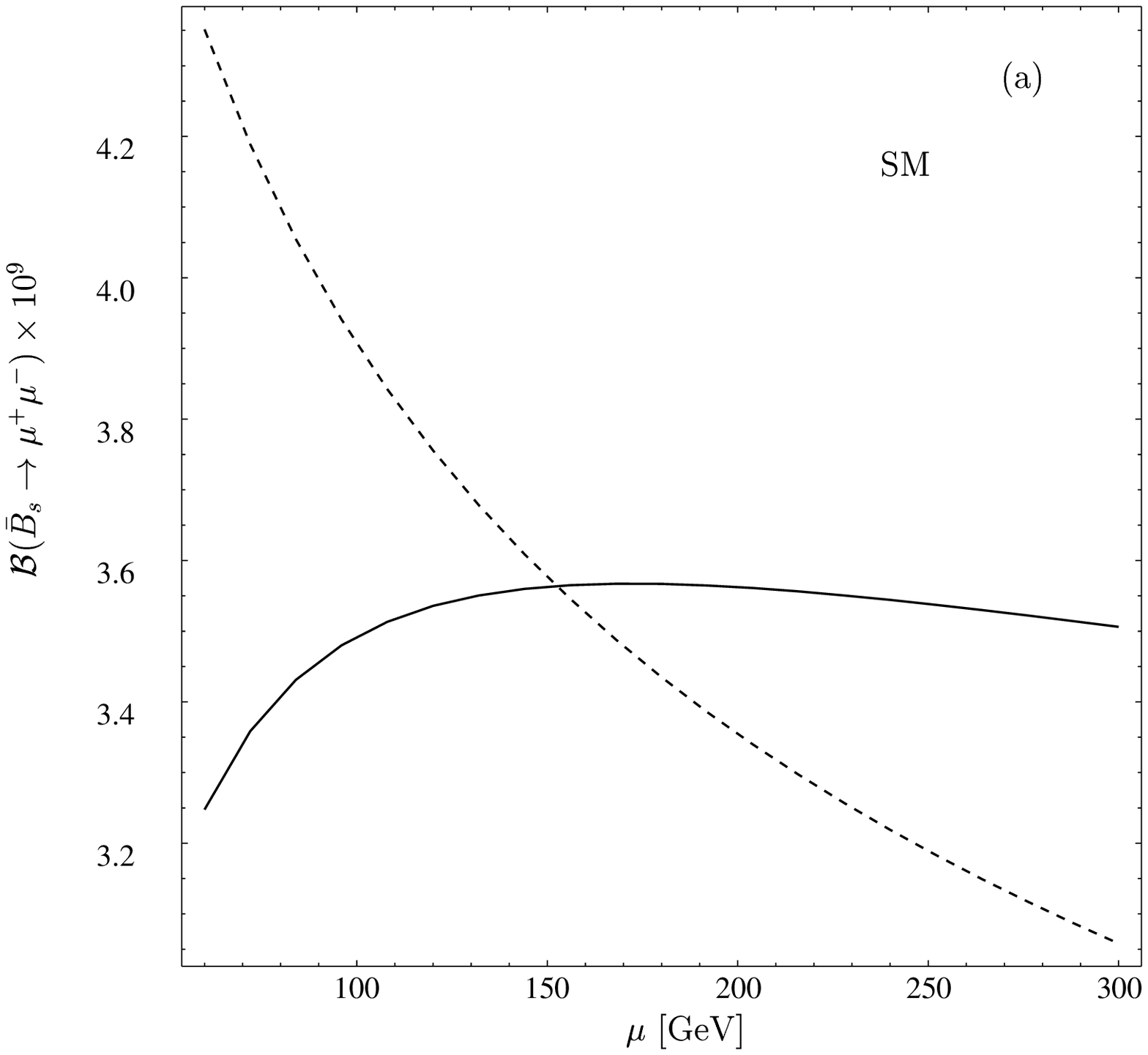,height=2.8in}\hspace{1em}
\epsfig{file=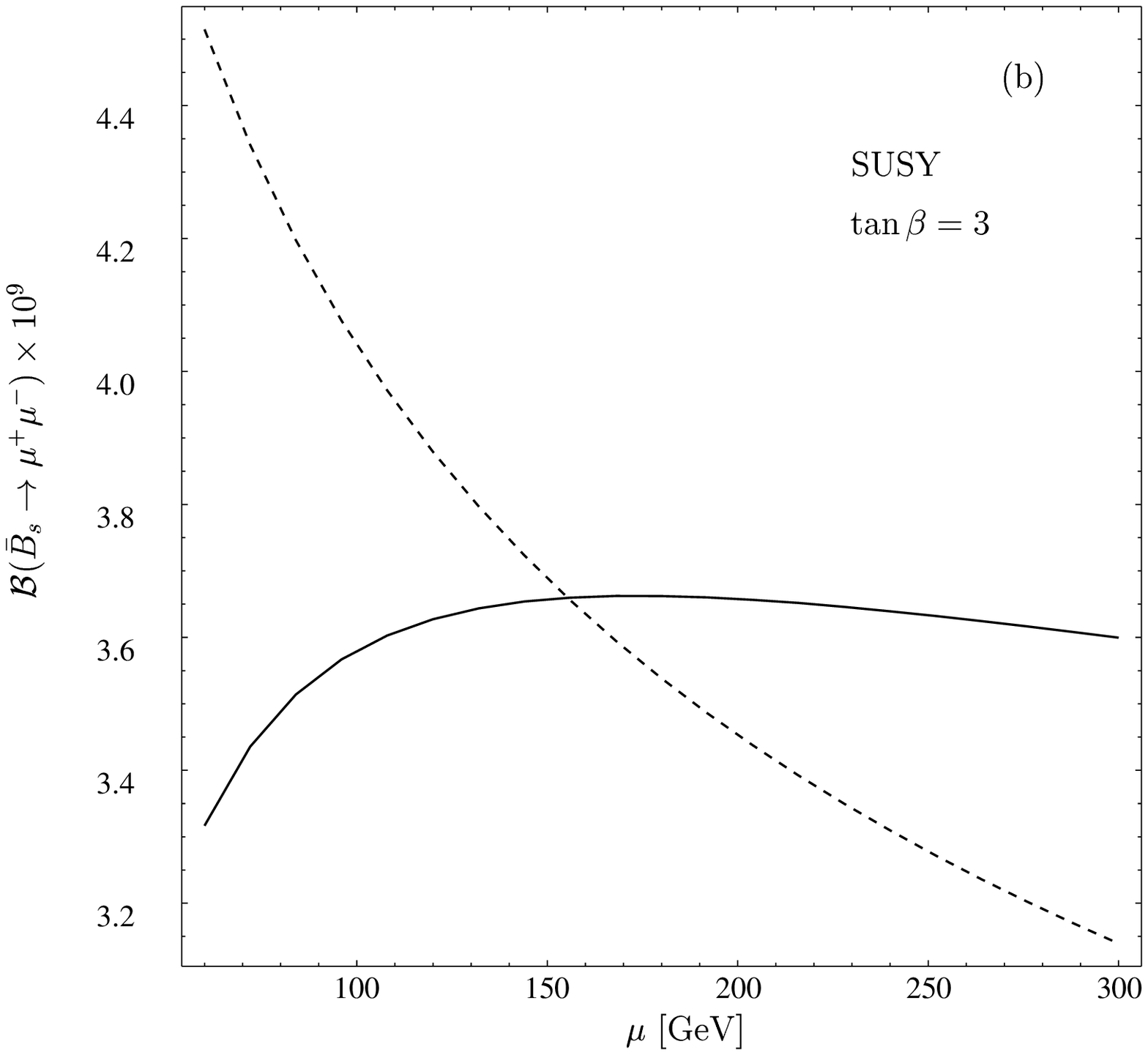,height=2.8in}
\caption{The $\mu$ dependence of the $\B_s\to \m^+\m^-$ branching ratio
for (a) the SM, and (b) SUSY in the low $\tan\beta$ regime, as defined in 
Eq.~(\ref{choice:parameters:low:right-handed}).
The solid  and dashed curves denote the predictions with  and without 
QCD corrections, respectively. 
\label{fig:branching:ratios:Bmumu}}
\end{center}
\end{figure}

\begin{figure}
\begin{center}
\epsfig{file=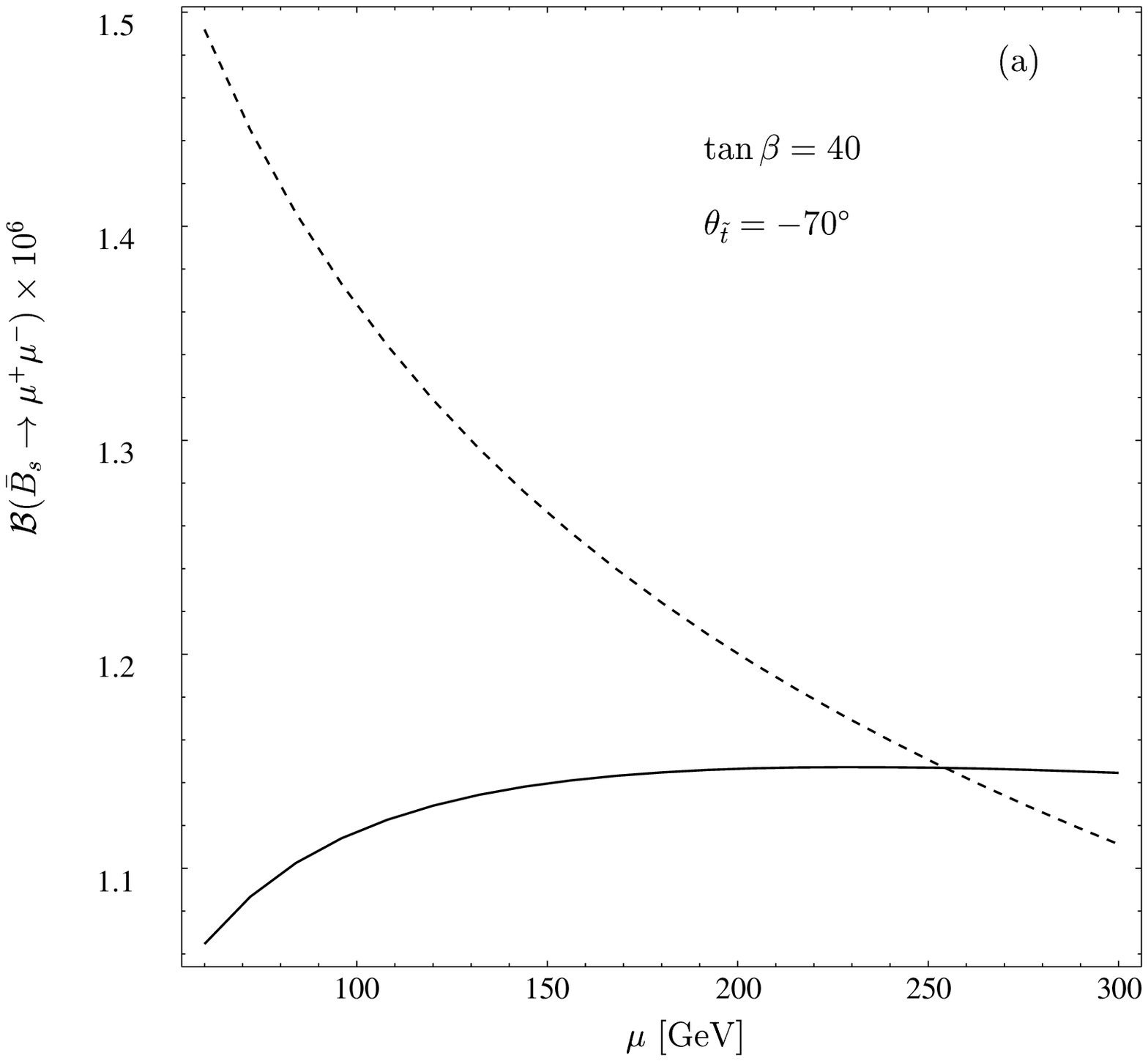,height=2.8in}\hspace{1em}
\epsfig{file=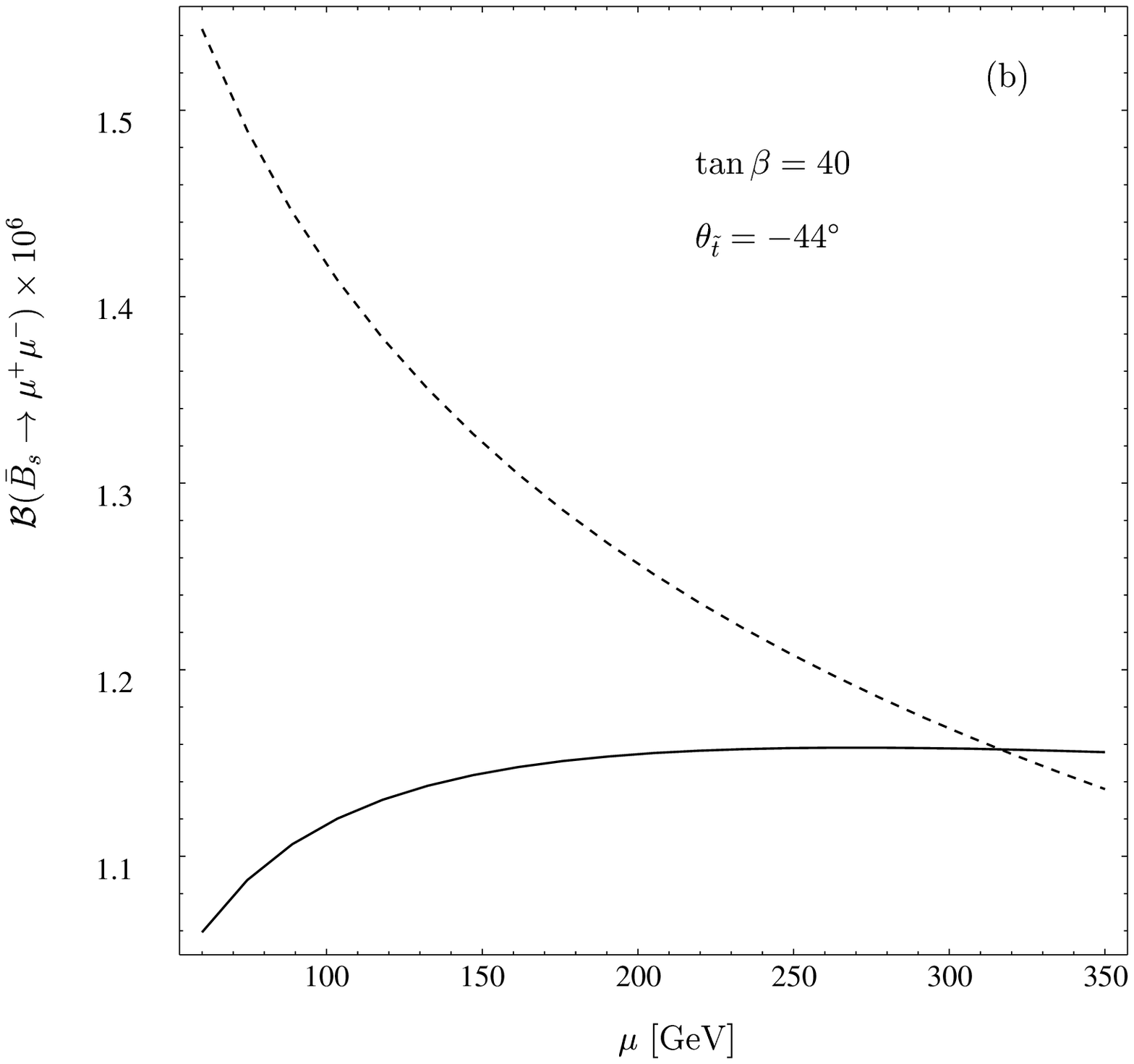,height=2.8in}
\caption{The $\mu$ dependence of the $\B_s\to \m^+\m^-$ branching ratio 
within SUSY at large $\tan\b$. 
The solid and  dashed curves denote the SUSY prediction with and without 
QCD corrections, respectively.
(a) For the case of a predominantly right-handed light stop quark, 
using the SUSY input parameters given in \eq{choice:parameters:high:right-handed}. (b) For the case of almost 
maximal mixing in the scalar top quark sector according to the 
parameter set given in \eq{choice:parameters:high:max-mixing}.
Note the order-of-magnitude enhancement of the branching ratio, compared to 
the SM and low $\tan\b$ SUSY predictions
in \fig{fig:branching:ratios:Bmumu}.    
\label{fig:branching:ratios:Bmumu:high}}
\end{center}
\end{figure}

Turning to the  $\B_s\to \m^+\m^-$ decay, 
we show in Figs.~\ref{fig:branching:ratios:Bmumu} and 
\ref{fig:branching:ratios:Bmumu:high}
the branching fraction for the low 
and high $\tan\b$ regime, respectively.
While in the former region the branching ratio in the SM 
is only mildly affected by the SUSY contributions 
(see \fig{fig:branching:ratios:Bmumu}), 
in the high $\tan\b$ regime the supersymmetric 
effects can be enormous. This can be seen from  
\fig{fig:branching:ratios:Bmumu:high} where we have plotted 
the branching ratio for the case of a predominantly 
right-handed light stop quark 
[\fig{fig:branching:ratios:Bmumu:high}(a)]  and for the scenario of 
maximal mixing in the scalar top-quark sector 
[\fig{fig:branching:ratios:Bmumu:high}(b)]. 
Note that the value of $\m$ at which the leading and 
next-to-leading order results for the branching ratio coincide depends on  
the choice of input parameters. 

We should mention that the large new-physics effects  
in $\B_s\to \m^+\m^-$  in the high $\tan\beta$ regime, 
compared to those in $\B\rightarrow X_{s} \nu\bar\nu$, are due to the fact 
that in the former decay mode the leading contribution scales roughly as
$\sim m_b m_\mu\tan^3\b$ while in the latter decay it behaves as 
$\sim m_b m_s \tan^2\b$.

\begin{figure}
\vspace{-1cm}
\begin{center}
\epsfig{file=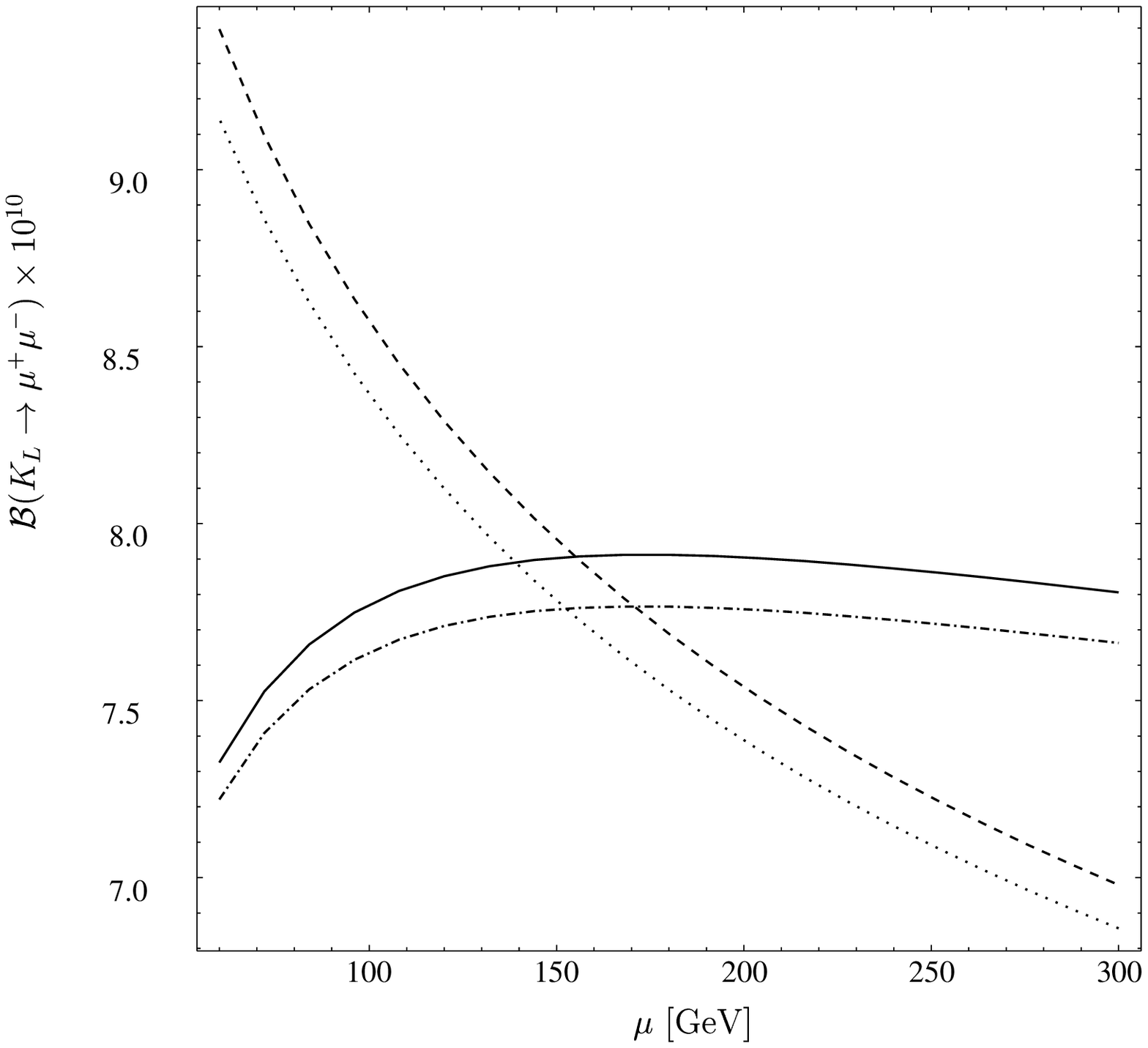,height=2.8in}\hspace{1em}
\epsfig{file=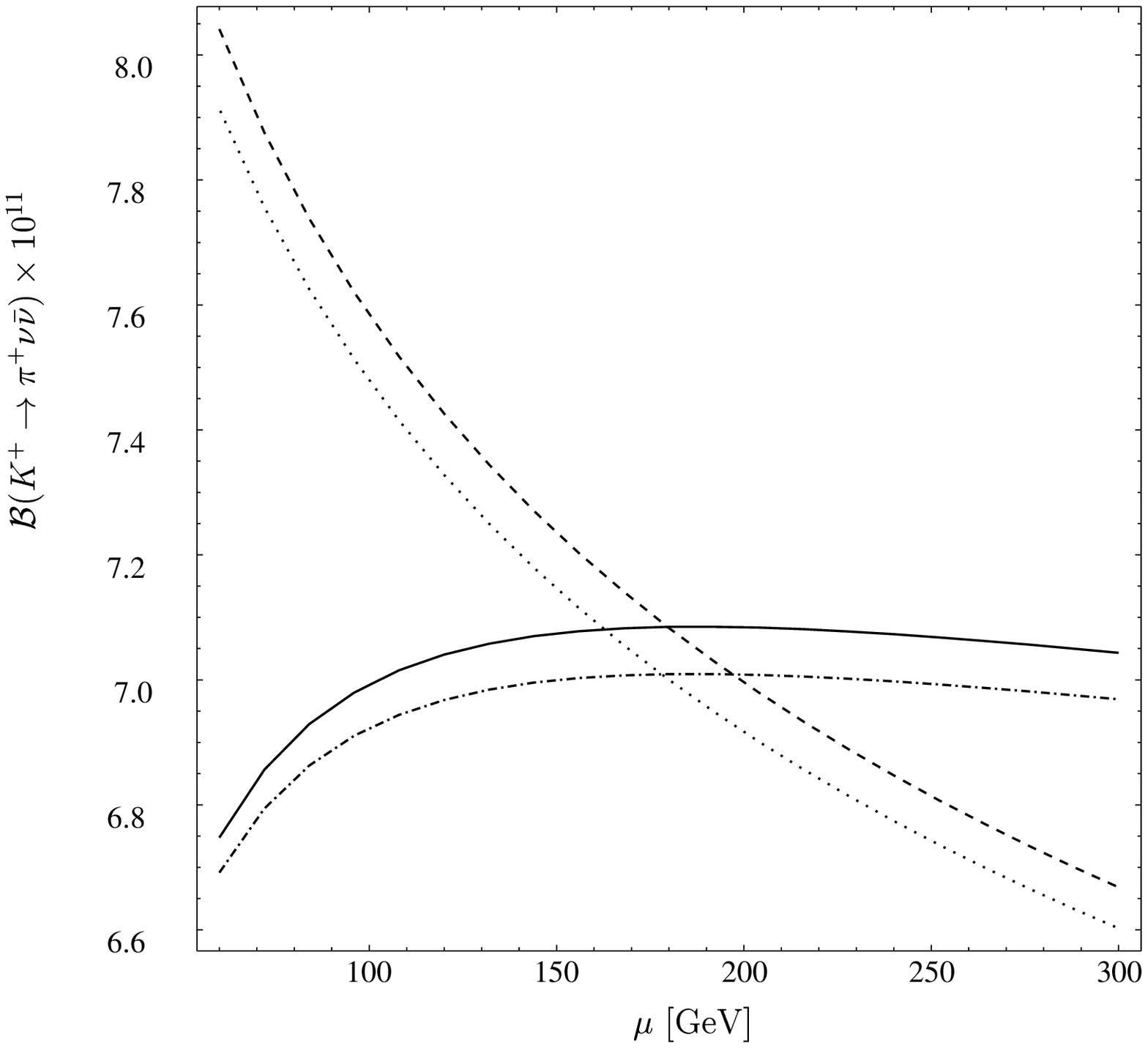,height=2.8in}\vspace{1.5em}
\epsfig{file=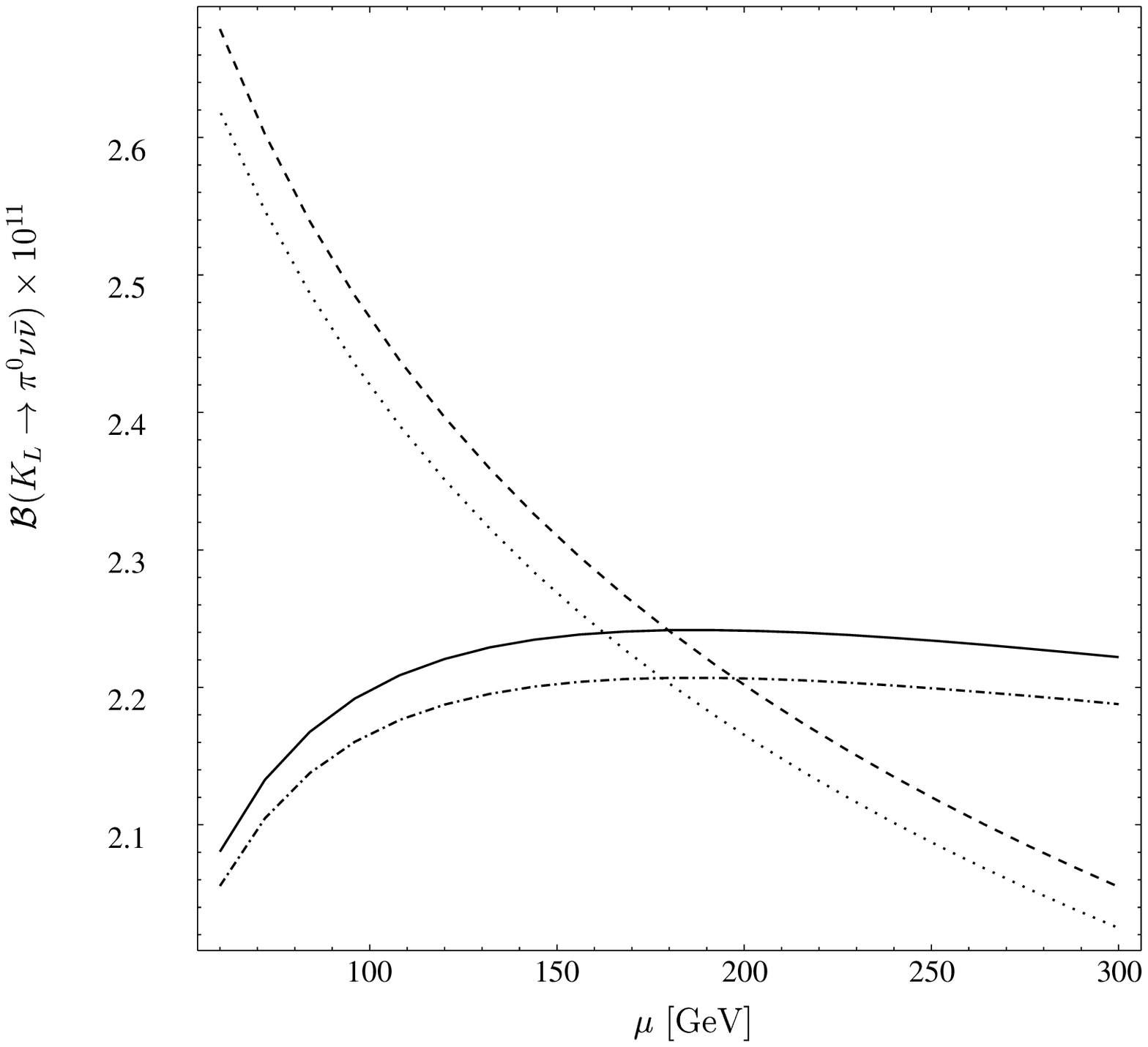,height=2.8in}
\caption{The $K_L\to \m^+\m^-$, $K^+\to \pi^+\nu\bar{\nu}$
and $\klpn$ branching ratios, as a function of the renormalization scale $\m$. 
The dash-dotted (dotted) 
curves correspond to the SM prediction with (without) QCD corrections while
the solid (dashed) curves denote the SUSY prediction with (without) QCD 
corrections.~We have chosen $\tan\b=3$ along with the  
SUSY parameter set displayed in  Eq.~(\ref{choice:parameters:low:right-handed}).\label{fig:branching:ratios:K}}
\end{center}
\end{figure}

In  \fig{fig:branching:ratios:K} we have plotted the branching ratios 
for the decays  $K_L\rightarrow\mu^+\mu^-$, $K^+\rightarrow\pi^+\nu\bar\nu$ 
and $K_L\rightarrow\pi^0\nu\bar\nu$
in the low $\tan\b$ regime. For large values of $\tan\b$
the SUSY contributions are  negligibly small. 
As far as new operators are concerned, they do not give any sizable 
contributions, due to the suppression of the light quark masses. 
%
%
%
%
%
%
%
%
%
%

\section{Corrections to the \bm$b$-quark mass at 
large \bm$\tan\b$}\label{SUSY:QCD:Yukawa}
Before summarizing, a few remarks are in order regarding
corrections to the $b$ quark Yukawa coupling, 
which may be important in the 
high $\tan\b$ regime \cite{first:papers:Bmumu,SUSY:large:tanbeta:bquark}.
As a matter of fact, sizable contributions to the down-type quarks, and 
hence to the CKM 
elements, may occur. Recently, it has been pointed out in 
\rfs{Bmumu:susy:epsilon,gino:retico:01,SUSY:large:tanbeta:Yukawa,bsgamma:paolo:etal} that the corrections to the $b$ quark Yukawa coupling, 
which are complementary to those 
presented in the preceding sections, can be substantial in rare 
$B$ decays like $\B_s\to \m^+\m^-$ and $b\to s \g$. Since our main 
emphasis has been on the $\m$ dependence of the various branching ratios,
rather than on their precise values, this issue will not be pursued here.  
\newsection{Summary and conclusions}\label{conclusions}
We have carried out, for the first time, a calculation of 
QCD corrections to the branching ratios of the decays 
$\B\rightarrow X_{d,s} \nu\bar\nu$, $\bar B_{d,s}\rightarrow
\mu^+\mu^-$, 
$K^+\rightarrow\pi^+\nu\bar\nu$, $K_L\rightarrow\pi^0\nu\bar\nu$ and
$K_L\rightarrow\mu^+\mu^-$ within SUSY. Our results are applicable  to
a version of the MSSM in which (a) the gluinos decouple; (b) flavour violation 
is governed exclusively  by the CKM matrix; and (c) the neutralinos do not 
contribute.

Our main results  can be summarized as follows:
\ben
\item We have provided a compendium of branching ratios and Wilson 
coefficients in the presence of supersymmetry, including all relevant 
dimension-six operators.~Our results are valid for arbitrary values of $\tan\b$ except for 
the neutral Higgs-boson contributions, which have been obtained in the high 
$\tan\b$ regime. 

\item The inclusion of QCD corrections  within the 
SM and SUSY leads to a significant 
reduction of the unphysical renormalization scale dependence, which is related
  to the running quark and squark masses, and which is unavoidably present
  in the existing  leading-order formulae. While in the leading-order 
 expression the 
scale uncertainty is typically $10$--$20\%$ in the branching ratio, 
it is reduced to a few per cent after $O(\alqcd)$ corrections are taken into 
account.
\item  For the set of SUSY parameters considered in this paper, 
it is possible to find a value of the renormalization scale $\mu$
for which the $O(\alpha_s)$ 
QCD corrections to the branching fractions are below, say, $2\%$, 
so that they can be neglected. In this case, one can estimate these 
QCD-corrected branching ratios merely by using the leading-order result
with the proper choice of $\m$.
Its actual value depends, 
of course,  on the process considered, 
as well as on the specific choice of the SUSY parameters 
(see our discussion in \Sec{numerical:analysis}).
\item As far as new operators are concerned, our analysis implies that
in the low $\tan\beta$ regime $2\leqslant \tan\beta\leqslant 5$, 
only the SM operators are relevant, with their Wilson coefficients 
modified by the presence of non-SM contributions. By contrast, 
in the large $\tan\beta$ region  $40\leqslant\tan\beta\leqslant 60$,
the effects of new operators can lead to an order-of-magnitude 
enhancement in the decays  $\bar B_{d,s}\rightarrow\mu^+\mu^-$, 
in accordance with previous studies 
\cite{first:papers:Bmumu,me:bqll:SUSY,bobeth:etal:0401,Bmumu:susy:epsilon,gino:retico:01}, and  also provide substantial contributions to 
$\Delta M_{B_s}$ \cite{gino:retico:01,BCRS01}.~As for the 
decays $\bar B\rightarrow X_{d,s}\nu\bar\nu$,
the  supersymmetric corrections due to new operators, 
although  enhanced by powers of $\tan\b$,  
are generally smaller since they are suppressed 
by the light quark masses $m_{d,s}$.~Finally, the corresponding corrections in kaon 
decays are completely negligible.
\een

\section*{Acknowledgements}
We would like to thank  Manuel Drees, Thorsten Ewerth, 
Miko\l aj Misiak and Janusz Rosiek
for useful discussions.~This
research was partially supported by the German `Bundesministerium f\"ur 
Bildung und Forschung' under contract 05HT1WOA3 and by the 
`Deutsche Forschungsgemeinschaft' (DFG) under contract Bu.706/1-1.

\appendix
\newsection{Standard model notation}\label{diff:notation}
In this Appendix we give the correspondence of our results with the notation 
commonly used in the literature within the context of the 
SM. In this case, the decays $K_L\to \p^0 \n\bar\n$, $\B\to X_q\n\bar\n$ and
$\B_q\to l^+l^-$ are described by the loop functions $X_\sm$ and $Y_\sm$ in 
the top-quark sector, which are defined as \cite{Bur01,BBL:review} 
\be\label{sm:x}
X_\sm (x)\equiv X_0(x) + \frac{\alqcd}{4\p}X_1(x)=C(x)-4B(x,1/2),
\ee
\be\label{sm:y}
Y_\sm (x)\equiv Y_0(x) + \frac{\alqcd}{4\p}Y_1(x)=C(x)-B(x,-1/2),
\ee
where $C$ and $B$ correspond to $Z^0$-penguin and box-type contributions,
respectively. In  terms of the loop functions given in Appendix 
\ref{loop-functions:app}, we find
\be\label{c:func}
C(x)=\frac{1}{4} 
     \Bigg\{f^{(0)}_1(x)+\frac{\alpha_s}{4\pi}\Bigg[f^{(1)}_1(x)
+ 8x \frac{\partial}{\partial x}f^{(0)}_1(x) \ln 
\Bigg(\frac{\m^2}{m_t^2}\Bigg) \Bigg]\Bigg\},
\ee
\be\label{b:12:func}
B(x,1/2)= \frac{1}{4}\Bigg\{f^{(0)}_2 (x) +\frac{\alpha_s}{4\pi}\Bigg[f^{(1)}_6 (x)
+ 8x \frac{\partial}{\partial x}f^{(0)}_2(x) \ln \Bigg(\frac{\m^2}{m_t^2}\Bigg)\Bigg]\Bigg\},
\ee
\be\label{b:m12:func}
B(x,-1/2)= \frac{1}{4} \Bigg\{ f_2^{(0)}(x) + 
    \frac{\alpha_s}{4\pi} \left[f_{10}^{(1)} (x)+8 x
      \frac{\partial}{\partial x} f_2^{(0)}(x)\ln \Bigg(\frac{\m^2}{m_t^2}\Bigg)\right]\Bigg\}.
\ee 
The expressions in \eqs{c:func}{b:m12:func} agree with the results found in 
\rfs{MU98,BB98}. Note that in these papers the explicit $\m$ 
dependence is 
given in terms of $\ln (\m^2/M_W^2)$.

\newsection{Auxiliary functions}\label{loop-functions:app}
Defining the dilogarithm $\text{Li}_2$ by 
\be
\Li{z} = - \int_0^z dt \frac{\ln(1-t)}{t},
\ee
the loop functions $f_p^{(0)}, f_{p'}^{(1)}$ appearing in the formulae of 
\Sec{effective:Hamiltonian} have the following form:
\be
f^{(0)}_1 (x) =-\frac{x(6-x)}{2(x - 1)}+\frac{x(2+ 3x)}{2(x - 1)^2}\ln x,
\ee
\be
f^{(0)}_2 (x) =-\frac{x}{x - 1} + \frac{x}{(x - 1)^2} \ln x,
\ee
\be
f^{(0)}_3 (x,y) = \frac{x \ln x}{(x - 1)(x - y)}+\frac{y 
\ln y}{(y - 1)(y - x)}, 
\ee
\be
f^{(0)}_4 (x,y) = \frac{x^2 \ln x}{(x - 1)(x - y)} + \frac{y^2 
\ln y}{(y - 1)(y - x)},
\ee 
\be
f^{(0)}_5(x,y,z) =
     \frac{x^2 \ln x}{(x - 1) (x - y) (x - z)} +(x\leftrightarrow y)+(x\leftrightarrow z),
\ee
\be
f^{(0)}_6(x,y,z) =
     \frac{x \ln x}{(x - 1) (x - y) (x - z)} +(x\leftrightarrow y)+
(x\leftrightarrow z),
\ee
\be
f^{(0)}_7(x,y)=\frac{x \ln x}{(x - 1)(x - y)}+\frac{x \ln y}{(y - 1)(y - x)},
\ee
\be
f^{(0)}_8(x) = \frac{x \ln x}{x - 1},
\ee
\be
f^{(0)}_{9}(w,x,y,z) =
  \frac{w^2 \ln w}{(w - 1) (w - x) (w - y) (w - z)} +
  (w\leftrightarrow x)+(w\leftrightarrow y)+(w\leftrightarrow z),
\ee
\be
f^{(0)}_{10}(w,x,y,z) =
     \frac{w \ln w}{(w - 1) (w - x) (w - y) (w - z)} +
     (w\leftrightarrow x)+(w\leftrightarrow y)+(w\leftrightarrow z),
\ee
\be
f^{(0)}_{11}(x,y) =
     \frac{x \ln x}{(x - y)}+
     \frac{x \ln y}{(y - x)}, 
\ee
\be
f^{(1)}_1 (x) = 
      \frac{4x (29+7 x + 4 x^2)}{3(x - 1)^2} -
      \frac{4x (23+14 x+3 x^2)}{3(x - 1)^3} \ln x
  -\frac{4x (4+x^2)}{(x - 1)^2} \Li{1-\frac1x}, 
\ee
\be
f^{(1)}_2 (x)  = 
      \frac{32 x ( 3 - x)}{3(x - 1)^2} - 
      \frac{8 x (11-3 x)}{3(x - 1)^3} \ln x 
     -\frac{8 x (2-x)}{(x - 1)^2} \Li{1-\frac1x}, 
\ee
\bea
f^{(1)}_3 (x,y) &=&
-\frac{28 y}{3 ( x - y )  ( y-1 ) } +\frac{2 x ( 11 x + 3 y )} 
     {3 ( x-1 )  {( x - y ) }^2}  \ln x + 
\frac{2 y [  x ( 25 - 11 y ) -y ( 11 + 3 y )  
     ]}{3 {( x - y ) }^2 {( y-1 ) }^2}   \ln y\nnu\\
 &+&\frac{4 ( 1 + y )}
    {( x-1 )  ( y-1 ) }  \Li{1 - \frac{1}{y}}+ 
    \frac{4 ( x + y ) }
    {( x-1 )  ( x-y ) }  \Li{1 - \frac{x}{y}},
\eea
\bea
f^{(1)}_4 (x,y) &=&
  \frac{59 x ( 1 - y )  - y( 59 - 3 y )}
     {6 ( y-1 )  ( x-y ) } +
     \frac{4 x ( 7 x^2 - 3 x y + 3 y^2 )}
     {3 ( x-1 )  {( x - y ) }^2}   \ln x +
  2 \ln^2 y\nnu\\
  &+&\frac{4 y^2 [x (18-11 y)-y(11-4 y)]}{3 
      {( x - y ) }^2 {( y-1 ) }^2}   \ln y\nnu\\
  &+& \frac{4 ( 1 + y^2 )  }
    {( x-1 )  ( y-1 ) } \Li{1 - \frac{1}{y}}+ 
   \frac{4 ( x^2 + y^2 )  }
    {( x-1 )  ( x - y ) } \Li{1 - \frac{x}{y}},
\eea
\bea
f_5^{(1)} (x,y)&=&
  -\frac{83 + 27 x ( y-1 )  - 27 y}
    {6 ( x-1 )  ( y-1 ) } -\Bigg\{
   \frac{4 x [ 1+ x ( 12 + y )  - y - 6 x^2  ]
      }{3 {( x-1 ) }^2 ( x - y ) } \ln x\nnu\\
 &-& \frac{2 [ 1 + 6 x^2 ( y-1 )  -
        3 x^3 ( y-1 )  + x ( 3 y -4 )  ]
        }{3 {( x-1 ) }^2 ( x - y )
      ( y-1 ) } \ln^2 x \nnu\\
 &+&
   \frac{4 y [ 3 x^2 ( y-1 )  +
   x y ( 3 - 2 y ) + y^2 ( y-2 )
   ] }{3 ( x-1 )
   {( x - y ) }^2 ( y-1 ) } \Li{1 -
   \frac{x}{y}} \nnu\\
 &+&
   \frac{4 [ 1 - 3 x - x^2 ( 3 - 6 y )- x^3 
   ]}{3 ( x-1 )
   ( x - y )  ( y-1 ) } \Li{1 -
   \frac{1}{x}} + (x\leftrightarrow y)\Bigg\}\nnu\\
 &+&4 \ln x\left( 1 + x \frac{\partial}{\partial x}+
   y\frac{\partial}{\partial y}\right) f_4^{(0)} (x,y),
\eea
\be
f^{(1)}_6 (x)=  
   \frac{2 x (29+3 x)}{3(x - 1)^2} - \frac{2 x (25+7 x)}{3(x - 1)^3}
      \ln x 
   -\frac{8 x}{(x - 1)^2} \Li{1-\frac1x} ,
\ee
\bea
  f_{7}^{(1)}(x) &=& 
    \frac{4 x [27 -11 x + (x-1)^2 \pi^2]}{3(x - 1)^2} 
   -\frac{4 x (37 - 33 x + 12 x^2)}{3(x - 1)^3} \ln x \nnu\\
  &-&\frac{8 x (2 - 2 x + x^2)}{(x - 1)^2} \Li{1 - \frac1x},
\eea
\bea
f^{(1)}_8 (x,y,z) &=&
  -\frac{28 y^2}
    {3 ( x - y )  ( y-1 )  ( y - z ) }
    + \Bigg[\frac{4 x ( 7 x^2 - 3 x y + 3 y^2 ) }
    {3 ( x-1)  {( x - y ) }^2 
      ( x - z ) }  \ln x + (x\leftrightarrow z)\Bigg]
\nnu\\
  &-& \frac{4 y^2 
      \{ x [4 y^2 + 18 z - 11 y (1+z) + y [3 y^2 -11 z + 4 y (1+z)] 
        \} }{3 {( x - y ) }^2 
      {( y-1 ) }^2 {( y - z ) }^2}  \ln y
    \nnu\\
  &-& 
   \frac{4 ( 1 + y^2 )}
    {( x-1 )  ( y-1 )  ( z-1
      ) }   \Li{1 - \frac{1}{y}}\nnu\\
  &+&
    \Bigg[\frac{4 ( x^2 + y^2 )  
      }{( x-1 )  
      ( x - y )  ( x - z ) } \Li{1 -
      \frac{x}{y}}+  (x\leftrightarrow z)\Bigg],
\eea
\bea
f^{(1)}_{9} (x,y,z) &=&
  -\frac{28 y}{3 ( x - y )  ( y-1 )  
      ( y - z ) } + \Bigg[\frac{2 x ( 11 x + 3 y )  
      }{3 ( x-1 )  {( x - y ) }^2 
      ( x - z ) } \ln x+ (x\leftrightarrow
      z)\Bigg] \nnu \\
  &+& \frac{2 y 
      \{ 
 x [3 y^2-25 z + 11 y (1+x)  ] + y [11 z - 17 y^2 + 3 y (1+z)]            
        \}}{3 {( x - y ) }^2 
      {( y-1 ) }^2 {( y - z ) }^2} \ln y\nnu\\
  &-& 
   \frac{4 ( 1 + y ) }
    {( x-1 )  ( y-1 )  ( z-1 ) }
    \Li{1 - \frac{1}{y}}\nnu\\
  &+&
    \Bigg[\frac{4 ( x + y )  }
    {( x-1 )  ( x - y )  ( x - z ) }
    \Li{1 - \frac{x}{y}} + (x\leftrightarrow
      z)\Bigg],
\eea
\be
f^{(1)}_{10} (x) =  
   {\frac{4 x( 19 -3 x )}{3 {( x-1 ) }^2} - 
      \frac{4 x (17 - x )  }
       {3 {( x-1 ) }^3} \ln x- 
      \frac{8 x}
       {{( x-1 ) }^2}} \Li{1 - \frac{1}{x}},
\ee
\bea
  f_{11}^{(1)}(x,y) &=&
    \frac{4 x [8 y + (x-1)(x-y)\pi^2]}{3y(x - 1)(x - y)} 
  -\frac{8x[x^2-7 y + 3 x (1+y)]}{3(x - y)^2(x - 1)^2} \ln x \nnu\\
   &-&\frac{8 x (3 x - 7 y)}{3(x - y)^2(y - 1)} \ln y
   -\frac{8 x}{y - 1} \Li{1 - \frac1x} 
  +\frac{8 x}{y(y - 1)} \Li{1 - \frac{y}{x}},
\eea
\bea
f^{(1)}_{12} (x,y,z) &=&
  -\frac{28 y^2}
   {3 ( x - y )  ( y-1 )  
   ( y - z ) } + 
   \Bigg[\frac{4 x^2 ( 6 x + y ) }
    {3 ( x-1 )  {( x - y ) }^2 
         ( x - z )}  \ln x + (x\leftrightarrow
           z)\Bigg]\nnu\\
  &-&\frac{4y^2 \{ 
 x [ 6 y^2 +20 z - 13 y (1+z)] + y [y^2-13 z + 6 y (1+z) ]\}}{3 {( x - y ) }^2 
         {( y-1 ) }^2 {( y - z ) }^2}  \ln y,
\eea
\bea
f^{(1)}_{13} (x,y,z) &=&
 -\frac{28 y}
       {3 ( x - y )  ( y-1 )  
         ( y - z ) } + \Bigg[
      \frac{4 x ( 6 x + y ) }
       {3 ( x-1 )  {( x - y ) }^2 
         ( x - z ) }  \ln x + (x\leftrightarrow
           z)\Bigg]\nnu\\ 
   &+&   \frac{4 y \{ x [y^2-13 z + 6 y (1+z) ] + y [y-8 y^2+6 z + y z ] 
           \} }{3 {( x - y ) }^2 
         {( y-1 ) }^2 {( y - z ) }^2}  \ln y,
\eea
\bea
f^{(1)}_{14} (x,y) &=&
\frac{32 x^2}{3 ( x-1 )  ( x - y ) } - 
      \frac{8 x^2 [7 x ( 1 + y ) - 11 y - 3 x^2 
           ]}{3 {( x-1 ) }^2 
         {( x - y ) }^2}   \ln x \nnu\\
 &-&  \frac{8 x y (3 x - 7 y )  }
       {3 {( x - y ) }^2 ( y-1) } \ln y
 -  \frac{8 x }{y-1} \Li{1 - \frac{1}{x}}+ 
      \frac{8 x }{y-1} \Li{1 - \frac{y}{x}},
\eea
\bea
f^{(1)}_{15} (x) &=& \frac{1-3 x}{x-1}+\frac{2 x}{(x-1)^2} \ln x +
\frac{2 x}{(x-1)} \Li{1-\frac{1}{x}},
\eea
\bea
f_{16}^{(1)}(x) &=& 
    \frac{28}{3 (x - 1)} - \frac{4 x (13 -6 x)}{3 (x - 1)^2} \ln x,
\eea
\bea
f_{17}^{(1)}(x,y) &=&
  -\frac{28}{3 ( x-1 )  ( y-1 ) } + 
  \frac{4  y ( 10 - 3  y )}
       {3 (x-y ) {( y-1 ) }^2} \ln y - 
   \frac{4  y }{( x-y )( y-1 )^2} {\ln^2 y} \nnu\\
 &+&\Bigg[ \frac{4 ( 13 x - 6 x^2 - 3 y - 7 x  y + 3 x^2  y ) }
    {3 {( x-1) }^2  (x-y )( y-1 ) } + 
      \frac{4  y \ln y}{( x-y )( y-1)^2}\Bigg]\ln x ,
\eea 
\bea
f_{18}^{(1)}(x,y) &=&
  -\frac{28  y}{3 ( x - y )(y-1)} + 
  \frac{4 x (6 x+y) }{3 (x-1)(x-y)^2} \ln x - 
  \frac{4  y [y( 6 +  y ) - x (13 - 6  y )]}
  {3 (x-y )^2 (y-1)^2}\ln y, \nnu\\
\eea
\bea
  f_{19}^{(1)}(x,y) &=&
  -\frac{28 [x (y - 1) + y]}{3(x - y)(y - 1)} + 
  \frac{4 x^2 (6 x + y)}{3(x - 1)(x - y)^2} \ln x
+\frac{4 y^2 [x (20 - 13 y) -y (13 -6 y)]}{3(x - y)^2(y - 1)^2} \ln y.\nnu\\
\eea

%
\end{document}